\documentclass[prx,twocolumn, amsmath,amssymb,superscriptaddress]{revtex4}
\usepackage{graphicx}
\usepackage{amssymb}
\usepackage{epstopdf}
\usepackage{makecell}
\usepackage{bbold}
\usepackage{soul}
\usepackage[dvipsnames]{xcolor}

\DeclareMathAlphabet\mathbfcal{OMS}{cmsy}{b}{n}
\usepackage{accents}
\DeclareRobustCommand{\triplecontract}{%
  \mathrel{\vbox{\baselineskip.65ex\lineskiplimit0pt\hbox{.}\hbox{.}\hbox{.}}}%
}
\newlength{\dhatheight}
\newcommand{\doublehat}[1]{%
    \settoheight{\dhatheight}{\ensuremath{\hat{#1}}}%
    \addtolength{\dhatheight}{-0.35ex}%
    \hat{\vphantom{\rule{1pt}{\dhatheight}}%
    \smash{\hat{#1}}}}
\newcommand{\dbar}{{\rm d}\hspace*{-0.08em}\bar{}\hspace*{0.1em}}

\begin{document}
\title{Theory of nematic and polar active fluid surfaces}
\author{Guillaume Salbreux}
\email{guillaume.salbreux@unige.ch}
\affiliation{University of Geneva, Quai Ernest Ansermet 30, 1205 Gen\`eve, Switzerland}
\affiliation{The Francis Crick Institute, 1 Midland Road, NW1 1AT, United Kingdom}
\author{Frank J\"ulicher}
\affiliation{Max Planck Institute for the Physics of Complex Systems, N\"othnitzer Str. 38, 01187 Dresden, Germany}
\author{Jacques Prost}
\affiliation{Institut Curie, PSL University 26 rue d’Ulm F-75248 Paris Cedex 05, France}
\affiliation{Mechanobiology Institute, NUS, 5A Engineering Dr 1, 117411, Singapore}
\author{Andrew Callan-Jones}
\affiliation{Laboratoire de Mati\`ere et Syst\`emes Complexes, Universit\'e de Paris, France}
\date{\today}

\begin{abstract}
We derive a fully covariant theory of the hydrodynamics of nematic and polar active surfaces, subjected to internal and external forces and torques. We study the symmetries of polar and nematic surfaces and find that in addition to 5 different types of in-plane isotropic surfaces, polar and nematic surfaces can be classified into 5 polar, 2 pseudopolar, 5 nematic and 2 pseudonematic types of surfaces. We give examples of physical realisations of the different types of surfaces we have identified. We obtain expressions for the equilibrium tensions, moments, and external forces and torques acting on a passive polar or nematic surface. We calculate the entropy production rate using the framework of thermodynamics close to equilibrium and find constitutive equations for polar and nematic active surfaces with different symmetries. We study the instabilities of a confined flat planar-chiral polar active layer and of a confined deformable polar active surface with broken up-down symmetry. 
\end{abstract}
\maketitle

\begin{figure}
 \centering
 \includegraphics[width=1.0\columnwidth]{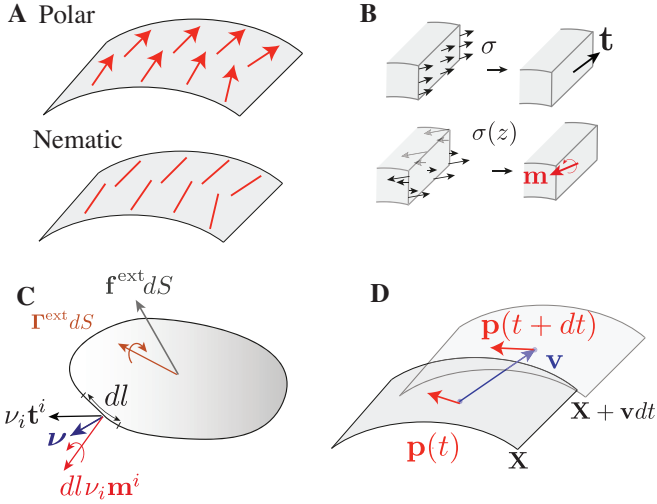}
 \caption{
 \label{fig:schematic}{\bf A}. In this manuscript, we consider active surfaces with a tangent polar or nematic order. {\bf B}. The distribution of stresses within a thin layer gives rise to tensions and torques acting within the surface, when integrated across the thickness of the layer. {\bf C}. Internal and external forces and torques act on a surface element with surface area $dS$. {\bf D}. The order parameter changes in time and is transported by the flow. Lagrangian time derivatives are taken by comparing the order parameter at two times $t$ and $t+dt$, following the flow of the surface.
 } 
\end{figure}

\section{Introduction}
Living systems contain a rich repertoire of dazzling and seemingly choreographed surface movements.  These underlie vital biological functions, ranging from the subcellular up to the organ scale.  Biological surfaces are frequently active, being driven by chemical reactions at the microscopic level.  They usually also possess internal degrees of freedom corresponding to in-plane order and that interplay with each other over large scales, giving rise to self-organized behavior.  For example, self-generated flow and nematic re-ordering in the actomyosin cortex drive cell surface deformations at the late stage of cell division~\cite{Reymann:2016aa}. At a larger scale, morphogenetic movements involve coherent flows in epithelia, thin tissues that grow and deform to give shape to organs.  The constituent cells often carry an in-plane polarity that dynamically couples to tissue flow to help establish patterns in the developing organism~\cite{Eaton:2011aa}. 

At the root of cell and tissue surface mechanics is the actomyosin cytoskeleton, a type of filamentous, soft active material~\cite{Marchetti_RMP}.
Theoretical descriptions of these active gels have been successfully applied to a number of problems involving three dimensional active flows and nematic or polar re-ordering~\cite{prost2015active}.  
Yet, surface movements pose special theoretical difficulties owing to the geometric nonlinearities inherent in them. Thus, transposing active gel physics from three to two dimensions is a challenging problem.  

A step in this direction was made in recent work 
establishing 
a theoretical framework for active isotropic surfaces~\cite{salbreux2009hydrodynamics}.  From the different types of broken symmetries compatible with isotropic surfaces, generic relations were derived linking the tensions and moments across a surface cut to variables such as the local curvature, velocity field, and chemical activity. This theory has been successfully used to describe mechano-chemical instabilities of isotropic active fluid surfaces \cite{mietke2019minimal, mietke2019self}. This and other continuum theories have also been used to address the deformations of epithelia, which have distinct apical and basal interfaces and can therefore be seen as surfaces with broken up-down symmetry \cite{messal2019tissue, haas2019nonlinear, morris2019active, al2021active}.  Despite these advances, understanding and systematically exploring how in-plane order couples to surface deformation and activity is beyond the reach of current theories. 

Numerous experimental observations have shown that broken symmetry variables play a key role in biological systems. Nematic order has been observed to emerge in the cell cortex \cite{Reymann:2016aa, spira2017cytokinesis}, but polar order could also exist. Long-range coherent patterns of cell planar polarity emerge during development \cite{goodrich2011principles, sagner2012establishment}. Epithelia cultured in vitro exhibit patterns of nematic order when the individual cells are elongated \cite{saw2018biological}. 
Recent experiments on reconstituted and in vivo systems have revealed how activity and topological defects in the surface nematic field can lead to dramatic surface shape changes~\cite{keber2014topology}. Coupling between nematic ordering and flows and deformations have been shown to play a key role in epithelial tissues \cite{blanch2021integer, blanch2021quantifying, doostmohammadi2021physics, guillamat2022integer} and can even explain regeneration in {\it Hydra}~\cite{Maroudas-Sacks:2021aa}.
Along these lines, coarse-grained simulations have recapitulated many of these findings and have shed some light on the basic physics~\cite{Alaimo:2017aa,Metselaar:2019aa}. 
Still, there is no comprehensive continuum framework yet that can be used to predict how in-plane polar or nematic order interacts with active surface deformation and flow.  We expect that an understanding of these effects will have broad relevance, including in living systems and in designing engineered materials \cite{manna2021harnessing}.  We propose here to develop such a framework. 

In this work we derive a covariant theory for polar and nematic surfaces which are driven out of equilibrium by chemical reactions. We classify these surfaces according to their symmetries, and find 7 polar surfaces and 7 nematic surfaces (Fig. \ref{fig:symmetries}, section \ref{subsection_symmetries}). We calculate equilibrium tensions and torques of a generic polar or nematic fluid surface, and obtain the entropy production rate close to equilibrium. We then obtain linear constitutive equations, out of equilibrium, for the different classes of polar and nematic surfaces (sections \ref{sec:polar_surfaces} and \ref{sec:nematic_surfaces}). We then discuss flows and deformation of a confined active polar film with broken planar-chiral or up-down symmetry (section \ref{sec:active_polar_film_confined}).

\section{Tensor variations}
We start by introducing notations and differential operators for vectors and tensors which will be convenient in the next sections. Notations for differential geometry are as in \cite{salbreux2017activesurfaces} and are summarized in Appendix \ref{appendix_diff_geom}. Briefly, we denote by latin indices $i,j$ the surface coordinates $s^1$, $s^2$. The position of the surface in three-dimensions is denoted by a vector $\mathbf{X}(s^1,s^2)$. The tangent vectors to the surface are denoted $\mathbf{e}_i=\partial_i \mathbf{X}$ and the unit normal vector $\mathbf{n} = \mathbf{e}_1 \times \mathbf{e}_2 /|\mathbf{e}_1 \times \mathbf{e}_2|$. The velocity field on the surface is denoted $\mathbf{v}$.
In the following we consider polar and nematic surfaces, whose local order is described respectively by a tangent polar vector $\mathbf{p}$ and a tangent symmetric traceless tensor $\mathbf{Q}$. To describe variations of vectors and tensors on the surface, we introduce a corotational operator for vectors and second-rank tensors varying together with the shape $\mathbf{X}$. For a given infinitesimal variation of the shape $d\mathbf{X}$, we denote
an infinitesimal rotation by $d\boldsymbol{\vartheta}=\frac{1}{2}\nabla\times d\mathbf{X}$ (Eq. \ref{def_rotation_rate_appendix}). Similarly, we introduce the vorticity $\boldsymbol{\omega}=\frac{1}{2}\nabla\times\mathbf{v}$ associated to the surface velocity field $\mathbf{v}$. The corotational variation of a vector $\mathbf{a}$ or second-rank tensor $\mathbf{B}$, not necessarily tangent to the surface, is then denoted with a capital $D$, and the corotational time derivative with a capital $D_t$:
\begin{align}
D\mathbf{a}&=d\mathbf{a}- d\boldsymbol{\vartheta}\times \mathbf{a}\label{def_corot_variation_p}~,\\
D_t \mathbf{a} &= \frac{{\rm d}\mathbf{a}}{{\rm d}t}-\boldsymbol{\omega}\times \mathbf{a} \label{def_corot_derivative_p}~,\\
D\mathbf{B}&=d\mathbf{B}- d\boldsymbol{\vartheta}\times_1 \mathbf{B}- d\boldsymbol{\vartheta}\times_2 \mathbf{B}\label{def_corot_variation_Q}~,\\
 D_t \mathbf{B} &= \frac{{\rm d}\mathbf{B}}{{\rm d}t}-\boldsymbol{\omega}\times_1 \mathbf{B}-\boldsymbol{\omega}\times_2 \mathbf{B} ~,\label{def_corot_derivative_Q}
\end{align} 
where the notation $\times_1$ and $\times_2$ indicates vector products with the first and second members, respectively, of the dyad forming $\mathbf{B}$ (Eq. \ref{eq:cross_product_tensor_notation}). The corotational operator $D$ applied to a vector/tensor describes the vector/tensor variation that does not arise from rotation of the surface in which the vector/tensor is embedded. 
 In Eqs. \ref{def_corot_derivative_p} and \ref{def_corot_derivative_Q}, the time derivatives of vectors $\mathbf{a}$ and second-rank tensors $\mathbf{B}$, $d\mathbf{a}/dt$ and $d \mathbf{B}/dt$, are Lagrangian time derivatives and are taken following the flow of material points on the surface (Fig. \ref{fig:schematic}D).  In the following, when considering time-dependent surface fields and surface deformations, we adopt an Eulerian perspective where the coordinates of the surface move along the normal to the surface, but do not vary with the tangential flow. In that case, for a vector field $\mathbf{a}(s^1,s^2,t)$ or a tensor field $\mathbf{B}(s^1,s^2,t)$ (Appendix \ref{appendix:subsec:corotational_time_derivatives}),
\begin{align}
\frac{{\rm d} \mathbf{a}}{{\rm d}t}&=\partial_t \mathbf{a} + v^i \partial_i \mathbf{a}\label{eq:advected_time_derivative_p}~,\\
\frac{{\rm d} \mathbf{B}}{{\rm d}t}&=\partial_t \mathbf{B} + v^i \partial_i \mathbf{B}\label{eq:advected_time_derivative_Q}~.
\end{align}
To keep notations compact, it is useful to introduce a covariant derivative operator which returns the components of the infinitesimal variations of a tangent vector or tensor. We therefore introduce the following notation for covariant variations of a  vector $\mathbf{a}$ and a second-rank tensor $\mathbf{B}$:
\begin{align}
\nabla a^i &= (d\mathbf{a})\cdot\mathbf{e}^i\label{def_covariant_diff_p}\\
 \nabla_j a^i &= \left(\partial_j \mathbf{a}\right)\cdot\mathbf{e}^i\\
  \nabla_t a^i &= \left(\frac{{\rm d} \mathbf{a}}{{\rm d}t}\right)\cdot\mathbf{e}^i\label{def_time_covariant_diff_p}\\
\nabla B^{ij} &= (d \mathbf{B}):(\mathbf{e}^i\otimes \mathbf{e}^j)\label{def_covariant_diff_Q}\\
\nabla_k B^{ij} &= (\partial_k \mathbf{B}):(\mathbf{e}^i\otimes \mathbf{e}^j)\\
\nabla_t B^{ij} &= \left(\frac{{\rm d} \mathbf{B}}{{\rm d}t}\right):(\mathbf{e}^i\otimes \mathbf{e}^j)\label{def_time_covariant_diff_Q}~,
\end{align}
which can be distinguished, e.g., from component variations such as $d a^{i}=d (\mathbf{a}
\cdot\mathbf{e}^i)$. The operator $\nabla_j$ is the standard covariant derivative with respect to the surface coordinates.  The operator $\nabla_t$ takes the Lagrangian time derivative of a vector or tensor and projects it on the surface tangent directions.

 The components of the corotational operator read for a vector $\mathbf{a}$ or a second-rank tensor $\mathbf{B}$:
\begin{align}
D a^i &=(D\mathbf{a})\cdot\mathbf{e}^i\label{Eq:corotational_variation_p}\\
 D_t a^i &=(D_t\mathbf{a})\cdot\mathbf{e}^i \label{Eq:corotational_time_derivative_p}\\
D B^{ij} &= (D\mathbf{B}):(\mathbf{e}^i\otimes\mathbf{e}^j )\label{Eq:corotational_variation_Q}\\
 D_t B^{ij}&= (D_t\mathbf{B}):(\mathbf{e}^i\otimes\mathbf{e}^j )\label{Eq:corotational_time_derivative_Q} ~.
\end{align}
The operators $\nabla$ and $D$ obey the product rule for derivatives; e.g. $D(a^i b^j)= (Da^i) b^j+ a^i (Db^j)$, and indices of components of the operators $\nabla$ and $D$ can be lowered or raised with the metric; for instance $D a_i =(D\mathbf{a})\cdot\mathbf{e}_i= g_{ij} D a^j$ (Appendix \ref{appendix_diff_geom}). These definitions can be generalized to higher-order tensors, using generalized cross-products between a vector and a tensor defined in Eq. \ref{eq:cross_product_tensor_notation_generalized}.

\section{Symmetries of polar and nematic surfaces}
\label{subsection_symmetries}

In this section we discuss symmetries of possible phases for nematic and polar surfaces. We present a classification of surfaces according to how a surface element transforms under mirror operations and rotations.  

\subsection{Symmetries, order parameters, and phases}
To discuss surface symmetries, we first introduce the signatures of scalars, vectors and tensors under a change of coordinates. We define the signature  $\epsilon_{A}^T$ of a tensorial quantity $A_{i_1...i_n}$ on the surface under a 3D linear transformation $\mathbf{T}$,  such that 
\begin{equation}
\label{def:tensorial_signature}
	A'_{i_1...i_n}=\epsilon_{A}^T T_{i_1}{}^{ k_1}...T_{i_n }{}^{k_n} A_{k_1...k_n}\,,
\end{equation}
i.e., $\epsilon_A^T=\pm 1$ is a factor that distinguishes a tensor from a pseudo-tensor. A true tensor has $\epsilon_A^T=1$ for all transformations $\{T\}$; a pseudo-tensor has $\epsilon_A^T=-1$ for a subset of transformations $\{T\}$. For a surface, the effect of a local change of coordinates can be decomposed into a tangential and a normal part:
\begin{align}
	\mathbf{T}^{\rm 3D}=\mathbf{T}^{\rm 2D}\mathbf{T}^n\,,
\end{align}
such that $\mathbf{T}^{\rm 2D}$ corresponds to a change of coordinates $(s^1, s^2)$ on the surface  and $\mathbf{T}^n$ is a change of coordinates perpendicular to the surface. Vectorial and tensorial quantities on the surface have their components transformed by $\mathbf{T}^{\rm 2D}$ only.
\begin{figure}
	\centering 
	\includegraphics[width=1.0\columnwidth]{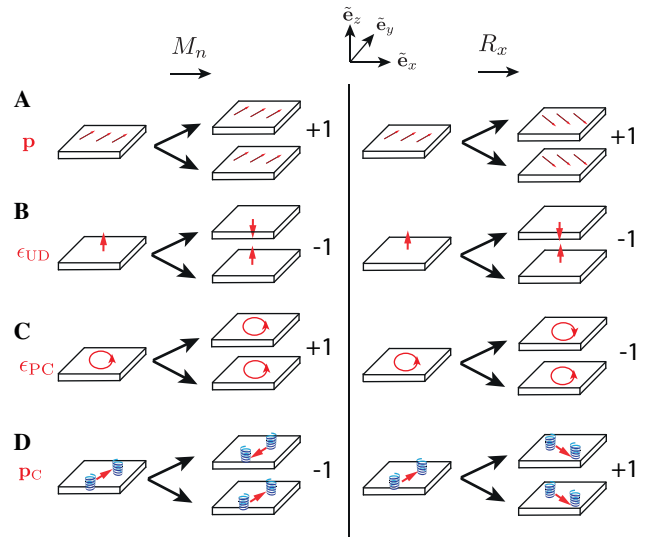}
	\caption{
		\label{fig:signatures}Examples of signatures of scalar and vectorial order parameters under linear transformations (Eq. \ref{def:tensorial_signature}). As an example we consider two transformations, a mirror operation with respect to the tangent plane $M_n$ and a rotation by $\pi$ around the $x$ axis, $R_x$ (Fig. \ref{fig:symmetries}A). For each transformation, the top schematic indicates how the surface as a whole is modified by the transformation, and the order parameter is redefined accordingly, while the lower schematic shows the order parameter modified by the tangential part of the transformation. The signature ($+1$ or $-1$ in the right-hand side) then indicates whether the two redefined order parameters are the same (signature $+1$) or not (signature $-1$). {\bf A}. A true tangent polar vector has signature $+1$ under all transformations. {\bf B}. The up-down pseudoscalar $\epsilon_{\rm UD}$ has signature $-1$ under $M_n$ and $R_x$. {\bf C}. $\epsilon_{\rm PC}$ corresponds to a direction of rotation on the surface, and is invariant under $M_n$ but ahs signature $-1$ under $R_x$. {\bf D}. The pseudopolar order parameter $\mathbf{p}_{\rm C}$ has signature $-1$ under a mirror symmetry, but signature $+1$ under $R_x$.} 
\end{figure}

The introduction of signatures of order parameters under linear transformations is helpful for the following reason. Scalar, tangent vectorial and tangent tensorial order parameters can be introduced to characterise the statistical distribution and symmetry properties of molecules within the surface. Under 3D spatial linear transformations, molecules within a surface element are modified and change their position and orientation. Order parameters can then be recalculated to characterize this transformed state, according to Eq.~\ref{def:tensorial_signature}. The recalculated order parameters may not simply correspond to a transformation of the original order parameters by the tangential part of the linear transformation  (Fig. \ref{fig:signatures}); i.e. order parameters do not necessarily behave as simple scalar, vectorial and tensorial quantities on the surface. Instead, they may acquire an additional sign change under a subset of transformations. The corresponding set of signatures allow to classify order parameters, as further described below.

\subsection{Isotropic surfaces}

\begin{figure*}
	\centering
	\includegraphics[width=1.0\textwidth]{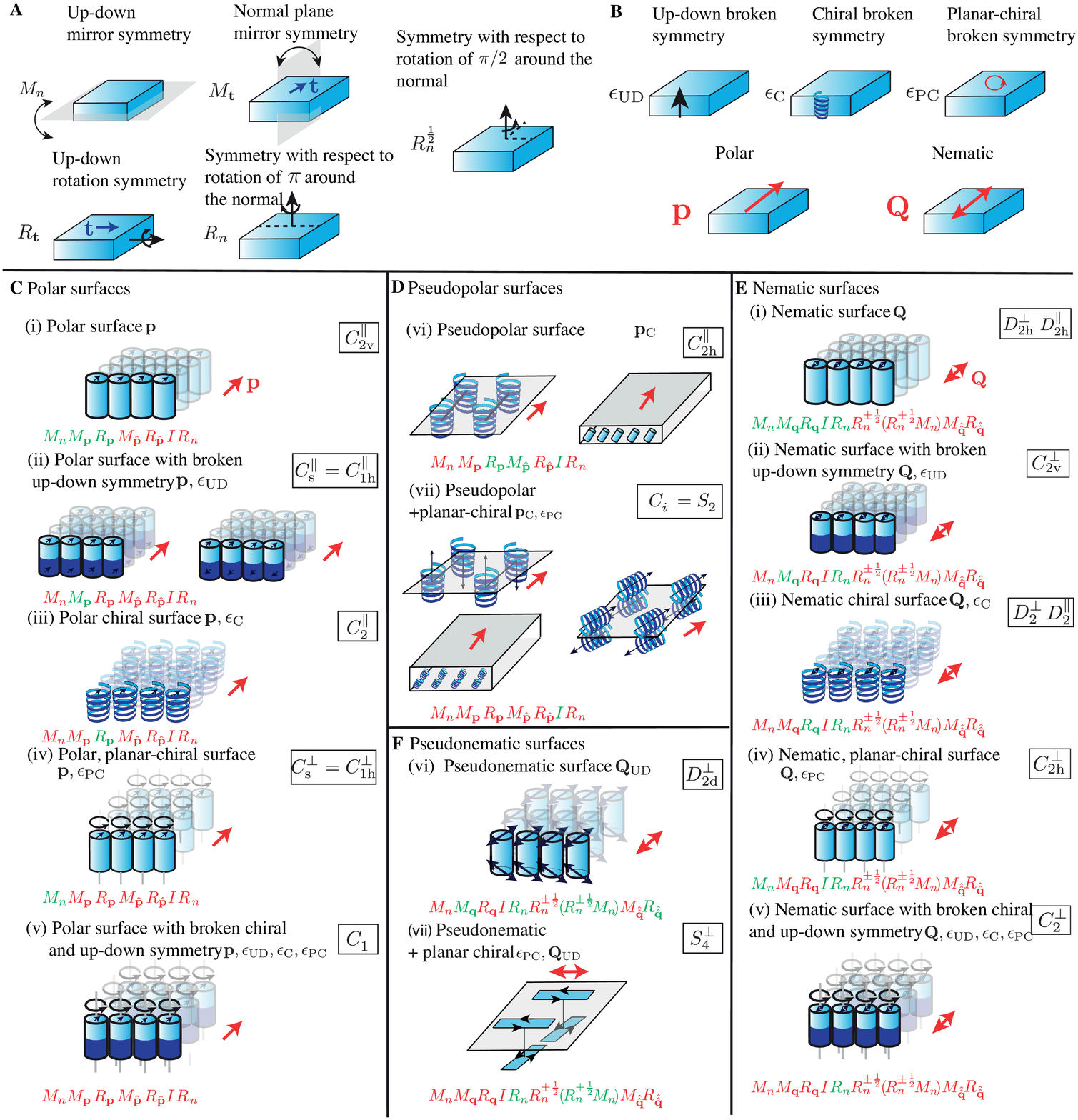}
	\caption{
		\label{fig:symmetries}Classification of polar and nematic surfaces. {\bf A}. We consider changes in the surface state under up-down mirror symmetry $M_n$,  mirror symmetry with respect to a plane perpendicular to the surface $M_t$, up-down rotation symmetry of angle $\pi$ $R_t$,  rotation by $\pi$ around the normal $R_n$, and rotation by $\pi/2$ around the normal $R_n^{\frac{1}{2}}$. Combinations of $R_n^{\frac{1}{2}}$ with other symmetries are considered but not represented. {\bf B}. Order parameters. We consider combinations of 3 pseudoscalar order parameters with polar or nematic order parameters tangent to the surface. {\bf C}, {\bf D}. Polar surfaces can be classified in 7 types according to their symmetries : 5 polar surfaces and 2 pseudopolar surfaces. {\bf E}, {\bf F}. Nematic surfaces can be categorised in 7 types according to their symmetries: 5 nematic surfaces and 2 pseudonematic surfaces. In panels C-F, schematics give examples of actual surfaces belonging to each category. Red and green letters indicate respectively broken and preserved symmetries. The Schoenflies notation for the surface symmetry is given for each surface type. When relevant, we add a $\parallel$ or $\perp$ superscript to the standard Schoenflies notation to indicate if the rotation axis, (or, for $C_{\rm s}=C_{\rm 1h}$, the normal to the mirror plane of symmetry, or for $D_{\rm 2d}$ and $S_4$, the axis of rotation-reflection), is tangent or perpendicular to the surface.} 
\end{figure*}

We first briefly discuss isotropic surfaces, which were considered in Ref. \cite{salbreux2017activesurfaces}.  For such a surface broken symmetries are described by pseudoscalar fields. To classify possible pseudoscalar fields, we consider their signatures under the following set of transformations (Fig. \ref{fig:symmetries}): $M_n$ (reflection operation with respect to a plane tangent to the surface), $M_t$ (reflection operation with respect to a plane normal to the surface and containing the tangent vector $\mathbf{t}$), $R_t$ (rotation by an angle $\pi$ around a tangent vector $\mathbf{t}$), $R_n$ (rotation by an angle $\pi$ around the normal $\mathbf{n}$), and $I$  (full inversion of space).  These transformations verify the composition relations $M_n R_n = I$, $M_t R_t = M_n$ and $M_n R_t=M_t$. As a result, a pseudoscalar field must break none, or at least two, of the symmetries within each set $\{M_n, R_n, I\}$ and $\{M_t, R_t, M_n\}$. For instance, a field that breaks the $M_t$ symmetry, but is invariant under $R_t$, also breaks the mirror symmetry $M_n$. Furthermore, a surface scalar field must be invariant under $R_n$. Taking these rules into account, one can define three distinct types of scalar fields for a surface that breaks various symmetries \cite{salbreux2017activesurfaces}. The signatures of these three pseudoscalars, denoted $\epsilon_{\rm UD}$ (for up-down broken symmetry), $\epsilon_{\rm C}$ (for chiral broken symmetry) and $\epsilon_{\rm PC}$ (for planar-chiral broken symmetry) are given in Table \ref{table_signatures}.
\begin{table}
\begin{center}
	\begin{tabular}{|c|c|c|c|c|c|}\hline
		& $M_n$ & $M_t$ & $R_t$  &$I$&$R_n$\\\hline
		$\epsilon_{\rm UD}$& -1 & 1 & -1 &-1&1\\\hline
		$\epsilon_{\rm C}$& -1 &- 1 & 1 &-1&1\\\hline
		$\epsilon_{\rm PC}$& 1 & -1 & -1 &1&1\\\hline
	\end{tabular} 
\end{center}
\caption{\label{table_signatures} Table of signatures of surface pseudoscalars.}
\end{table}
These pseudoscalars are order parameters that distinguish different types, or phases, of isotropic surface. We note that the 2D Levi-Civita tensor has the same signature as $\epsilon_{\rm PC}$, corresponding to a planar-chiral surface, while the 3D Levi-Civita tensor has the same signature as $\epsilon_{\rm C}$, corresponding to a chiral surface.

By considering how pseudoscalar order parameters can combine, we find 5 phases of isotropic surfaces which are listed in Table \ref{isotropic_surfaces_table}. There, we also give the corresponding Schoenflies notation \cite{ashcroft1976solid} which describes the symmetry properties of the corresponding phase, as well as the orientation of the rotation axis relative to the tangent plane of the surface.
\begin{table*}
\begin{center}
	\begin{tabular}{|c|c|c|c|c|c|c|c|c|c|}\hline
		\makecell{Schoenflies \\notation}&\makecell{Rotation\\ axis }& \makecell{Order\\ parameters }&$M_n$ & $M_{t}$ & $R_{t}$ &$I$&$R_n$\\\hline
		$D_{\infty \rm h}$ &$\perp$&$\emptyset$& 1 & 1 & 1 &1&1 \\\hline
		$C_{\infty\rm v} $&$\perp$&$\epsilon_{\rm UD}$& X & 1 & X &X&1\\\hline
		$D_{\infty} $&$\perp$ &$\epsilon_{\rm C}$& X &X & 1&X&1 \\\hline
		$C_{\infty\rm  h} $&$\perp$&$\epsilon_{\rm PC}$& 1 & X & X &1&1\\\hline
		$ C_{\infty}$ &$\perp$&$\{\epsilon_{\rm UD}$, $\epsilon_{\rm C}$, $\epsilon_{\rm PC}\}$& X& X & X &X&1\\\hline
	\end{tabular} 
\end{center}
\caption{\label{isotropic_surfaces_table}Schoenflies notation, order parameters, and conserved and broken symmetries for the 5 types of isotropic surfaces. An ``X'' in the Table stands for a broken symmetry and a ``1'' for a conserved symmetry. Here, a broken symmetry is a transformation which changes the order parameter, which does not imply a signature $-1$ of the order parameter for the transformation. The second column refers to the orientation, with respect to the surface tangent plane, of the main rotation axis (here leaving the system invariant by any angle) associated to the symmetry group described by the Schoenflies notation.}
\end{table*}
More complex surface phases can be characterized by (pseudo-) vectorial and (pseudo-) tensorial order parameters, as described further below.

\subsection{Polar and pseudopolar surfaces}
We now consider polar surfaces, that is, surfaces with a broken rotational symmetry characterized by a tangent vector, $\mathbf{p}$. Such a vector transforms under a change of coordinates $\mathbf{T}^{\rm 3D}=\mathbf{T}^{\rm 2D}\otimes \mathbf{T}^n$ as:
\begin{align}
p'_i= \epsilon_{\mathbf{p}}^T {T^{\rm 2D}}_i{}^j p_j~,
\end{align}
where $\epsilon_{\mathbf{p}}^T  $ is the vector signature. If $\mathbf{p}$ is a true tangent vector, it has signature $\epsilon_{\mathbf{p}}^T=1$ for all types of transformations. In addition, one can define three pseudovectors $\mathbf{p}_{\rm UD}$, $\mathbf{p}_{\rm C}$, $\mathbf{p}_{\rm PC}$, which have a $-1$ signature under at least one of the transformations $M_n$, $M_t$, $R_t$, $R_n$, $I$. The corresponding signatures of vectors and pseudo vectors are listed in Table \ref{polar_signatures}. Subscripts of the pseudovectors $\mathbf{p}_{\rm UD}$, $\mathbf{p}_{\rm C}$, $\mathbf{p}_{\rm PC}$ are given in analogy between their signatures in Table \ref{polar_signatures} and signature of the pseudoscalars $\epsilon_{\rm UD}$, $\epsilon_{\rm C}$ and $\epsilon_{\rm PC}$ in Table \ref{table_signatures}.
\begin{table}
\begin{center}
	\begin{tabular}{|c|c|c|c|c|c|}\hline
		& $M_n$ & $M_t$ & $R_t$  &$I$&$R_n$\\\hline
		$\mathbf{p}$& 1 & 1 & 1 &1&1\\\hline
		$\mathbf{p}_{\rm UD}$& -1 & 1 & -1 &-1&1\\\hline
		$\mathbf{p}_{\rm C}$& -1 &- 1 & 1 &-1&1\\\hline
		$\mathbf{p}_{\rm PC}$& 1 & -1 & -1 &1&1\\\hline
	\end{tabular} 
\end{center}
\caption{\label{polar_signatures}Signatures of polar and pseudopolar order parameters.}
\end{table}

We note that a rotation of $\mathbf{p}$ by $\pm\pi/2$ about the unit normal vector to the surface $\mathbf{n}$, that is, $\hat{\mathbf{p}}=\mathbf{n} \times \mathbf{p}$ or $-\hat{\mathbf{p}}=-\mathbf{n}  \times \mathbf{p}$, defines an alternative order parameter.  This corresponds to a different convention to describe the ordering state of the surface, which does not change the physics. In component notation this is equivalent to defining $\hat{p}^i = -\epsilon^i{}_j p^j$ or $-\hat{p}^i = \epsilon^i{}_j p^j$. Here,
	$\epsilon_{ij}$ is the Levi-Civita tensor (Eq.~\ref{DefinitionEpsilonij}) and has the same signatures under the transformations $M_n$, $M_t$, $R_t$, $R_n$ as $\epsilon_{\rm PC}$.  For this reason $\mathbf{p}_{\rm PC}$ does not describe a different phase than a vector $\mathbf{p}$, as $\hat{\mathbf{p}}_{\rm PC}=\mathbf{n} \times \mathbf{ p}_{\rm PC}$ has the same signatures as a true vector $\mathbf{p}$.  Similarly, the pseudovectors $\mathbf{p}_{\rm UD}$ and $\mathbf{p}_{\rm C}$ describe the same phase, as $\hat{\mathbf{p}}_{\rm UD} = \mathbf{n}\times \mathbf{p}_{\rm UD} $ has the same signatures as the pseudovector $\mathbf{p}_{\rm C}$. Overall, there are therefore only two distinct phases characterized only by a vectorial order parameter: a simple polar surface (e.g. with polar order $\mathbf{p}$) or a simple pseudopolar surface (e.g. with pseudopolar order $\mathbf{p}_{\rm C}$).
\begin{table*}
\begin{center}
	\begin{tabular}{|c|c|c|c|c|c|c|c|c|c|c|c|c|}\hline
		\makecell{Schoenflies \\notation}&\makecell{Rotation axis or\\mirror plane normal\\orientation }& \makecell{Order\\ parameters }&$M_n$ & $M_{\mathbf{p}}$ & $R_{\mathbf{p}}$ & $M_{\hat{\mathbf{p}}}$& $R_{\hat{\mathbf{p}}}$ &$I$&$R_n$&\makecell{number \\in Fig. \ref{fig:symmetries}C-D}\\\hline
		$C_{2\rm  v}$&$\parallel$&$\mathbf{p}$& 1 & 1 & 1 &X&X&X&X&(i) \\\hline
		$C_{\rm s} =C_{\rm 1h} $&$\parallel$&$\{\epsilon_{\rm UD}$, $\mathbf{p}\}$& X & 1 & X &X&X&X&X&(ii)\\\hline
		$C_{2\rm }$ &$\parallel$ &$\{\epsilon_{\rm C}$, $\mathbf{p}\}$& X &X & 1 &X&X&X&X&(iii) \\\hline
		$C_{\rm s} =C_{\rm 1h}$&$\perp$&$\{\epsilon_{\rm PC}$, $\mathbf{p}\}$& 1 & X & X &X&X&X&X&(iv)\\\hline
		$C_{\rm 1} $&&$\{\epsilon_{\rm UD}$, $\epsilon_{\rm C}$, $\epsilon_{\rm PC}$, $\mathbf{p}\}$& X& X & X&X&X &X&X&(v)\\\hline
		$C_{2\rm  h }$ &$\parallel$&$\mathbf{p}_{\rm C}$& X&X & 1 &1&X&1&X&(vi)\\\hline
		$C_{\rm i}=S_2$&& $\{\epsilon_{\rm PC}$, $\mathbf{p}_{\rm C}\}$ & X & X & X &X&X&1&X&(vii)\\\hline
	\end{tabular} 
\end{center}
\caption{\label{polar_surfaces_table}Schoenflies notation, order parameters, and conserved and broken symmetries for the 7 types of polar surfaces. When relevant, the second column indicates whether the axis of rotation in the Schoenflies notation (or, for $C_{\rm s}= C_{\rm 1h}$, the normal to the mirror plane of symmetry) is parallel ($\parallel$) or orthogonal ($\perp$) to the tangent plane of the surface. The first 5 types are polar surfaces, while the 2 last types are pseudopolar surfaces. In the table, a ``X'' stands for a broken symmetry and a ``1'' for a conserved symmetry, and $\hat{\mathbf{p}}=\mathbf{n}\times\mathbf{p}$ is a tangent vector orthogonal to $\mathbf{p}$. See Fig. \ref{fig:symmetries}C-D for examples of surfaces of different types.}
\end{table*}

Scalar and vectorial order parameters can also exist simultaneously in a surface, resulting in more symmetries being broken. To obtain a complete list of surface polar phases, we consider any combination of the vectorial order parameters $\mathbf{p}$, $\mathbf{p}_{\rm UD}$, $\mathbf{p}_{\rm C}$, $\mathbf{p}_{\rm PC}$ and the scalar order parameters $\epsilon_{\rm C}$, $\epsilon_{\rm PC}$, and $\epsilon_{\rm UD}$, but eliminate combinations which are redundant as they describe the same phase. For instance, the combination  $\{\epsilon_{\rm C}, \mathbf{p}_{\rm C}\}$ is redundant with the combination $\{\epsilon_{\rm C}, \mathbf{p}\}$, as $\epsilon_{\rm C} \mathbf{p}_{\rm C}$ has the same signature under symmetry operations as a true vector $\mathbf{p}$. Furthermore, those vectorial order parameters, which when rotated by $\pm \pi/2$ have the same symmetry properties as other vectorial order parameters, are redundant. For instance the combination of order parameters $\{\epsilon_{\rm C}, \mathbf{p}_{\rm UD}\}$ can be redefined as $\{\epsilon_{\rm C}, \hat{ \mathbf{p}}_{\rm UD}\}$ which has the same signatures as $\{\epsilon_{\rm C},\mathbf{p}_{\rm C}\}$. The set $\{\epsilon_{\rm C},\epsilon_{\rm C} \mathbf{p}_{\rm C}\}$ can be obtained from the order parameters $\{\epsilon_{\rm C},\mathbf{p}_{\rm C}\}$, and has the same symmetry properties as the polar chiral surface $\{\epsilon_{\rm C}, \mathbf{p}\}$. Therefore surfaces described by $\{\epsilon_{\rm C}, \mathbf{p}_{\rm UD}\}$ and $\{\epsilon_{\rm C}, \mathbf{p}\}$ are actually the same phase. 

Listing all resultant independent combinations, we find 7 distinct surface phases which are listed in Table \ref{polar_surfaces_table}, together with the Schoenflies notation \cite{ashcroft1976solid} which characterize the symmetry properties of the corresponding phases. There, we also indicate whether the order parameter is invariant under a subset of orthogonal transformations. In addition to the transformations $M_n$, $R_n$, $I$ introduced earlier, we consider the transformations $M_\mathbf{p}$, $M_{\hat{\mathbf{p}}}$ (mirror operations in a plane containing $\mathbf{n}$ and $\mathbf{p}$ and $\mathbf{n}$ and $\hat{\mathbf{p}}$, respectively)
 and $R_{\mathbf{p}}$, $R_{\hat{\mathbf{p}}}$ (rotation of $\pi$ around the axis $\mathbf{p}$ or $\hat{\mathbf{p}}$). These transformations satisfy the composition relations $ M_{\mathbf{p}}R_{\mathbf{p}}=M_{\hat{\mathbf{p}}}R_{\hat{\mathbf{p}}}  =M_n$, $  M_{\mathbf{p}}  M_{\hat{\mathbf{p}}}=R_{\mathbf{p}}  R_{\hat{\mathbf{p}}} =R_n$, $   M_{\mathbf{p}}  R_{\hat{\mathbf{p}}}= R_{\mathbf{p}}  M_{\hat{\mathbf{p}}} =I$. These relations constrain the set of possible conserved and broken symmetries for each surface phase, since invariance with respect to two transformations implies invariance with respect to their composition. For instance, all polar phases break the symmetry $R_n$; therefore none of the phases can be invariant under both $M_{\mathbf{p}} $ and $M_{\hat{\mathbf{p}}}$, or under both $R_{\mathbf{p}} $ and $R_{\hat{\mathbf{p}}}$. It is possible to find additional combinations of conserved and broken symmetries compatible with the multiplicative table of transformations discussed above, which are not listed in Table \ref{polar_surfaces_table}: indeed we find that for these cases, the redefinition of the polar order parameter $\hat{\mathbf{p}}=\mathbf{n}\times\mathbf{p}$ allows to recover one of the categories already listed in Table \ref{polar_surfaces_table} (Appendix \ref{appendix_order_parameters}).

Examples of the polar surface phases listed in Table \ref{polar_surfaces_table} are given in Fig. \ref{fig:symmetries}C-D. There, we denote a surface to be polar if its set of order parameters includes a true vector $\mathbf{p}$, and pseudopolar otherwise.

\subsection{Nematic and pseudonematic surfaces}
We now discuss the symmetries of nematic surfaces, with tangent nematic order. Following the same reasoning as for polar surfaces, we find that there are 5 nematic surfaces and 2 pseudonematic surfaces.  Nematic order is characterized by a traceless symmetric second-rank tangent tensor $\mathbf{Q}$ ($Q_k{}^k=0$ and $Q_{ij}=Q_{ji}$). Such a tensor transforms under a change of coordinates $\mathbf{T}^{\rm 3D}=\mathbf{T}^{\rm 2D}\otimes \mathbf{T}^n$ as:
\begin{align}
	Q'_{ij} =\epsilon_{\mathbf{Q}}^T {T^{\rm 2D}}_{i}{}^{k} {T^{\rm 2D}}_{j}{}^{l} Q _{kl}~,
\end{align}
where $\epsilon_{\mathbf{Q}}^T$ is the tensor signature. A true tangent tensor has signature $\epsilon_{\mathbf{Q}}^T=1$ under all transformations $\mathbf{T}^{\rm 3D}$. As for polar surfaces, three different types of pseudonematic tensors $\mathbf{Q}_{\rm UD}$, $\mathbf{Q}_{\rm C}$, $\mathbf{Q}_{\rm PC}$ can be defined, according to their signatures under reflections and rotations. The signatures of nematic and pseudonematic order parameters are given in Table \ref{nematic_signatures}.
\begin{table}
\begin{center}
	\begin{tabular}{|c|c|c|c|c|c|c|c|}\hline
		& $M_n$ & $M_t$ & $R_t$  &$I$&$R_n$\\\hline
		$\mathbf{Q}$& 1 & 1 & 1 &1&1\\\hline
		$\mathbf{Q}_{\rm UD}$& -1 & 1 & -1 &-1&1\\\hline
		$\mathbf{Q}_{\rm C}$& -1 &- 1 & 1 &-1&1\\\hline
		$\mathbf{Q}_{\rm PC}$& 1 & -1 & -1 &1&1\\\hline
	\end{tabular} 
\end{center}
\caption{\label{nematic_signatures}Signatures of nematic and pseudonematic order parameters.}
\end{table}
The transformation of a nematic tensor $\mathbf{Q}\rightarrow \doublehat{\mathbf{Q}}$, with $\doublehat{Q}_{ij}=-\epsilon_i{}^k Q_{kj}$ or $-\doublehat{Q}_{ij}=\epsilon_i{}^k Q_{kj}$, corresponds to a rotation by $\pm\pi/4$ around the normal to the surface of the eigenvectors of the order parameter. The tensors $\mathbf{Q}$ and $\doublehat{\mathbf{Q}}$ correspond to two different conventions to quantify nematic order, but describe the same nematic phase. As the 2D antisymmetric tensor $\epsilon_{ij}$ has the same signatures under transformations as $\epsilon_{\rm PC}$, not all pseudonematic tensors describe physically distinct phases. Specifically, $\mathbf{Q}_{\rm C}$ and $\mathbf{Q}_{\rm UD}$ describe surfaces with the same symmetry. Similarly, $\mathbf{Q}_{\rm PC}$ and $\mathbf{Q}$ also describe surfaces with the same symmetry. This can be seen from the fact, for instance, that the order parameter $\doublehat{\mathbf{Q}}_{\rm PC}$ is a true tensor with the same signatures under transformations as $\mathbf{Q}$ in Table \ref{nematic_signatures}. Overall, there are only two distinct phases characterized only by a nematic order parameter: a simple nematic surface (e.g. with nematic order $\mathbf{Q}$) or a simple pseudonematic surface (e.g. with nematic order $\mathbf{Q}_{\rm UD}$).

As for polar surfaces, to obtain the full set of nematic surfaces, we consider the combination of a nematic or pseudonematic order parameter $\mathbf{Q}$, $\mathbf{Q}_{\rm UD}$, $\mathbf{Q}_{\rm C}$, $\mathbf{Q}_{\rm PC}$ with the pseudoscalar order parameters $\epsilon_{\rm UD}$, $\epsilon_{\rm PC}$, $\epsilon_{\rm C}$, and eliminate redundant combinations. The combination of the nematic order parameter $\mathbf{Q}$ with the pseudoscalar order parameters $\epsilon_{\rm UD}$, $\epsilon_{\rm PC}$, $\epsilon_{\rm C}$, gives 5 different types of nematic surfaces. Further combinations of pseudoscalar and pseudonematic order parameters can also correspond to surfaces with the same symmetry using the transformation $\mathbf{Q}\rightarrow \doublehat{\mathbf{Q}}$: for instance the set $\{\epsilon_{\rm PC}$, $\mathbf{Q}_{\rm UD}\}$ describes a surface with the same symmetries as $\{\epsilon_{\rm PC}$, $\mathbf{Q}_{\rm C}\}$.

 Listing all combinations which are not redundant, we obtain a list given in Table \ref{nematic_surfaces_table}. For each order parameter, we indicate whether the order parameter is conserved under a subset of orthogonal transformations. Nematic order parameters are all invariant under the rotation by $\pi$ around the normal, $R_n$, and are all modified by the rotation of $\pm\pi/2$ around the normal, $R_n^{\pm \frac{1}{2}}$.
 In Table  \ref{nematic_surfaces_table}, pseudonematic surfaces are distinguished from nematic surfaces by their behaviour under the transformations $M_{\doublehat{\mathbf{q}}}$ (a mirror operation with respect to a plane perpendicular to the surface and containing $\doublehat{\mathbf{q}}$), $R_{\doublehat{\mathbf{q}}}$ (a rotation of $\pi$ around the axis $\doublehat{\mathbf{q}}$) and the composition $R_n^{\pm \frac{1}{2}} M_n$. Here, $\doublehat{\mathbf{q}}=\mathbf{R}_n^{\frac{1}{4}} \mathbf{q}$ is the vector obtained by rotation of $\pi/4$ around the normal to the surface $\mathbf{n}$, of the eigenvector of $\mathbf{Q}$ with positive eigenvalue, $\mathbf{q}$. These transformations satisfy the composition relations
 $M_{\doublehat{\mathbf{q}}}    R_{\doublehat{\mathbf{q}}} =  R_{\doublehat{\mathbf{q}}}  M_{\doublehat{\mathbf{q}}} =M_n$, $ M_{\doublehat{\mathbf{q}}} M_{\mathbf{q}} =     R_{\doublehat{\mathbf{q}}} R_{\mathbf{q}}= R_n^{\frac{1}{2}}$, $  M_{\mathbf{q}} M_{\doublehat{\mathbf{q}}} =R_{\mathbf{q}}   R_{\doublehat{\mathbf{q}}} = R_n^{-\frac{1}{2}}$, $M_{\doublehat{\mathbf{q}}} R_{\mathbf{q}} =R_{\doublehat{\mathbf{q}}} M_{\mathbf{q}} =R_n^{\frac{1}{2}} M_n$ and $
M_{\mathbf{q}} R_{\doublehat{\mathbf{q}}} =R_{\mathbf{q}} M_{\doublehat{\mathbf{q}}} =R_n^{-\frac{1}{2}} M_n$. As for polar surfaces, these multiplication relations restrict the set of possible conserved and broken symmetries for each surface phase, as invariance with respect to two transformations implies invariance with respect to their composition. As for polar surfaces, additional combinations of broken and conserved symmetries not listed in Table \ref{nematic_surfaces_table} are possible; we find that these combinations arise from the redefinition $\doublehat{Q}_{ij}=-\epsilon_i{}^k Q_{kj}$ of one of the already listed surface phases (Appendix \ref{appendix_order_parameters}).

 Examples of the nematic surface phases listed in Table \ref{nematic_surfaces_table} are given in Fig. \ref{fig:symmetries}E-F. There, we denote a surface to be nematic if its set of order parameter includes a true tensor $\mathbf{Q}$, and pseudonematic otherwise.
\begin{table*}
\begin{center}
	\begin{tabular}{|c|c|c|c|c|c|c|c|c|c|c|c|c|c|c|c|}\hline
		\makecell{Schoenflies \\notation}&\makecell{ Rotation or \\rotation-reflection\\ axis }& \makecell{Order\\ parameters }&$M_n$ & \makecell{$M_{\mathbf{q}}$\\ or  $M_{\hat{\mathbf{q}}} $}&\makecell{$R_{\mathbf{q}}$\\ or$R_{\hat{\mathbf{q}}}$} &$I$&$R_n$  & $R_n^{\pm\frac{1}{2}}$ &$R_n^{\pm\frac{1}{2}} M_n$ & $M_{\doublehat{\mathbf{q}}}$ & $ R_{\doublehat{\mathbf{q}}}$& \makecell{number \\in Fig. \ref{fig:symmetries}E-F}\\\hline
		\makecell{$D_{\rm 2h}$\\ $D_{\rm 2 h}$}&\makecell{$\perp$ \\$\parallel$}&$\mathbf{Q}$& 1 & 1 & 1 &1&1&X&X&X&X&(i) \\\hline
		$C_{\rm 2v} $&$\perp$&$\{\epsilon_{\rm UD}$, $\mathbf{Q}\}$& X & 1 & X &X&1&X&X&X&X&(ii)\\\hline
		\makecell{$D_{\rm 2} $\\$D_{2}$}&\makecell{$\perp$ \\$\parallel$} &$\{\epsilon_{\rm C}$, $\mathbf{Q}\}$& X &X & 1 &X&1&X&X&X&X&(iii) \\\hline
		$C_{\rm 2h} $&$\perp$&$\{\epsilon_{\rm PC}$, $\mathbf{Q}\}$& 1 & X & X &1&1&X&X&X&X&(iv)\\\hline
		$C_{\rm 2} $&$\perp$&$\{\epsilon_{\rm UD}$, $\epsilon_{\rm C}$, $\epsilon_{\rm PC}$, $\mathbf{Q}\}$& X& X & X &X&1&X&X&X&X&(v)\\\hline
		$D_{\rm 2d}$ &$\perp$&$\mathbf{Q}_{\rm UD}$& X&1 & X &X&1&X&1&X&1&(vi)\\\hline
		 $S_4$& $\perp$&$\{\epsilon_{\rm PC}$, $\mathbf{Q}_{\rm UD}\}$&X&X&X&X&1&X&1&X&X&(vii)\\\hline
	\end{tabular} 
\end{center}
\caption{\label{nematic_surfaces_table}
	Schoenflies notation, order parameters, and conserved and broken symmetries for the 7 types of nematic surfaces. An ``X'' stands for a broken symmetry and a 1 for a conserved symmetry. $\mathbf{q}$ denotes a vector going along the eigenvector of $\mathbf{Q}$ with the largest eigenvalue, $\hat{\mathbf{q}}$ its rotation by $\pi/2$ around the
	normal, $\hat{\hat{\mathbf{q}}}$ its rotation by $\pi/4$ around the normal. When relevant, the second column indicates whether the axis of rotation in the Schoenflies notation (or, for $D_{\rm 2d}$ and $S_4$, the rotation-reflection axis) is parallel ($\parallel$) or orthogonal ($\perp$) to the tangent plane of the surface. See Fig. \ref{fig:symmetries}E-F for examples of surfaces of different types.}
\end{table*}

We note that the Schoenflies notation we use here refers to the symmetry properties of the phase, rather than of individual molecules. Due to the statistical arrangements of molecules within a surface element, these symmetry properties differ in general. For instance, molecules belonging to the symmetry group $C_{2\rm v}$ in the Schoenflies notation, with a uniform orientation and a mirror plane of symmetry constrained to be tangent to the surface, give rise to a $C_{\rm 2 v}^{\parallel}$ polar surface. Letting the orientation axis of the molecules tilt towards one side of the surface gives rise instead to a polar up-down surface $C_{\rm s}^{\|}$. 

\section{Force and torque balance and virtual work}

The surface mechanics is described by a tension tensor $\mathbf{t}^i$ and a moment tensor $\mathbf{m}^i$, which, respectively, characterize momentum flux and angular momentum flux in the surface. The force $\mathbf{f}$ and torque $\boldsymbol{\Gamma}$ acting on an infinitesimal surface line element with length $dl$ and unit vector $\boldsymbol{\nu}$, oriented perpendicular to the line element and tangent to the surface, are then given by $\mathbf{f}=dl \nu_i \mathbf{t}^i$ and $\boldsymbol{\Gamma}=dl \nu_i \mathbf{m}^i$. Ignoring the moment of inertia tensor for simplicity, conservation of momentum and of angular momentum result in the force and torque balance expressions (Appendix \ref{appendix_ForceBalance_conservation}):
\begin{eqnarray}
\nabla_i \mathbf{t}^i&=&-\mathbf{f}^{\rm{ext}}+\rho \mathbf{a}
\label{ForceBalanceEquation},\\
\nabla_i \mathbf{m}^i&=&\mathbf{t}^i\times\mathbf{e}_i-\boldsymbol{\Gamma}^{\rm{\rm{ext}}}\label{TorqueBalanceEquation},
\end{eqnarray}
where $\mathbf{a}={\rm d} \mathbf{v}/{{\rm d}t}$ is the local center-of-mass acceleration, $\mathbf{f}^{\rm ext}$ is the external force density and $\boldsymbol{\Gamma}^{\rm ext}$ is the external torque density acting on the surface (Fig. \ref{fig:schematic}C). 

The force and torque balance equations can be expressed in terms of the components of the tension and moment tensors (Appendix \ref{appendix_ForceBalance_conservation}):
\begin{eqnarray}
\nabla_i t^{ij}+C_{i}{}^j t_{n}^i&=&-f^{{\rm ext},j}+\rho a^j\label{ForceBalanceTangential},\\
\nabla_i t_{n}^i-C_{ij} t^{ij}&=&-f^{\rm{ext}}_{n}+\rho a_{n}\label{ForceBalanceNormal},\\
\nabla_i m^{ij}+C_{i}{}^j m_{n}^i &=&\epsilon_{i}{}^{j}   t_{n}^i-\Gamma^{{\rm ext},j} \label{TorqueBalanceTangential},\\
\nabla_i m_{n}^i-C_{ij} m^{ij}&=&-\epsilon_{ij}t^{ij}-\Gamma^{\rm{ext}}_{n}\label{TorqueBalanceNormal}~,
\end{eqnarray}
where $C_{ij}$ is the surface curvature tensor.

We now write the infinitesimal virtual work for an element of surface $\mathcal{S}$ with closed contour $\mathcal{C}$, associated to an infinitesimal virtual surface displacement $d\mathbf{X}$,
 as follows:
\begin{align}
	dW&=d W_{\rm ext} +d W_{\rm contour}	\label{eq:VirtualWorkDef}\\
	d W_{\rm ext}&=\int_{\mathcal{S}} dS\,\left[\left(\mathbf{f}^{\rm ext}-\rho \mathbf{a}\right)\cdot d \mathbf{X}+\boldsymbol{\Gamma}^{\rm ext}\cdot d \boldsymbol{\vartheta}\right]	\label{eq:VirtualWorkextDef}\\
	d W_{\rm contour}&=\oint_{\mathcal{C}} dl \nu_i\left(\mathbf{t}^i\cdot d \mathbf{X}+\mathbf{m}^i\cdot d\boldsymbol{\vartheta}\right)	\label{eq:VirtualWorkcontourDef}\,.
\end{align} 
where we follow the form given in Ref.~\cite{salbreux2017activesurfaces}. The infinitesimal rotation associated with the infinitesimal shape change $d\mathbf{X}$ is $d\boldsymbol{\vartheta}=\frac{1}{2}\nabla\times d\mathbf{X}$ (Eq. \ref{def_rotation_rate_appendix}). The first contribution $d W_{\rm ext}$ corresponds to the effect of external forces and torques acting on the surface, while the second contribution $d W_{\rm contour}$ corresponds to the work associated to forces and torques internal to the surface, acting on the contour of the element of surface considered here. Here the vector $\boldsymbol{\nu}$ is a unit vector tangent to the surface and normal to the contour $\mathcal{C}$.

Using the force and torque balance equations \ref{ForceBalanceEquation} and \ref{TorqueBalanceEquation}, the infinitesimal virtual work can be re-written (see Appendix \ref{sec_Appendix_virtual_work_polar_surface}):
\begin{align}
	\label{eq:VirtualWork}
d W
&=\int_\mathcal{S} dS\left[\overline{t}^{ij}\frac{d g_{ij}}{2}  + \overline{m}^{ij}
D C_{ij}+m^{i}_{n} (\partial_id\boldsymbol{\vartheta}) \cdot\mathbf{n}\right]~,
\end{align}
where expressions for the infinitesimal metric variation $dg_{ij}$ and the corotational curvature infinitesimal variation $DC_{ij}$ are given in Eqs. \ref{VariationMetricCovariant}  and \ref{Dcurvature_components}. In Eq. \ref{eq:VirtualWork}, we have introduced the modified tension and moment tensors, which are coupled in the virtual work respectively to the change of metric and the corotational curvature change \cite{salbreux2017activesurfaces}:
\begin{align}
\overline{t}^{ij}&=t^{ij}_s+\frac{1}{2}\left(\overline{m}^{ik}C_k{}^j+\overline{m}^{jk}C_k{}^i\right)\label{eq:tbardef}\\
\overline{m}^{ij}&=-m^{ik}\epsilon_{k}{}^{j}\label{eq:mbardef}~,
\end{align}
where the $s$ subscript denotes the symmetric part of the tensor (Eq. \ref{TensorSymmetricAntisymmetricDecomposition}). We note that with the definitions introduced above for the tension and moment tensors, the modified moment tensor $\bar{m}^{ij}$ can always be symmetrized at the cost of a redefinition of the tension tensor $\mathbf{t}^i$ (Appendix \ref{sec_Appendix_virtual_work_polar_surface}).

\section{Polar active fluid surface}
\label{sec:polar_surfaces}

We now turn to a description of equilibrium and non-equilibrium thermodynamics of polar active fluid surfaces. Here we use concepts from irreversible thermodynamics to derive constitutive equations for a curved polar fluid. We restrict ourselves to the case of a polar vector tangent to the surface, $\mathbf{p}$. We consider a fluid surface, consisting of several species $\alpha=1...N$ with concentrations $c^{\alpha}$.  The local mass density is given by $\rho=\sum_{\alpha} m^{\alpha} c^{\alpha}$ with $m^{\alpha}$ the molecular mass of species $\alpha$. The corresponding mass conservation equations are given in Appendix \ref{subsec:appendix:conservation_equations}.

In the following, we derive equilibrium tensions and torques from a generic expression for a polar surface virtual work. We then obtain the entropy production rate of the surface close to equilibrium, and obtain linear constitutive equations for the non-equilibrium generalized fluxes.

\subsection{Free energy and external potential}

We consider here a region of surface $\mathcal{S}$ of a polar fluid membrane, with free energy given by
\begin{align}
F=\int_{\mathcal{S}} dS f(\mathbf{v}, c^{\alpha}, \mathbf{C}, \mathbf{p}, \boldsymbol{\nabla}\otimes \mathbf{p})~,
\end{align}
 with $f$ the total free energy density, which has two contributions:
\begin{equation}
\label{TotalFreeEnergyDensity}
f(\mathbf{v}, c^{\alpha}, \mathbf{C}, \mathbf{p}, \boldsymbol{\nabla}\otimes \mathbf{p})=\frac{1}{2}\rho v^2 + f_0(c^{\alpha}, \mathbf{C}, \mathbf{p}, \boldsymbol{\nabla}\otimes \mathbf{p})~,
\end{equation}
where the kinetic energy is given by $\frac{1}{2}\rho v^2=\frac{1}{2}\rho \left[v_i v^i+(v_n)^2\right] $ and $f_0$ is the free energy density in the rest frame. The tensor $\boldsymbol{\nabla}\otimes \mathbf{p}=\mathbf{e}^k\otimes\partial_k \mathbf{p}$ is the gradient of polarity (Eq. \ref{def:appendix:vector_gradient}). Here we consider for simplicity a situation at fixed temperature and where the polar vector $\mathbf{p}$ is tangent to the surface.
 The differential of $f_0$ can be written:
\begin{equation}
\label{FluidMembraneFreeEnergyDensity}
df_0=\mu^{\alpha} d c^{\alpha}+\mathbf{K} :d\mathbf{C} - \mathbf{h}_0 \cdot d \mathbf{p} +\boldsymbol{\pi} : d(\boldsymbol{\nabla}\otimes\mathbf{p}) ,\\
\end{equation}
where $\mu^{\alpha}$ is the chemical potential of component $\alpha$, $\mathbf{K}$ is the passive bending moment, $\mathbf{h}_0$ and $\boldsymbol{\pi}$ are the forces conjugate to the polarity field and to its gradient on the surface. These conjugate forces are obtained by taking partial derivatives of the free energy density $f_0$:
\begin{align}
\label{eq:definitions_thermodynamic_forces_polar}
\mu^{\alpha}=&\frac{\partial f_0}{\partial c^{\alpha}},~\boldsymbol{K} =\frac{\partial f_0}{\partial\boldsymbol{C}}, \nonumber\\
\mathbf{h}_0=&-\frac{\partial f_0}{\partial \mathbf{p}},~ \boldsymbol{\pi} =\frac{\partial f_0}{\partial (\boldsymbol{\nabla}\otimes\mathbf{p})}~.
\end{align}
We assume that $f_0$ is such that $\mathbf{K}$ and $\mathbf{h}_0$ are tangent to the surface, that $\mathbf{K}$ is symmetric, and that $\boldsymbol{\pi}$ can be written $\boldsymbol{\pi} =\mathbf{e}_i\otimes \boldsymbol{\pi}^i $.
Because of invariance by rotation of the free energy (Appendix \ref{AppendixGibbsDuhem} and Eq. \ref{GibbsDuhemRotation}), the differential of the free energy density can also be written:
\begin{equation}
\label{FluidMembraneFreeEnergyDensity_Corotational}
df_0=\mu^{\alpha} d c^{\alpha}+\mathbf{K} :  D\mathbf{C} - \mathbf{h}_0 \cdot D \mathbf{p} +\boldsymbol{\pi} :D(\boldsymbol{\nabla}\otimes\mathbf{p}).\\
\end{equation}
As $\mathbf{C}$ and $\mathbf{p}$ are tangent to the surface, $D\mathbf{C}$ and $D\mathbf{p}$ are also tangent to the surface (Eq. \ref{eq:appendix:normal_part_corotational_variation}). 
We also introduce the total molecular field $\mathbf{h}$, which is decomposed in two contributions:
\begin{align}
\label{eq:def_total_molecular_field_h}
\mathbf{h}&=-\frac{\delta F}{\delta \mathbf{p}}=\mathbf{h}_0+\nabla_j \boldsymbol{\pi}^{j}~,
\end{align}
and is not necessarily tangent to the surface. Note that we define here the functional derivative of the functional $F[\phi]$ through $F[\phi+d\phi]-F[\phi] = \int_{\mathcal S} dS \frac{\delta F}{\delta \phi} d\phi$ to first order in $d\phi$.  

 In the following we also assume that the surface is subjected to an external potential:
\begin{align} 
U=\int_{\mathcal{S}} dS c^{\alpha} u^{\alpha}(\mathbf{X}, \mathbf{n}, \mathbf{p}),\label{eq:external_potential_definition_polar}
\end{align}
 where $u^{\alpha}$ is an external potential density acting on species $\alpha$. Such a potential could be for instance generated by an external magnetic field.

\subsection{Invariance of free energy by translation and rotation}

Using the fact that the free energy density $f_0(c^{\alpha}, \mathbf{C}, \mathbf{p}, \boldsymbol{\nabla}\otimes \mathbf{p})$ is only a function of the surface variables and does not explicitly depend on space, one obtains the Gibbs-Duhem relation (Appendix \ref{AppendixGibbsDuhem})
\begin{eqnarray}
\label{GibbsDuhemTranslation}
\nabla_j \left[ (f_0 - \mu^{\alpha} c^{\alpha})g_{i}{}^{j} - K^{jk} C_{ik} - \boldsymbol{\pi}^{j}\cdot \partial_i \mathbf{p}\right]\nonumber\\
+C_{ik} \nabla_j K^{jk} = -(\partial_i \mu^{\alpha})c^{\alpha} - \mathbf{h} \cdot \partial_i \mathbf{p} ~.
\end{eqnarray}

A similar calculation using that the free energy density is invariant by local solid rotations, leads to the generalized Gibbs-Duhem relation (see Appendix \ref{AppendixGibbsDuhem}):  
\begin{align}
\label{GibbsDuhemRotation}
\boldsymbol{\pi}^i \times\partial_i \mathbf{p}+( \boldsymbol{\pi}^i\cdot\partial_j \mathbf{p})\epsilon_{i}{}^{j} \mathbf{n} =\mathbf{h}_0\times\mathbf{p}+\mathbf{C}\times_1 \mathbf{K}  +\mathbf{C}\times_2 \mathbf{K}~.
\end{align}
where the notations for cross-product of tensors is introduced in Appendix \ref{appendix_diff_geom}.
These relations are valid at and out of equilibrium, and reflect a mathematical relation between the surface variables and their conjugated forces in the free energy.  These arise from the form of the free energy, which does not depend on the position or orientation of the surface in space.

\subsection{Equilibrium relations}

To obtain equilibrium relations and expressions for the equilibrium tensions and torques, we consider changes of free energy associated with an infinitesimal virtual surface displacement $d\mathbf{X}$, virtual change of polarity $d \mathbf{p}$ and change of concentration $d c^{\alpha}$ of the component $\alpha$. Therefore, we write that for a surface patch $\mathcal{S}$ enclosed by the contour $\mathcal{C}$, the variation of $F_0=\int_{\mathcal{S}} dS f_0$ is:
\begin{align}
d F_0 = d W +\int_{\mathcal{S}} dS [-\mathbf{h}\cdot D\mathbf{p} +\mu^{\alpha} \dbar c^{\alpha}] +\oint_{\mathcal{C}} dl \nu_i\boldsymbol{\pi}^i\cdot D\mathbf{p}\label{eq:equilibrium_variations_W}~,
\end{align}
where $dW$ is the mechanical differential work associated with the surface deformation (Eqs. \ref{eq:VirtualWorkDef} and \ref{eq:VirtualWork}). The corotational differential $D\mathbf{p}$ has been defined in Eq. \ref{def_corot_variation_p}. In this equation, the molecular field $\mathbf{h}$ couples to the corotational variation of the tangent polarity vector $\mathbf{p}$, while $\boldsymbol{\pi}^i$ is the internal molecular field coupled to corotational variation of $\mathbf{p}$ at the boundary of the patch $\mathcal{S}$. The chemical potential $\mu^{\alpha}$ couples to the variation of concentration of molecular species $\alpha$ which does not arise from dilution, $\dbar c^{\alpha}$ defined in Eq. \ref{eq:variation_c_alpha_appendix}. For surfaces perturbations where the polarity rotates with the surface, $D\mathbf{p}=0$, and the concentration fields change only by geometric dilution, $\dbar c^{\alpha}=0$, the variation of free energy is equal to the work $dW$ associated with the surface deformation.

Similarly, the variation of the external potential $U$ has a mechanical contribution, corresponding to the work $dW_{\rm ext}$ (Eq. \ref{eq:VirtualWorkDef}), and contributions from the variations in the polar field $D\mathbf{p}$ and the concentration fields $\dbar c^{\alpha}$:
\begin{align}
d U=-dW_{\rm ext}+\int_{\mathcal{S}} dS [\mathbf{h}^{\rm ext} \cdot D\mathbf{p} -\mu^{{\rm ext},\alpha} \dbar c^{\alpha}]\label{eq:equilibrium_variations_Wext}~,
\end{align}
where external molecular field $\mathbf{h}^{\rm ext}$ and external chemical potential $\mu^{{\rm ext},\alpha}$ are defined by:
\begin{align}
\mathbf{h}^{\rm ext}=&\frac{\delta U}{\delta\mathbf{p}}~,\label{eq:def_hext}\\
\mu^{{\rm ext},\alpha}=&-\frac{\delta U}{\delta c^{\alpha}}~.
\end{align}
We assume that $U$ is such that $\mathbf{h}^{\rm ext}$ is tangent to the surface.
Eqs \ref{eq:equilibrium_variations_W} and \ref{eq:equilibrium_variations_Wext} can be combined such that:
\begin{align}
d F_0+d U=&\int_{\mathcal{S}} dS  \left[(\mathbf{h}^{\rm ext}-\mathbf{h}) \cdot D\mathbf{p} +(\mu^{\alpha}-\mu^{{\rm ext},\alpha}) \dbar c^{\alpha}\right]\nonumber\\
&+dW_{\rm contour}+\oint_{\mathcal{C}} dl \nu_i\boldsymbol{\pi}^i\cdot D\mathbf{p}\label{eq:equilibrium_variations_total}~.
\end{align}

The equilibrium tension and moment tensors can be obtained by calculating the infinitesimal change of surface free energy \ref{FluidMembraneFreeEnergyDensity} under a surface deformation, and using the expression of the virtual work, Eq. \ref{eq:VirtualWork} together with the equilibrium condition \ref{eq:equilibrium_variations_W}. This procedure then gives (Appendix \ref{AppendixEquilibriumPolar:internal_free_energy}):
\begin{align}
t_e^{ij} =&(f_0-\mu^{\alpha} c^{\alpha}) g^{ij} -K^{ik}C_k{}^j- \boldsymbol{\pi}^i\cdot\partial^j \mathbf{p}~,\label{eq:equilibriumtangentialtension}\\
t_{e,n}^i=&\nabla_j K^{ji}+\epsilon^i{}_j\left[\Gamma_e^{{\rm ext}, j} - (\mathbf{h}\times\mathbf{p})\cdot\mathbf{e}^j \right]~,\label{eq:equilibriumtne}\\
\mathbf{m}_e^{i}=&K^{ik}\epsilon_k{}^j \mathbf{e}_j + \mathbf{p}\times\boldsymbol{\pi}^i ~.\label{FluidEquilibriumNormalTorqueTensor}
\end{align}
The last term in Eq. \ref{eq:equilibriumtangentialtension} is the surface tension tensor arising from distortion of the polar field; $\gamma g^{ij} - \boldsymbol{\pi}^i\cdot\partial^j \mathbf{p}$ with $\gamma=f_0-\mu^{\alpha} c^{\alpha}$ is the equivalent, for a surface, of the Ericksen stress tensor in a liquid crystal \cite{degennesprost, napoli2010equilibrium}. The expression for the normal component of the equilibrium tension tensor $t^i_{e, n}$ arises from the tangential torque balance Eq. \ref{TorqueBalanceTangential}, and involves the external tangential torque density acting on the surface. 

To obtain external forces, external torques and molecular field deriving from an external potential $U$, we obtain the differential $d U$ from the potential $U$ defined in Eq. \ref{eq:external_potential_definition_polar}, and use the equilibrium condition \ref{eq:equilibrium_variations_Wext} and external virtual work \ref{eq:VirtualWorkextDef} to identify $\mathbf{f}^{\rm ext}_e$, $\boldsymbol{\Gamma}^{\rm ext}_e$, $\mathbf{h}^{\rm ext}$, and $\mu^{{\rm ext},\alpha}$. One then obtains (Appendix \ref{appendix_force_torque_external_potential}):
\begin{align}
\mathbf{f}^{\rm ext}_e &= -c^{\alpha}\frac{ \partial u^{\alpha}}{\partial \mathbf{X}}~,\label{eq:external_force_density_polar}\\
\boldsymbol{\Gamma}^{\rm ext}_e&=c^{\alpha}\left(\frac{\partial u^{\alpha}}{\partial \mathbf{n}}\times \mathbf{n}
+\frac{\partial u^{\alpha}}{\partial \mathbf{p}}\times \mathbf{p}\right)~,
\label{eq:external_torque_density_polar}\\
\mathbf{h}^{\rm ext}&=c^{\alpha} \frac{\partial u^{\alpha}}{\partial \mathbf{p}}~,\label{eq:external_molecular_field_polar}\\
\mu^{{\rm ext}, \alpha}&= -u^{\alpha}~.
\end{align}

Next, minimising the total free energy $F_{\rm tot}=F_0+U$ with respect to the polarity field $\mathbf{p}$ and the concentration fields $c^{\alpha}$ at fixed surface shape, results in the equilibrium relations for the tangential part of the molecular field and for the chemical potential, using 
Eq. \ref{eq:equilibrium_variations_total}:
\begin{align}
h^i =&h^{{\rm ext},i} \label{eq:polar_balance_equilibrium}~,\\
\mu^{\alpha} = &\mu^{{\rm ext}, \alpha}\label{eq:polar_chemical_potential_balance}~.
\end{align}

We now discuss relations between equilibrium tensions and torques and invariance of the surface properties under a rigid translation or rotation, Eqs. \ref{GibbsDuhemTranslation} and \ref{GibbsDuhemRotation}. Using the expressions for the equilibrium tensors \ref{eq:equilibriumtangentialtension}-\ref{FluidEquilibriumNormalTorqueTensor} and the relations \ref{appendix:equilibrium_normal_torque_polar} and \ref{appendix:gradient_external_chemical_potential_polar}, the Gibbs-Duhem relations \ref{GibbsDuhemTranslation} and \ref{GibbsDuhemRotation}, obtained by invariance of the free energy under solid translations and rotations, can be re-written in the simpler form
\begin{align}
\nabla_i t_e^{ij}+C_{i}{}^j t_{e,n}^i +f_e^{{\rm ext}, j}=&(h_i^{{\rm ext}}-h_i) \nabla^j p^i \nonumber\\
&+c^{\alpha}\partial^j(\mu^{{\rm ext}, \alpha}- \mu^{\alpha})  ~,\\
\nabla_i m_{n,e}^i-C_{ij} m_e^{ij}+\epsilon_{ij}t_e^{ij}+\Gamma_{e,n}^{\rm ext}=&\epsilon^{ij}(h_i^{{\rm ext}} -h_i) p_j~.
\end{align}
The right-hand sides of these two equations vanish when the equilibrium equations for the concentration field (Eq. \ref{eq:polar_chemical_potential_balance}) and for the polar field (Eq. \ref{eq:polar_balance_equilibrium}) are satisfied. In that case the left-hand sides of these two equations also vanish, corresponding to the tangential force balance Eq. \ref{ForceBalanceTangential} and normal torque balance Eq. \ref{TorqueBalanceNormal}, applied to equilibrium tension, moments, external force and torque densities.

Overall, at equilibrium the tangential force balance equation and normal torque balance equations are satisfied by expressing the chemical potential balance $\mu^{\alpha}=\mu^{\rm ext,\alpha}$ and that the polar field is at equilibrium, Eq. \ref{eq:polar_balance_equilibrium}. The tangential torque balance expression gives the expression of the normal tension $t_n^i$, Eq. \ref{eq:equilibriumtne}. The remaining normal force balance equation following from Eq. \ref{ForceBalanceNormal}:
\begin{align}
\nabla_i t_{e,n}^i-C_{ij} t_e^{ij}&=-f^{\rm{ext}}_{e,n}~,\label{ForceBalanceNormal_polar_equilibrium}
\end{align}
together with the equilibrium relation for the tension tensor $\mathbf{t}_{e}^i$, Eqs. \ref{eq:equilibriumtangentialtension} and \ref{eq:equilibriumtne}, and for the external force density Eq. \ref{eq:external_force_density_polar}, is an equation for the equilibrium shape of the surface.

\subsection{Entropy production rate}
We can now calculate the entropy production, by calculating the time derivative of the surface free energy. We consider a region of surface $\mathcal{S}$ enclosed 
by a contour $\mathcal{C}$, which can deform in three dimensions. We consider a situation where the contour $\mathcal{C}$ is deforming with the surface. We also consider that the total external force density $\mathbf{f}^{\rm ext}$ and external torque density $\boldsymbol{\Gamma}^{\rm ext}$ have a conservative part arising from an external potential $U$, which are written $\mathbf{f}^{\rm ext}_e$, $\boldsymbol{\Gamma}^{\rm ext}_e$, with corresponding definitions given in Eqs. \ref{eq:external_force_density_polar} and \ref{eq:external_torque_density_polar}. We consider that species $\alpha$ can change due to the surface tangent flux $\mathbf{J}^{\alpha}$ and due to a source term $r^{\alpha}$ arising from chemical reactions. Corresponding balance equations for the concentration of species $\alpha$ in the surface are discussed in Appendix \ref{subsec:appendix:conservation_equations}. The total flux $\mathbf{J}^{\alpha}$ can be separated into a part that depends on the center-of-mass velocity, $c^{\alpha} \mathbf{v}^{\alpha}$, and a relative flux $\mathbf{j}^{\alpha}$. The rate of the total free energy change $F_{\rm tot}=F+U$ is calculated in Appendix~\ref{AppendixEntropyProductionRatePolar}, and we obtain:
\begin{align}
	\dot{F}_{\rm tot}=&\int_\mathcal{S} dS\,
	\Big\{-\left(\overline{t}^{ij}-\overline{t}_e^{ij}\right) v_{ij}
	-(\overline{m}^{ij}-\overline{m}_e^{ij})D_t C_{ij}\nonumber\\
	&-(m_n^i-m_{n,e}^i)\omega_{in}-(\mathbf{h}-\mathbf{h}^{\rm ext})\cdot D_t\mathbf{p}\nonumber \\
		&+(\mu^\alpha -\mu^{{\rm ext},\alpha})r^\alpha+j^{\alpha,i}\partial_i(\overline{\mu}^{\alpha}-\overline{\mu}^{{\rm ext},\alpha})\nonumber\\
	&+(\mathbf{f}^\mathrm{ext}-\mathbf{f}^{\rm ext}_e)\cdot\mathbf{v}+(\boldsymbol{\Gamma}^\mathrm{ext}-\boldsymbol{\Gamma}^\mathrm{ext}_e)\cdot\boldsymbol{\omega}\Big\}\nonumber \\	
	&+\oint_\mathcal{C} dl\nu_i \left[\mathbf{t}^i\cdot \mathbf{v}+\mathbf{m}^i\cdot\boldsymbol{\omega}+\boldsymbol{\pi}^i \cdot D_t \mathbf{p}\right.\nonumber\\
	&\left.\hspace{1.5cm}-(\overline{\mu}^\alpha -\overline{\mu}^{\rm ext,\alpha})j^{\alpha,i}\right]~,
	\label{eq:Fdot}
\end{align}
where we have introduced the gradient of flow $v_{ij}$, the vorticity $\boldsymbol{\omega}$, the normal vorticity gradient $\omega_{in}$ and the corotational derivative of the curvature tensor $D_t C_{ij}$, which are given by:
\begin{align}
\label{eq:expression_vij}
	v_{ij}=&\frac{1}{2}(\nabla_i v_j+\nabla_j v_i)+C_{ij} v_n~, \\
	\boldsymbol{\omega}=&\epsilon^{ij}(\partial_j v_n -C_{jk} v^k)\mathbf{e}_i+\frac{1}{2}\epsilon^{ij}(\nabla_i v_j)\mathbf{n}~,\label{eq:expression_vorticity}\\
	\omega_{in} =& (\partial_i \boldsymbol{\omega})\cdot\mathbf{n}=\partial_i\omega_n-C_i{}^j\omega_j~,\\
	\label{eq:expression_Dt_Cij}
	D_t C_{ij}=&-\nabla_i(\partial_j v_n)-v_n C_{ik}C^k{}_j+v^k \nabla_k C_{ij} \nonumber\\
	&+\omega_n(\epsilon_i{}^k C_{kj}+\epsilon_j{}^k C_{ki})\,,
	\end{align}
	and
\begin{equation}
\overline{\mu}^\alpha=\mu^\alpha-\frac{m^\alpha}{m^1}\mu^1, \overline{\mu}^{{\rm ext},\alpha}=\mu^{{\rm ext},\alpha}-\frac{m^\alpha}{m^1}\mu^{{\rm ext},1}~,
\end{equation}
are the relative chemical potential of species $\alpha$. Here we have made the arbitrary choice of taking chemical potentials relative to the chemical potential of species $\alpha=1$, which in practice can be chosen to be the solvent. As a result, the sum over species $\alpha$ in Eq. \ref{eq:Fdot} can be taken for $\alpha=2...N$. If one considers $I=1..M$ chemical reactions, the contribution $(\mu^{\alpha}-\mu^{{\rm ext}, \alpha}) r^{\alpha}$ can be rewritten as a sum over chemical reactions (Appendix \ref{subsec:appendix:conservation_equations}):
\begin{align}
(\mu^{\alpha}-\mu^{{\rm ext}, \alpha}) r^{\alpha}=-\sum_I \Delta \mu^I r^I~,
\end{align}
where $\Delta \mu^I = \sum_I a^{\alpha, I} (\mu^{\alpha}-\mu^{{\rm ext}, \alpha}) $ with $a^{\alpha,I}$ the stoichiometric coefficients for $\alpha$ in reaction $I$, and $r^I$ the rate of reaction $I$ as defined in Eq. \ref{eq:appendix_def_reaction_rate}.
In the following we consider a single chemical reaction between a fuel and its product; such that $(\mu^{\alpha}-\mu^{{\rm ext}, \alpha}) r^{\alpha}=-r\Delta \mu$ with $r$ the rate of fuel consumption and $\Delta\mu$ the chemical potential of conversion of fuel to the product.  

For an isothermal surface, the free energy density evolves according to the balance equation (Eq. \ref{BalanceFreeEnergy}):
\begin{align}
\label{eq:balance_free_energy}
\partial_t f+\nabla_i (f v^i+j^{f,i})+v_n C_i{}^i f&=J^f_n-T\theta~,
\end{align}
where $J_n^f$ and $j^{f,i}$ are respectively the normal flux of free energy and tangential flux relative to the center of mass of free energy, and the entropy production rate within the surface is denoted $\theta$. Using Eq. \ref{eq:appendix_free_energy_variation}, identification with the rate of change of free energy then gives the rate of entropy production within the surface:
\begin{align}
\label{eq:entropy_production_polar_surface}
T \theta=&\left(\overline{t}^{ij}-\overline{t}_e^{ij}\right) v_{ij}
	+(\overline{m}^{ij}-\overline{m}_e^{ij})D_t C_{ij}\nonumber\\
	&+(m_n^i-m_{n,e}^i)\omega_{in}+(\mathbf{h}-\mathbf{h}^{\rm ext})\cdot D_t\mathbf{p}\nonumber \\
	&+r\Delta \mu-j^{\alpha,i}\partial_i(\overline{\mu}^{\alpha}-\overline{\mu}^{{\rm ext},\alpha})~,
\end{align}
and the flux of free energy with normal and tangential components:
\begin{align}
J_n^f&=(\mathbf{f}^\mathrm{ext}-\mathbf{f}^{\rm ext}_e)\cdot\mathbf{v}+(\boldsymbol{\Gamma}^\mathrm{ext}-\boldsymbol{\Gamma}^\mathrm{ext}_e)\cdot\boldsymbol{\omega}~,\\
j^{f,i}&= -\mathbf{t}^i\cdot \mathbf{v}-\mathbf{m}^i\cdot\boldsymbol{\omega}-\boldsymbol{\pi}^i \cdot D_t \mathbf{p}+(\overline{\mu}^\alpha -\overline{\mu}^{\rm ext,\alpha})j^{\alpha,i}.
\end{align}
Here we have assumed that external forces and torques do not contribute to the surface entropy production.
The conjugate fluxes and forces can be read from the entropy production rate Eq. \ref{eq:entropy_production_polar_surface} and are listed in Table \ref{table:conjugate_force_fluxes_polar}.
\begin{table*}
\begin{center}
\begin{tabular}{l@{\hskip 1in}c@{\hskip 0.5in}}
\hline\hline\\[-.35cm]
Flux & Force\\[.08cm] \hline\\[-.3cm]
 \makecell{In-plane deviatoric tension tensor \\$\overline{t}^{ij}_d=\overline{t}^{ij}-\overline{t}^{ij}_e$} &In-plane shear tensor $v_{ij}$   \\
 \makecell{In-plane deviatoric bending moment tensor\\ $\overline{m}^{ij}_d= \overline{m}^{ij}-\overline{m}^{ij}_e$  } &Bending rate tensor $D_t C_{ij}$ \\
\makecell{ Normal deviatoric moment \\$m_{n,\,d}^i= m_n^i-m_{n,e}^i$  }&Vorticity gradient $\omega_{in}=(\partial_i \boldsymbol{\omega})\cdot \mathbf{n}$\\
\makecell{Convected, corotational derivative of \\polarity $D_t p^i$} & \makecell{Deviatoric molecular field \\$h^i_d=h^i-h^{\rm ext, i}$ }\\
Rate of fuel consumption $r$ & Fuel hydrolysis chemical potential $\Delta\mu$\\
Flux of species $\alpha=2.,.N$, $j^{\alpha, i}$ & \makecell{Deviatoric gradient of relative chemical potential \\ of species $\alpha=2...N$, $-\partial_i \bar{\mu}_d^{\alpha}=-\partial_i (\bar{\mu}^{\alpha}-\bar{\mu}^{{\rm ext},\alpha})$}\\
\end{tabular}
\end{center}
\caption{\label{table:conjugate_force_fluxes_polar} List of conjugate forces and fluxes for an active fluid polar surface.}
\end{table*}

\subsection{Constitutive equations}

To obtain constitutive equations, we perform a linear expansion of thermodynamic fluxes into thermodynamic forces listed in Table \ref{table:conjugate_force_fluxes_polar} \cite{de2013non}. We discuss coupling terms that are allowed for different types of polar surfaces classified in section \ref{subsection_symmetries}.

\subsubsection{Polar surfaces}
We first consider true polar surfaces, associated with a vectorial order parameter $\mathbf{p}$. The deviatoric parts of the stress and moment tensors, the convected time derivative of the polarity field, the flux of species $\alpha$, relative to the centre of mass and the rate of fuel consumption can be decomposed as
\begin{align}
\overline{t}^{ij}_d&=\overline{t}^{ij}_0+\epsilon_{\rm UD} \overline{t}^{ij}_{\rm{UD}}+\epsilon_{\rm C} \overline{t}^{ij}_{\rm{C}}+\epsilon_{\rm PC}  \overline{t}^{ij}_{\rm{PC}},\nonumber\\
\overline{m}^{ij}_d&=\overline{m}^{ij}_0+\epsilon_{\rm UD} \overline{m}^{ij}_{\rm{UD}}+\epsilon_{\rm C}  \overline{m}^{ij}_{\rm{C}}+\epsilon_{\rm PC}  \overline{m}^{ij}_{\rm{PC}},\nonumber\\
m^i_{n,d}&=m^i_{n0}+\epsilon_{\rm UD} m^i_{n\rm{UD}}+\epsilon_{\rm C}m^i_{n\rm{C}}+\epsilon_{\rm PC}   m^i_{n\rm{PC}}\nonumber\\
D_t p^i&=P_0^i+\epsilon_{\rm UD} P_{\rm UD}^i+\epsilon_{\rm C} P_{\rm C}^i+\epsilon_{\rm PC} P_{\rm PC}\nonumber\\
j^{\alpha, i}&=j_0^{\alpha, i} + \epsilon_{\rm UD} j_{\rm UD}^{\alpha, i} + \epsilon_{\rm C} j_{\rm C}^{\alpha, i}+\epsilon_{\rm PC}  j_{\rm PC}^{\alpha, i}\nonumber \\
r&=r_0 + \epsilon_{\rm UD} r_{\rm UD} + \epsilon_{\rm C} r_{\rm C}+\epsilon_{\rm PC}  r_{\rm PC} ~,
\label{GeneralTensorDecompositionSymmetry}
\end{align}
where $\overline{t}^{ij}_0$ is the part of the stress tensor that exists for any surface, $\overline{t}^{ij}_{\rm{UD}}$ correspond to terms present when the surface breaks up-down symmetry, $\overline{t}^{ij}_{\rm{C}}$ exist for chiral surfaces, and $\overline{t}^{ij}_{\rm{PC}}$ for planar-chiral surfaces. Similar rules apply for the decomposition of other thermodynamic fluxes.

To express constitutive equations for each of the components, one can then write possible terms of the expansion at linear thermodynamic fluxes into generalized thermodynamic forces, and ask whether the corresponding terms break the symmetries $M_n$, $M_t$, $R_t$, $I$, according to the signatures given in section \ref{subsection_symmetries}.  To check which symmetries are broken or satisfied by couplings in the constitutive equations, we note that $\overline{t}^{ij}$ is a true tensor, whereas $\overline{m}^{ij}$ is a pseudotensor with  signature $-1$ under $M_n, R_t, I$ and signature $+1$ under $M_t, R_n$. Indeed, $\overline{m}^{ij}$ is the product of $m^{ij}$ and the tensor $\epsilon^{ij}$ (Eq. \ref{eq:mbardef}). $m^{ij}$ is proportional to a torque (Eq. \ref{DefinitionsMoment}) and has therefore signature $-1$ under the mirror and inversion symmetries $M_n$, $M_t$, $I$. The tensor $\epsilon^{ij}$ has signature $-1$ under $M_t, R_t$ and $+1$ under $M_n, R_n, I$. Also, $\mathbf{m}_n=m_n^i \mathbf{e}_i$ is a pseudo-vector with signature $-1$ under $M_t, R_t$ and $+1$ under $M_n, R_n, I$, since $\mathbf{n}$ is a pseudo-vector with signature $-1$ under $M_n, R_t, I$ and $+1$ under $M_t, R_n$. 
These observations can be summarized in the Table of transformation \ref{forces_signature_table}.
\begin{table} 
\begin{center}
\begin{tabular}{|c|c|c|c|c|c|c|c|}\hline
& $M_n$ & $M_t$ & $R_t$  &$I$&$R_n$\\\hline
$\overline{t}^{ij}$& 1 & 1 & 1 &1&1\\\hline
$m^{ij}$& -1 &-1 &1 &-1&1\\\hline
$\overline{m}^{ij}$& -1 &1 &- 1 &-1&1\\\hline
$m_n^i$& 1 & -1 & -1 &1&1\\\hline
\end{tabular} 
\caption{\label{forces_signature_table} Signatures of tension and moment tensors under a set of linear transformations.}
\end{center}
\end{table}

The corresponding constitutive equations for polar surfaces are given in Appendix \ref{appendix:full_constitutive_equations}. Because many coupling terms are possible, for simplicity we have imposed some restrictions on the coupling terms considered. We limit ourselves to an expansion of phenomenological coefficients to first order in the curvature tensor, the polarity vector and its associated nematic tensor $q_{ij}=p_i p_j -\frac{1}{2} p^2 g_{ij}$. Viscous coupling terms between $\bar{t}^{ij}_d$, $\bar{m}^{ij}$, $m_{n,d}^i$ and $v_{ij}$, $D_tC_{ij}$, $\omega_{in}$ and diagonal coupling terms between a flux and its conjugated force that are dependent on the polarity vector and on the curvature tensor are not listed. We do not include cross-couplings between the various mechanical tensors and the molecular field $\mathbf{h}$ that depend on the curvature tensor. Also, we do not include cross-coupling terms involving the gradient of the chemical potential, $\partial_i \bar{\mu}_d^{\alpha}$, except for active couplings of the relative flux $\mathbf{j}^{\alpha}$ with the fuel hydrolysis chemical potential, $\Delta \mu$. Finally, we do not write active terms which are proportional to the gradient of the polarity, $\partial_i \mathbf{p}$. 

We give below the contributions to constitutive equations which depend on the polarity vector. Viscous terms and curvature-coupling terms for in-plane isotropic, chiral and planar-chiral surfaces, which do not depend on the polarity vector, were derived in Ref. \cite{salbreux2017activesurfaces}. The polarity-dependent contributions to the deviatoric contributions to the tension tensor then read
\begin{align}
\overline{t}^{ij}_0=&\frac{\nu}{2}(p^i h_d^j+p^j h_d^i -h_d^k p_k g^{ij}) +\nu' h_d^k p_k g^{ij}+\zeta_{\rm n} \Delta \mu q^{ij}\nonumber\\[.2cm]
\overline{t}^{ij}_{\rm{UD}}=&\tilde{\zeta}_{\rm n}  \Delta \mu q^{kl} C_{kl} g^{ij}  +\tilde{\zeta}'_{\rm n} q^{ij} C_{k}{}^k  \Delta \mu
 \nonumber \\[.2cm]
\overline{t}^{ij}_{\rm{C}}=&\zeta_{\rm{C n}}\Delta\mu(\epsilon^{ik} q_{k}{}^l C_l{}^j+\epsilon^{jk} q_{k}{}^{l} C_l{}^i)\nonumber\\
&+\tilde{\zeta}_{\rm{C n}}\Delta\mu(\epsilon^{ik} C_{k}{}^l q_l{}^j+\epsilon^{jk} C_k{}^l q_l{}^i)  \nonumber \\[.2cm]
\overline{t}^{ij}_{\rm{PC}}=&\frac{\nu_{{\rm PC}}}{2}(\epsilon^{i}{}_{k}h_d^k p^j+\epsilon^{j}{}_{k} h_d^k p^i-\epsilon_{lk} h_d^k p^l  g^{ij}  )\nonumber\\
&+ \nu_{\rm PC}' \epsilon_{kl} p^k h_d^l g^{ij}  +\zeta_{\rm PC{\rm n}}\Delta \mu \epsilon^{ik} q_k{}^j ,
\label{ConstitutiveEquationtij}
\end{align}
where we use the notation $h_d^i=h^i-h^{{\rm ext}, i}$ for the deviatoric molecular field, and we have introduced the nematic tensor that can be constructed from the polar order parameter $\mathbf{p}$:
\begin{align}
q^{ij}=p^i p^j -\frac{1}{2} p^2 g^{ij}~.
\end{align}
We do not restrict ourselves here to $|\mathbf{p}|=1$; although we note that the modulus of $\mathbf{p}$ is not in general a hydrodynamic variable, except close to a polar/non-polar transition. In the linear expansion of thermodynamic fluxes into thermodynamic forces, we have used that the tensor $\epsilon^{ik} q_{k}{}^j$ is symmetric (Eq. \ref{eq:appendix_product_epsilon_S_identity}) and that $q_i{}^k \epsilon_{kj} =-\epsilon_{ik}q^k{}_j $ (Eq. \ref{eq:appendix_product_epsilon_Q_identity}). To avoid introducing redundant couplings, we have also used that $[\epsilon^{ik}p_k h_d^j]_s =[\epsilon^{i}{}_{k}h_d^k p^j]_s- \epsilon_{kl} p^k h_d^l g^{ij}$ as well as the identities which follow from Eqs \ref{eq:appendix_product_S_Q_identity}-\ref{eq:appendix_product_Q_S_identity}, $[C_{ik} q^k{}_j]_s=[q_{ik} C^k{}_j]_s=\frac{1}{2}C_k{}^k q_{ij} + \frac{1}{2} C_{kl}q^{kl} g_{ij}$. Here we use the notation $[\cdot]_s$ for the symmetric part of a tensor, as defined in Eq. \ref{TensorSymmetricAntisymmetricDecomposition}. We find only two independent active terms in the contribution $\bar{t}_{\rm C}^{ij}$ proportional to $[\boldsymbol{\epsilon} \mathbf{q} \mathbf{C}]_s$ and $[\boldsymbol{\epsilon} \mathbf{C} \mathbf{q}]_s$, due to the tensor product identities $\mathbf{q}\boldsymbol{\epsilon} \mathbf{C} =-\boldsymbol{\epsilon} \mathbf{q} \mathbf{C} $, $\mathbf{C} \boldsymbol{\epsilon} \mathbf{q} = -\mathbf{C} \mathbf{q}\boldsymbol{\epsilon}$, $[\mathbf{C}\mathbf{q}\boldsymbol{\epsilon}]_s=-[\boldsymbol{\epsilon} \mathbf{q}\mathbf{C}]_s$ and $[\mathbf{q}\mathbf{C}\boldsymbol{\epsilon}]_s=-[\boldsymbol{\epsilon} \mathbf{C}\mathbf{q}]_s$, as well as the two relations which follow from Eqs. \ref{eq:appendix_product_S_Q_identity} and \ref{eq:appendix_product_Q_S_identity}:
\begin{align}
\label{eq:epsilon_C_q_identity}
[\epsilon_{ik}C^{kl} q_{lj}]_s =\frac{1}{2} C_k{}^k \epsilon_{ik} q^{k}{}_{j} -\frac{1}{2} \epsilon_{kl} q^{lm} C_{m}{}^{k} g_{ij}~,\\
\label{eq:epsilon_q_C_identity}
[\epsilon_{ik}q^{kl} C_{lj}]_s = \frac{1}{2}C_k{}^k \epsilon_{ik} q^{k}{}_{j} +\frac{1}{2} \epsilon_{kl} q^{lm} C_{m}{}^{k} g_{ij}~.
\end{align}

The deviatoric contributions to the moment tensor $\bar{m}_d^{ij}$ which depend on the polarity vector $\mathbf{p}$ read:
\begin{align}
\overline{m}^{ij}_0=&\tilde{\zeta}_{c{\rm n}} \Delta \mu  q^{kl} C_{kl} g^{ij} +\tilde{\zeta}'_{c{\rm n}} \Delta \mu q^{ij}C_{k}{}^k    \nonumber\\
\overline{m}^{ij}_{\rm{UD}}=&\frac{\beta}{2} (p^i h_d^j+p^j h_d^i-p_k h_d^k g^{ij})+ \beta' p_k h_d^k g^{ij}+\zeta_{c{\rm n}} \Delta \mu q^{ij}\nonumber \\
\overline{m}^{ij}_{\rm{C}}=&
\frac{\beta_{\rm{C}}}{2} (\epsilon^{i}{}_{k} h_d^k p^j+\epsilon^{j}{}_{k} h_d^k p^i- \epsilon_{lk} h_d^k p^l  g^{ij}  )+\beta_{\rm{C}}'  \epsilon^{k}{}_{l} p_k h_d^l g^{ij} \nonumber\\&
+\zeta_{c{\rm C}{\rm n}}\Delta \mu\epsilon^{ik} q_k{}^j
\nonumber\\
\overline{m}^{ij}_{\rm{PC}}=& \zeta_{c\rm{PC n}}\Delta\mu(\epsilon^{ik} q_{k}{}^l C_l{}^j+\epsilon^{jk} q_k{}^l C_l{}^i)\nonumber\\
&+\tilde{\zeta}_{c\rm{PC n}}\Delta\mu(\epsilon^{ik} C_{k}{}^l q_l{}^j+\epsilon^{jk} C_k{}^{l} q_l{}^i) ~,
 \label{ConstitutiveEquationmij}
\end{align}
where to list independent terms, we have used the same relations as for the tension tensor $\bar{t}_{ij}$.
In Eq. \ref{ConstitutiveEquationmij}, we also have only introduced symmetric contributions to the bending moment tensor $\bar{m}_{ij}$.  We find the following contributions to the tensor $m_n$:
\begin{align}
m^i_{n0}&=\chi_{\rm p}\Delta \mu  \epsilon^{ij} p_j
+\psi\, \epsilon^{i}{}_{j} h_d^j \nonumber\\
m^i_{n\rm{UD}}&=(\overline{\chi}_{\rm p} \epsilon^{i}{}_{k} C^{kj} +\overline{\chi}'_{\rm p}  \epsilon^{j}{}_{k} C^{ki} ) \Delta \mu p_j  \nonumber\\
m^i_{n\rm{C}}&=(\chi_{\rm{C}p} C^{ij}  +\chi_{\rm{C}p}' C_k{}^k g^{ij})\Delta \mu p_j  \nonumber \\
m^i_{n\rm{PC}}&=\chi_{\rm{PC}\rm{p}}  \Delta \mu  p^i + \psi_{\rm PC} h_d^i~.
\label{ConstitutiveEquationmn}
\end{align}
We have used the relation $\epsilon^i{}_k C^{kj} = \epsilon^j{}_k C^{ki}+C_k{}^k \epsilon^{ij}$ (Eq. \ref{eq:appendix_product_epsilon_S_identity}) to avoid redundant couplings in the equation for the normal moment tensor $\mathbf{m}_{n\rm{UD}}$.

In the equations above, active coupling coefficients with the nematic tensor $q_{ij}$ are denoted with a subscript ``${\rm n}$'', while active polar coupling coefficients with the polar vector $\mathbf{p}$ are denoted with a subscript ``${\rm p}$''. Terms contributing to the dynamics of the polarity field reads:
\begin{align}
P_0^i=&\frac{1}{\gamma} h_d^i-\nu \tilde{v}^{ij} p_j-\nu' v_k{}^k p^i+\lambda \Delta\mu p^i+\psi\,\epsilon^{ij}\omega_{jn}\nonumber\\[.2cm]
P_{\rm  UD}^i=&
-\beta p_j D_t \tilde{C}^{ij} - \beta' p^i D_t C_k{}^k \nonumber\\
&+\lambda_{\rm UD}\Delta\mu \,C^{ij}p_j 
+\lambda'_{\rm UD}\Delta\mu \, C^{j}{}_jp^i\nonumber\\[.2cm]
P_{\rm  C}^i=&\beta_{\rm{C}} \epsilon^i{}_k D_t \tilde{C}^{kj} p_j +\beta_{\rm{C}}' \epsilon^{ij} p_j D_t C_k{}^k\nonumber\\
&+(\lambda_{\rm C} \epsilon^{i}{}_{k} C^{kj} +\lambda_{\rm C}'\epsilon^{j}{}_{k} C^{ki} ) p_j \Delta \mu
\nonumber\\
P_{\rm  PC}^i=&\frac{1}{\gamma_{\rm PC}}\epsilon^{i}{}_{j} h_d^j+\nu_{\rm{PC}} \epsilon^i{}_k \tilde{v}^{kj}p_j+\nu_{\rm{PC}}'v_k{}^k \epsilon^{ij} p_j\nonumber\\
&+\lambda_{\rm{PC}}\Delta\mu \epsilon^{ij} p_j -\psi_{\rm PC}\omega^{i}_{n} \,,
\label{constitutiveequation_dpdt}
\end{align}
where we have introduced the notation $\tilde{A}_{ij}=A_{ij}-\frac{1}{2}A_{k}{}^k g_{ij}$ for the traceless part of a tensor $A_{ij}$.
Here $1/\gamma_{\rm PC}$ is an odd inverse rotational viscosity, analogous to odd or Hall viscosities that can arise in surfaces 
with broken planar-chiral symmetry (Appendix \ref{appendix:full_constitutive_equations}).
Note that $1/\gamma_{\rm PC}$ can be nonzero in systems close to equilibrium only if time-reversal symmetry is broken, 
for example in the presence of a magnetic field. If time-reversal symmetry holds, Onsager symmetry require this
odd inverse rotational viscosity to vanish.

The flux of species $\alpha=2...N$, relative to the centre of mass has different contributions which are dependent on the polarity vector and are given by:
\begin{align}
j^{\alpha, i}_{0}&=\kappa_{\rm p}^{\alpha} p^i \Delta \mu  \nonumber\\
j^{\alpha, i}_{\rm{UD}}&=(\kappa_{\rm{UD}p}^{\alpha} C^{ij}  +\kappa_{\rm{UD}p}'^{\alpha} C_k{}^k g^{ij})p_j  \Delta \mu\nonumber \\
j^{\alpha,  i}_{\rm{C}}&=(\kappa_{\rm Cp}^{\alpha} \epsilon^{i}{}_{k} C^{kj} +\kappa_{\rm Cp}'^{\alpha}  \epsilon^{j}{}_{k} C^{ki} ) p_j \Delta \mu \nonumber\\
j^{\alpha, i}_{\rm PC}&=\kappa_{\rm PCp}^{\alpha} \epsilon^{ij} p_j\Delta \mu ~,
\label{ConstitutiveEquationfluxj}
\end{align}
where we have used the relation $\epsilon^i{}_k C^{kj} = \epsilon^j{}_k C^{ki}+C_k{}^k \epsilon^{ij}$ to avoid redundant couplings in the equation for $\mathbf{j}_{\rm C}$. 
Finally, the rate of fuel consumption has the following contributions which depend on the polarity vector:
\begin{align}
r_0=&-\zeta_{\rm n} v^{ij} q_{ij}-(\tilde{\zeta}_{c\rm n} q^{kl} C_{kl} g^{ij} + \tilde{\zeta}'_{c \rm n} C_k{}^k q^{ij}  ) D_t C_{ij}\nonumber\\
&-\chi_{\rm p} \epsilon^{ij}  p_j \omega_{in} +\lambda p_i h_d^i- \kappa_{\rm p}^{\alpha} p^i\partial_i \bar{\mu}_d^{\alpha}\\
r_{\rm UD}=&-(\tilde{\zeta}_{\rm n} q^{kl} C_{kl} g^{ij} +\tilde{\zeta}_{\rm n}' q^{ij} C_k{}^k  )v_{ij} -\zeta_{c\rm n} q^{ij} D_t C_{ij}\nonumber\\
&-(\overline{\chi}_{\rm p} \epsilon^i{}_k C^{kj} +\overline{\chi}'_{\rm p}\epsilon^j{}_k C^{ki} )p_j\omega_{in} \nonumber\\
&+(\lambda_{\rm UD} C^{ij}  +\lambda'_{\rm UD} C_k{}^k g^{ij}) p_j h_{d,i}\nonumber\\
&-( \kappa_{\rm UDp}^{\alpha}C^{ij}+\kappa_{\rm UDp}'^{\alpha} C_{k}{}^k g^{ij})p_j\partial_i \bar{\mu}_d^{\alpha}\\
r_{\rm C}=&-( 2 \zeta_{\rm Cn} \epsilon^{ik} q_{kl} C^{lj}   + 2 \tilde{\zeta}_{\rm Cn} \epsilon^{ik} C_{kl} q^{lj}  )v_{ij} \nonumber\\
& -   \zeta_{c \rm Cn}\epsilon^{ik} q_k{}^j   D_t C_{ij}\nonumber\\
&-(\chi_{\rm Cp}C^{ij}+\chi_{\rm Cp}' C_{k}{}^k g^{ij}  )p_j \omega_{in} \nonumber\\
&+(\lambda_{\rm C}\epsilon^{i}{}_{k} C^{kj} +\lambda_{\rm C}' \epsilon^{j}{}_{k} C^{ki}) p_j h_{d,i}\nonumber\\
&-(\kappa_{\rm Cp}^{\alpha} \epsilon^i{}_k C^{kj} +{\kappa'}_{\rm Cp}^{\alpha}\epsilon^j{}_k C^{ki}  )p_j\partial_i \bar{\mu}_d^{\alpha}\\
r_{\rm PC}=&- \zeta_{\rm PC n} \epsilon^{ik} q_k{}^j v_{ij} \nonumber\\
&-(2 \zeta_{c\rm PCn} \epsilon^{ik} q_k{}^l C_l{}^j  +2 \tilde{\zeta}_{c\rm PCn} \epsilon^{ik} C_k{}^l q_l{}^j )D_t C_{ij}\nonumber\\
&- \chi_{\rm PCp} p^i \omega_{in} + \lambda_{\rm PC}\epsilon^{ij} p_j h_{d,i}- \kappa_{\rm PCp}^{\alpha} \epsilon^{ij} p_j\partial_i \bar{\mu}_d^{\alpha}
\label{ConstitutiveEquationr}.
\end{align}

In the equations above, coefficients containing the letters $\nu$, $\beta$, $\zeta$, $\chi$ and $\psi$ are all reactive. Coefficients containing the letters $\gamma$, $\lambda$ and $\kappa$ are dissipative. The Onsager symmetry relations impose symmetry relations between couplings which are reactive or dissipative. Couplings which relate flux and forces with the same signature under time reversal are reactive and are antisymmetric. Conversely, couplings which relate flux and forces with opposite signatures under time reversal are dissipative and are symmetric. 

\subsubsection{Pseudopolar surfaces}

We now discuss pseudopolar surfaces. As discussed in section \ref{subsection_symmetries}, several combinations of pseudoscalar and pseudovector order parameters can be used to describe the order of a surface, such that only two physically distinct pseudopolar surfaces exist. Here we consider surfaces described by the order parameter $\mathbf{p}_{\rm C}$ and by the combination $\{\epsilon_{\rm PC}, \mathbf{p}_{\rm C}\}$. We note that for surfaces whose order is described by a pseudovector $\mathbf{p}_{\rm C}$, a true nematic tensor $q^{ij}=p_{\rm C}^i p_{\rm C}^j-\frac{1}{2} p_{{\rm C},k}~ p_{\rm C}^k g^{ij}$ can be defined. We then obtain the following constitutive equations, writing here only terms that depend on the pseudopolar vector (other terms are given in Appendix \ref{appendix:full_constitutive_equations}):
\begin{align}
\overline{t}_d^{ij}=&\zeta_{\rm n} \Delta \mu q^{ij}+\frac{\nu}{2}(p_{\rm C}^i h_{{\rm C},d}^j+p_{\rm C}^j h_{{\rm C},d}^i- h_{{\rm C},d}^k p_{{\rm C},k} g^{ij}) \nonumber\\
&+\nu' h_{{\rm C},d}^k p_{{\rm C},k} g^{ij}+\epsilon_{\rm PC}\left[\zeta_{\rm PC{\rm n}}\Delta \mu \epsilon^{ik} q_k{}^j \right.\nonumber\\
&\left.+\frac{\nu_{{\rm PC}}}{2}(\epsilon^{i}{}_{k}h^k_{{\rm C},d} p_{\rm C}^j+\epsilon^{j}{}_{k} h_{{\rm C},d}^k p_{\rm C}^i- \epsilon_{lk} h_{{\rm C},d}^k p_{\rm C}^l  g^{ij}) \right.\nonumber\\
&\left.+ \nu_{\rm PC}' \epsilon_{kl} p_{\rm C}^k h_{{\rm C},d}^l g^{ij}\right]\\
\overline{m}_d^{ij}=&\tilde{\zeta}_{c{\rm n}} q^{kl} C_{kl} g^{ij} \Delta \mu +\tilde{\zeta'_{c{\rm n}}} q^{ij}C_{k}{}^k  \Delta \mu \nonumber\\
&+\epsilon_{\rm PC}\left[\zeta_{\rm{cPC n}}\Delta\mu(\epsilon^{ik} q_{k}{}^l C_l{}^j+\epsilon^{jk} q_k{}^l C_l{}^i)\right.\nonumber\\
&\left.+\tilde{\zeta}_{\rm{cPC n}}\Delta\mu(\epsilon^{ik} C_{k}{}^l q_l{}^j+\epsilon^{jk} C_k{}^{l} q_l{}^i) \right]\\
m^i_{n,d}&=(\chi_{\rm{C}p} C^{i}{}_{j}  +\chi_{\rm{C}p}' C_k{}^k g^i{}_{j})p_{\rm C}^j  \Delta \mu \nonumber\\
&+\epsilon_{\rm PC}(\overline{\chi}_{\rm p} \epsilon^{ik} C_{kj} +\overline{\chi}'_{\rm p}  \epsilon_{jk} C^{ki} ) p_{\rm C}^j \Delta \mu \\
D_t p_{\rm C}^i =&\frac{1}{\gamma} h_{{\rm C},d}^i-\nu \tilde{v}^{ij} p_{{\rm C},j}-\nu' v_k{}^k p_{\rm C}^i+\lambda \Delta\mu p_{\rm C}^i\nonumber\\
&+\epsilon_{\rm PC}\left[\frac{1}{\gamma_{\rm PC}}\epsilon^{i}{}_{j} h_{{\rm C},d}^j+\nu_{\rm{PC}} \epsilon^{ik} \tilde{v}_{kj}p_{\rm C}^j\right.\nonumber\\
&\left.+\nu_{\rm{PC}}'v_k{}^k \epsilon^{ij} p_{{\rm C},j}+\lambda_{\rm{PC}}\Delta\mu \epsilon^{i}{}_{j} p_{\rm C}^j \right]\\
j^{\alpha, i}=&(\kappa^{\alpha}_{\rm Cp} \epsilon^{ik} C_{kj} +\kappa_{\rm Cp}'^{\alpha}   \epsilon_{jk} C^{ki} ) p_{\rm C}^j \Delta \mu  \nonumber\\
&+\epsilon_{\rm PC}(\kappa^{\alpha}_{\rm{UD}p} C^{i}{}_{j}  +\kappa_{\rm{UD}p}'^{\alpha} C_k{}^k g^{i}{}_{j})p_{\rm C}^j  \Delta \mu\\
r=&-\zeta_{\rm n} v^{ij} q_{ij}-(\tilde{\zeta}_{c{\rm n}} q^{kl} C_{kl} g^{ij} + \tilde{\zeta}'_{c\rm n} C_k{}^k q^{ij}  ) D_t C_{ij}\nonumber\\
&-(\chi_{\rm Cp}C^{ij}+\chi_{\rm Cp}' C_{k}{}^k g^{i}{}_{j}  )p_{\rm C}^j \omega_{in} \nonumber\\
&+\lambda p_{{\rm C},i} h_{{\rm C}d}^i
-(\kappa_{\rm Cp}^{\alpha} \epsilon^{ik} C_{kj} 
+\kappa_{\rm Cp}'^{\alpha}\epsilon_{jk} C^{ki}  )p_{\rm C}^j\partial_i \bar{\mu}_d^{\alpha}\nonumber\\
&- \epsilon_{\rm PC}\left[-\lambda_{\rm PC}\epsilon_{ij} p_{\rm C}^j h_{{\rm C},d}^{i} + \zeta_{\rm PC n} \epsilon^{ik} q_k{}^j v_{ij}\right.\nonumber\\
&\left. +(\overline{\chi}_{\rm p} \epsilon^i{}_k C^{kj} +\overline{\chi}'_{\rm p}\epsilon^j{}_k C^{ki} )p_j\omega_{in}\right.\nonumber\\
&\left.+( \kappa^{\alpha}_{\rm UDp}C^{ij}+\kappa_{\rm UDp}'^{\alpha} C_{k}{}^k g^{ij})p_{{\rm C},j}\partial_i \bar{\mu}_d^{\alpha}\right.\nonumber\\
&\left. +(2 \zeta_{c\rm PCn} \epsilon^{ik} q_k{}^l C_l{}^j  +2 \tilde{\zeta}_{c\rm PCn} \epsilon^{ik} C_k{}^l q_l{}^j )D_t C_{ij} \right]~.
\end{align}
In the above, we have used the notation for the molecular field associated to the pseudovector, $\mathbf{h}_{\rm C}=-\frac{\delta F}{\delta \mathbf{p}_{\rm C}}$.
All the couplings in these equations were already introduced for polar surfaces, and only a subset of couplings for polar surfaces are permitted for pseudopolar surfaces.

\section{Nematic active fluid surfaces}
\label{sec:nematic_surfaces}

We discuss here nematic surfaces by introducing a traceless nematic order parameter $\mathbf{Q}$ on the surface, tangent to the surface. Our analysis follows the same presentation as the previous section for polar surfaces. The force and torque balance equations \ref{ForceBalanceEquation} and \ref{TorqueBalanceEquation} also apply here for a nematic surface.

\subsection{Free energy and external potential}

We consider here a region of surface $\mathcal{S}$ of a multicomponent nematic fluid membrane, with free energy given by
\begin{align}
F=\int_{\mathcal{S}} dS f(\mathbf{v}, c^{\alpha}, \mathbf{C}, \mathbf{Q}, \boldsymbol{\nabla}\otimes \mathbf{Q})~,
\end{align}
 where the total free energy density $f$ has two contributions:
\begin{equation}
\label{TotalFreeEnergyDensity_nematic}
f(\mathbf{v}, c^{\alpha}, \mathbf{C}, \mathbf{Q}, \boldsymbol{\nabla}\otimes \mathbf{Q})=\frac{1}{2}\rho v^2 + f_0(c^{\alpha}, \mathbf{C}, \mathbf{Q}, \boldsymbol{\nabla}\otimes \mathbf{Q})~,
\end{equation}
with $f_0$ is the free energy density in the rest frame. The tensor $\boldsymbol{\nabla}\otimes \mathbf{Q}=\mathbf{e}^k\otimes\partial_k \mathbf{Q}$ is the gradient of nematic tensor (Eq. \ref{def:appendix:tensor_gradient}).
We introduce conjugated fields to the thermodynamic variables, such that the differential of the free energy density $f_0$ reads:
\begin{align}
\label{eq:free_energy_differential_nematic}
df_0 = \mu^{\alpha} dc^{\alpha}+\mathbf{K} :d\mathbf{C} - \mathbf{H}_0 :d\mathbf{Q} + \boldsymbol{\Pi} \triplecontract d (\boldsymbol{\nabla}\otimes \mathbf{Q})~,
\end{align} 
where in addition to the chemical potential $\mu^{\alpha}$ and the passive bending moment tensor $\mathbf{K}$, we have introduced the force conjugate to the nematic order parameter, $\mathbf{H}_0$, and the force conjugate to the gradient of the nematic order parameter, $\boldsymbol{\Pi}$. These conjugate forces are obtained by taking partial derivatives of the free energy density $f_0$:
\begin{align}
\mu^{\alpha}=&\frac{\partial f_0}{\partial c^{\alpha}},~\mathbf{K} =\frac{\partial f_0}{\partial\mathbf{C}}, \nonumber\\
\mathbf{H}_0=&-\frac{\partial f_0}{\partial \mathbf{Q}},~ \boldsymbol{\Pi} =\frac{\partial f_0}{\partial (\boldsymbol{\nabla}\otimes\mathbf{Q})}~.
\end{align}\
We assume that $f_0$ is such that $\mathbf{K}$ and $\mathbf{H}_0$ are tangent to the surface, that $\mathbf{H}_0$ is symmetric traceless, and that one can write $\boldsymbol{\Pi} =\mathbf{e}_i\otimes \boldsymbol{\Pi}^i $, with the tensor $\boldsymbol{\Pi}^k$ taken to satisfy $\Pi^{k}{}_{ij}=\Pi^{k}{}_{ji}$, $\Pi^{k}{}_{ni}=\Pi^{k}{}_{in}$ and $\Pi^k{}_{nn}=0$.

Because of rotation invariance of the free energy (Appendix \ref{AppendixGibbsDuhem} and Eq. \ref{eq:Gibbs_Duhem_rotation_nematic}), the differential of the free energy density can also be written:
\begin{equation}
\label{FluidMembraneFreeEnergyDensity_nematic}
df_0=\mu^{\alpha} d c^{\alpha}+\mathbf{K} :  D\mathbf{C} - \mathbf{H}_0 : D \mathbf{Q} +\boldsymbol{\Pi} \triplecontract 
D(\boldsymbol{\nabla}\otimes\mathbf{Q}).\\
\end{equation}
As $\mathbf{C}$ and $\mathbf{Q}$ are tangent to the surface, $D\mathbf{C}$ and $D\mathbf{Q}$ are also tangent to the surface (Appendix \ref{appendix:corot_inf_var}). 
We note that as $\mathbf{Q}$ is tangent, symmetric and traceless, $D\mathbf{Q}$ is also tangent, symmetric and traceless (Appendix \ref{appendix:corot_inf_var}).
 

The total force conjugate to the nematic field, $\mathbf{H}=-\tilde{\frac{\delta F}{\delta \mathbf{Q}}}$, reads
\begin{align}
\label{eq:def_total_molecular_field_nematic}
\mathbf{H}=\mathbf{H}_0 + \nabla_i \boldsymbol{\Pi}^{i} - \frac{1}{2}(\nabla_i \Pi^i{}^j{}_j )\mathbf{e}_k\otimes\mathbf{e}^k~.
\end{align}
where we have imposed that the tangential part $H^{ij}$ is symmetric traceless, and we have used that $\mathbf{H}_0$ is tangent, symmetric and traceless, and that the tangent part of the tensor $\nabla_k \boldsymbol{\Pi}^k$ is symmetric since $(\nabla_k \boldsymbol{\Pi}^k)\cdot(\mathbf{e}_i\otimes\mathbf{e}_j)=\nabla_k \Pi^k{}_{ij} + \Pi^k{}_{nj} C_{ki} + \Pi^k{}_{in} C_{kj}$.

As for polar surfaces, we assume that the surface is subjected to an external potential:
\begin{align} 
U=\int_{\mathcal{S}} dS c^{\alpha} u^{\alpha}(\mathbf{X}, \mathbf{n}, \mathbf{Q}),
\end{align}
 where $u^{\alpha}$ is an external potential density acting on species $\alpha$. Here again, the external potential could be provided by an external magnetic field.

\subsection{Invariance of free energy by rotation and translation}

Using the fact that the free energy density $f_0(c^{\alpha}, \mathbf{C}, \mathbf{Q}, \boldsymbol{\nabla}\otimes \mathbf{Q})$ is only a function of the surface variables and does not explicitly depend on space, one obtains the Gibbs-Duhem relation (Appendix \ref{AppendixGibbsDuhem})
\begin{align}
\label{eq:Gibbs_Duhem_translation_nematic}
\nabla_j \left[(f_0 - \mu^{\alpha} c^{\alpha})g_{i}{}^{j} - K^{jk} C_{ik} - \boldsymbol{\Pi}^j :\partial_i \mathbf{Q}\right]
+C_{ik} \nabla_j K^{jk} \nonumber\\
= -(\partial_i \mu^{\alpha})c^{\alpha}  - \mathbf{H} : \partial_i \mathbf{Q}~.
\end{align}
Invariance by rotation of the free energy also leads to the following identity, which can be seen as a generalized Gibbs-Duhem relation:
\begin{align}
\label{eq:Gibbs_Duhem_rotation_nematic}
 \boldsymbol{\Pi}^k\times_1 \partial_k \mathbf{Q} + \boldsymbol{\Pi}^k\times_2  \partial_k \mathbf{Q}+\epsilon_i{}^j (\boldsymbol{\Pi}^i  :\partial_j\mathbf{Q}) \mathbf{n}=\nonumber\\
 \mathbf{H}_0 \times_1\mathbf{Q}+ \mathbf{H}_0 \times_2\mathbf{Q}+\mathbf{C}\times_1\mathbf{K}+\mathbf{C}\times_2\mathbf{K} ~,
\end{align}
where we use a notation for the cross-product of tensors introduced in Appendix \ref{appendix_diff_geom}.

\subsection{Equilibrium relations}

To obtain equilibrium relations and expressions for the equilibrium tensions and torques, we consider changes of free energy associated to an infinitesimal virtual surface displacement $d\mathbf{X}$, virtual change of nematic tensor $d \mathbf{Q}$ and change of concentration $d c^{\alpha}$ of the component $\alpha$. Therefore, we write that for a patch of surface $\mathcal{S}$ enclosed by the contour $\mathcal{C}$,  the variation of $F_0=\int_{\mathcal{S}} dS f_0$ is:
\begin{align}
d F_0 = d W +\int_{\mathcal{S}} dS [-\mathbf{H} : D\mathbf{Q} +\mu^{\alpha}  \dbar c^{\alpha}] +\oint_{\mathcal{C}} dl \nu_i\boldsymbol{\Pi}^i : D\mathbf{Q}\label{eq:equilibrium_variations_W_nematic}~.
\end{align}
The corotational differential $D\mathbf{Q}$ has been defined in Eq. \ref{def_corot_variation_Q}.  In this equation, the molecular field $\mathbf{H}$ couples to the tangent nematic tensor $\mathbf{Q}$, while $\boldsymbol{\Pi}$ is the internal molecular field coupled to $\mathbf{Q}$ on the contour of the surface patch, $\mathcal{C}$. As for polar surfaces, the chemical potential $\mu^{\alpha}$ couples to the variation of concentration of molecular species $\alpha$ which does not arise from geometric dilution, $\dbar  c^{\alpha}$ defined in Eq. \ref{eq:variation_c_alpha_appendix}. For surfaces perturbations where the nematic tensor rotates with the surface, $D\mathbf{Q}=0$, and where the concentration fields change only by geometric dilution, $\dbar c^{\alpha}=0$, the variation of free energy is equal to the work associated with the surface deformation, $dW$.
	
Similarly, the variation of the external potential $U$ has a mechanical contribution, corresponding to the work $dW_{\rm ext}$ (Eq. \ref{eq:VirtualWorkDef}), and contributions from the variations in the nematic field $D\mathbf{Q}$ and the concentration field $\dbar c$:
\begin{align}
d U=-dW_{\rm ext}+\int_{\mathcal{S}} dS [\mathbf{H}^{\rm ext}: D\mathbf{Q} -\mu^{{\rm ext}, \alpha} \dbar c^{\alpha}]\label{eq:equilibrium_variations_Wext_nematic}~,
\end{align}
where external molecular field $\mathbf{H}^{\rm ext}$ and external chemical potential $\mu^{\alpha, {\rm ext}}_e$ are defined by:
\begin{align}
\mathbf{H}^{\rm ext}=&\frac{\delta U}{\delta \mathbf{Q}}~,\label{eq:def_Hext}\\
\mu^{\alpha, \rm ext}=&-\frac{\delta U}{\delta c^{\alpha}}~.
\end{align}
We assume that $U$ is such that that $\mathbf{H}^{\rm ext}$ is traceless, symmetric and tangent to the surface.
As a result the variation of the total free energy $F_0+U$ reads:
\begin{align}
dF_0+dU =& \int_{\mathcal{S}} dS [(\mathbf{H}^{\rm ext}-\mathbf{H}): D\mathbf{Q} +(\mu^{\alpha}-\mu^{{\rm ext}, \alpha}) \dbar c^{\alpha}]\nonumber\\
&+dW_{\rm contour}+\oint_{\mathcal{C}} dl \nu_i\boldsymbol{\Pi}^i : D\mathbf{Q}~.
\end{align}

For a nematic surface with the free energy differential introduced in Eq. \ref{eq:free_energy_differential_nematic}, we then obtain (Appendix \ref{sec:appendix_eq_nematic_fluid_membrane}):
\begin{align}
t_e^{ij} =& (f_0-\mu^\alpha c^\alpha) g^{ij} -K^{ik}C_k{}^j-\boldsymbol{\Pi}^i :\partial^j\mathbf{Q}\label{eq:tangent_tension_equilibrium_nematic}\\
t_{e,n}^i=&\nabla_j K^{ji}+\epsilon^i{}_j\left[\boldsymbol{\Gamma}_e^{{\rm ext}} - (\mathbf{H}\times_1\mathbf{Q}+\mathbf{H}\times_2\mathbf{Q}) \right]\cdot\mathbf{e}^j\label{eq:normal_tension_equilibrium_nematic}\\
\mathbf{m}_e^{i}=& K^{ik} \epsilon_{k}{}^{j} \mathbf{e}_j- \boldsymbol{\Pi}^i\times_1\mathbf{Q}-\boldsymbol{\Pi}^i\times_2\mathbf{Q}~.
\end{align}
The expression for the normal component of the equilibrium tension tensor $t^i_{e,n}$ arises from the tangential torque balance Eq. \ref{TorqueBalanceTangential}, and involves the external tangential torque density acting on the surface.

The external forces, external torques, molecular field $\mathbf{H}$ and external molecular potential $\mu^{\rm ext,\alpha}$ deriving from the external potential $U$ read (Appendix \ref{appendix_force_torque_external_potential_nematic}):
\begin{align}
	\mathbf{f}^{\rm ext}_e &= -c^{\alpha} \frac{\partial u^{\alpha}}{\partial \mathbf{X}}\label{eq:externalforcedensitynematic}\\
	\boldsymbol{\Gamma}^{\rm ext}_e&=c^{\alpha}\left(\frac{\partial u^{\alpha}}{\partial \mathbf{n}}\times \mathbf{n}
+\frac{\partial u^{\alpha}}{\partial \mathbf{Q}} \times_1 \mathbf{Q}+
	\frac{\partial u^{\alpha}}{\partial \mathbf{Q}} \times_2 \mathbf{Q}\right)
	\label{eq:externaltorquedensitynematic}\\
	\mathbf{H}^{\rm ext}&=c^{\alpha} \frac{\partial u^{\alpha}}{\partial \mathbf{Q}}\\
	\mu^{{\rm ext}, \alpha} &= -u^{\alpha}~.
\end{align}

Imposing that the total free energy $F_0+U$ is minimised at equilibrium also results in the equilibrium relations for the tangential part of the molecular field and for the chemical potential: 
\begin{align}
H^{ij} =&H^{{\rm ext}, ij} \label{eq:nematic_balance_equilibrium}~,\\
\mu^{\alpha} = &\mu^{{\rm ext}, \alpha}\label{eq:nematic_chemical_potential_balance}~.
\end{align}
Using these expressions for the equilibrium tensors, the Gibbs-Duhem relations \ref{eq:Gibbs_Duhem_translation_nematic} and \ref{eq:Gibbs_Duhem_rotation_nematic}, obtained by invariance of translations and rotations, can be rewritten in the simpler form, using Eqs. \ref{eq:appendix:external_normal_torque_nematic} and \ref{appendix:gradient_external_chemical_potential_nematic}:
\begin{align}
\label{eq:equilibrium_tangential_force_balance}
\nabla_i t_e^{ij}+C_{i}{}^j t_{e,n}^i +f_e^{{\rm ext}, j}=&(H^{{\rm ext}}_{ik}-H_{ik}) \nabla^j Q^{ik} \nonumber\\
&+c^{\alpha}\partial^j(\mu^{{\rm ext}, \alpha}- \mu^{\alpha})  ~,\\
\label{eq:equilibrium_normal_torque_balance}
\nabla_i m_{n,e}^i-C_{ij} m_e^{ij}+\epsilon_{ij}t_e^{ij}+\Gamma_{e,n}^{\rm ext}=&2(H^{{\rm ext}, ij} - H^{ij})Q_{ik} \epsilon_j{}^k ~.
\end{align}
At equilibrium, the conditions \ref{eq:nematic_chemical_potential_balance} and \ref{eq:nematic_balance_equilibrium} imply that the right-hand side of these two equations cancel; such that Eqs. \ref{eq:equilibrium_tangential_force_balance} and \ref{eq:equilibrium_normal_torque_balance} are then equivalent to the tangential force balance and normal torque balance.

\subsection{Entropy production}
For a nematic surface, the rate of change of the total free energy $F_{\rm tot}=F+U$ reads:
\begin{align}
\label{eq:total_rate_free_energy_nematic}
	\dot{F}_{\rm tot}=&\int_\mathcal{S} dS\,
	\Big\{-\left(\overline{t}^{ij}-\overline{t}^{ij}_e\right) v_{ij}
	-(\overline{m}^{ij}-\overline{m}^{ij}_e)D_t C_{ij}	\nonumber \\
	&-(m_n^i-m_{n,e}^i) \omega_{in}
 - (\mathbf{H}-\mathbf{H}^{{\rm ext}} ):D_t \mathbf{Q} \nonumber\\
	&
+(\mu^\alpha -\mu^{{\rm ext},\alpha})r^\alpha+j^{\alpha,i}\partial_i(\overline{\mu}^{\alpha}-\overline{\mu}^{{\rm ext},\alpha})\nonumber\\
	&+(\mathbf{f}^\mathrm{ext}-\mathbf{f}_e^\mathrm{ext})\cdot\mathbf{v}+(\boldsymbol{\Gamma}^\mathrm{ext}-\boldsymbol{\Gamma}_e^\mathrm{ext})\cdot\boldsymbol{\omega}	\Big\}\nonumber\\
	&+
	\oint_\mathcal{C} dl\nu_i \left[\mathbf{t}^i\cdot \mathbf{v}+\mathbf{m}^i\cdot\boldsymbol{\omega}+\boldsymbol{\Pi}^i : D_t\mathbf{Q} \right.\nonumber\\
	&\left.\hspace{1.5cm}-(\overline{\mu}^\alpha -\overline{\mu}^{\rm ext,\alpha})j^{\alpha,i}\right]\,.
\end{align}
As for polar surface, we consider a single chemical reaction between a fuel and its product, such that we rewrite the term $(\mu^{\alpha}-\mu^{{\rm ext}, \alpha}) r^{\alpha}=-r\Delta \mu$ with $r$ the rate of fuel consumption and $\Delta\mu$ the chemical potential of conversion of fuel to product.  
Using Eq. \ref{eq:appendix_free_energy_variation}, we then identify the entropy production rate introduced in Eq. \ref{eq:balance_free_energy}:
\begin{align}
\label{eq:entropy_production_rate_nematic}
T\theta =& \left(\overline{t}^{ij}-\overline{t}^{ij}_e\right) v_{ij}
	+(\overline{m}^{ij}-\overline{m}_e^{ij}) D_t C_{ij}\nonumber\\
&	+ (m_n^i-m_{n,e}^i)\omega_{in}+(H^{ij}-H^{{\rm ext},ij} )D_t Q_{ij} \nonumber\\
&+ r \Delta \mu-j^{\alpha,i}\partial_i(\overline{\mu}^{\alpha}-\overline{\mu}^{{\rm ext},\alpha})~,
\end{align} 
and the normal and tangential fluxes of free energy:
\begin{align}
J_n^f=&(\mathbf{f}^\mathrm{ext}-\mathbf{f}^{\rm ext}_e)\cdot\mathbf{v} + (\boldsymbol{\Gamma}^\mathrm{ext}-\boldsymbol{\Gamma}_e^\mathrm{ext})\cdot\boldsymbol{\omega}\\
j^{f,i}=&-\mathbf{t}^i\cdot \mathbf{v}-\mathbf{m}^i\cdot\boldsymbol{\omega}-\Pi^{ijk} D_t Q_{jk}  \nonumber\\
&+(\overline{\mu}^\alpha -\overline{\mu}^{\rm ext,\alpha})j^{\alpha,i}~.
\end{align}
Here we have assumed that external forces and torques do not contribute to the surface entropy production.
In Table \ref{Table:nematic_forces_fluxes}, we list the corresponding conjugate fluxes and forces that can be read from the expression for the entropy production rate, Eq. \ref{eq:entropy_production_rate_nematic}. There we introduce the deviatoric molecular field, the tangent tensor $H_d^{ij}=H^{ij} - H^{{\rm ext}, ij}$, and the deviatoric chemical potential, $\mu_d^{\alpha}= \mu^{\alpha}- \mu^{{\rm ext},\alpha}$. 
\begin{table*}[th]
\begin{center}
\begin{tabular}{l@{\hskip 1in}c@{\hskip 0.5in}}
\hline\hline\\[-.35cm]
Flux & Force\\[.08cm] \hline\\[-.3cm]
In-plane deviatoric tension tensor $\overline{t}^{ij}_d$  &In-plane shear tensor $v_{ij}$  \\
 In-plane deviatoric bending moment tensor $\overline{m}^{ij}_d$   &Bending rate tensor $D_t C_{ij}$\\
Normal deviatoric moment $m_{n,\,d}^i$  & Vorticity gradient $\omega_{in}=(\partial_i \boldsymbol{\omega})\cdot \mathbf{n}$\\
Convected, co-rotational derivative of nematic tensor $D_t Q_{ij}$ & Deviatoric molecular field $H_{d, ij}=H_{ij}-H^{\rm ext}_{ij}$ \\
Rate of fuel consumption $r$ & Fuel hydrolysis chemical potential $\Delta\mu$\\
Flux of species $\alpha=2...N$, $j^{\alpha, i}$ & \makecell{Deviatoric gradient of relative chemical potential\\ of species $\alpha=2...N$, $-\partial_i \overline{\mu}_d^{\alpha}=-\partial_i (\overline{\mu}^{\alpha}-\overline{\mu}^{{\rm ext},\alpha})$}\\
\end{tabular}
\end{center}
\caption{\label{Table:nematic_forces_fluxes} List of conjugate forces and fluxes for an active fluid nematic surface.}
\end{table*}

\subsection{Constitutive equations}

We now obtain constitutive equations for different types of nematic surfaces, as illustrated in Fig.~\ref{fig:symmetries}.

\subsubsection{Nematic surfaces}
The deviatoric part of the stress and moment tensor can be decomposed as in Eq.~\ref{GeneralTensorDecompositionSymmetry}.
In writing down these relations, the nematic couplings already present in Eqs.~\ref{ConstitutiveEquationtij}-\ref{ConstitutiveEquationmij} carry over here, with $Q_{ij}$ replacing $q_{ij}$. Whenever possible we give the same name to phenomenological coefficients which resemble those already introduced for a polar surface. As for polar surfaces, for the sake of simplicity we make a number of restrictions on couplings: we do not include couplings between the various mechanical tensors and the deviatoric molecular field $H^{ij}_d$ that depend on the curvature tensor. Couplings are included to linear order in the curvature tensor $\mathbf{C}$ and in the nematic tensor $\mathbf{Q}$, except for diagonal coupling terms whose dependency in $\mathbf{C}$ and $\mathbf{Q}$ is not written. Viscous coupling terms between $\bar{t}^{ij}_d$, $\bar{m}^{ij}$, $m_{n,d}^i$ and $v_{ij}$, $D_tC_{ij}$, $\omega_{in}$ that are dependent on the nematic and curvature tensors are not listed. We do not include cross-coupling terms in the gradient of the chemical potential, $\partial_i \bar{\mu}_d^{\alpha}$ or the flux $j^{\alpha,i}$, except for active couplings with the fuel hydrolysis chemical potential, $\Delta \mu$.

Below we only list terms which involve the nematic tensor $Q^{ij}$ and the deviatoric molecular field $H^{ij}_d$; the complete constitutive equations are given in Appendix \ref{appendix:full_constitutive_equations}.
The contributions to the tension tensor which depend on the nematic tensor $\mathbf{Q}$ read:
\begin{align}
\overline{t}^{ij}_0=&\nu H_d^{ij}+\nu' Q_{kl} H_d^{kl} g^{ij}+\zeta_{\rm n} \Delta \mu Q^{ij}\nonumber\\
\overline{t}^{ij}_{\rm{UD}}=&\tilde{\zeta}_{\rm n} \Delta \mu Q^{kl} C_{kl} g^{ij}   +\tilde{\zeta}'_{\rm n}  \Delta \mu Q^{ij} C_{k}{}^k 
\nonumber \\[.2cm]
\overline{t}^{ij}_{\rm{C}}=&\zeta_{\rm{C n}}\Delta\mu(\epsilon^{ik} Q_{k}{}^l C_l{}^j+\epsilon^{jk} Q_{k}{}^{l} C_l{}^i)\nonumber\\
&+\tilde{\zeta}_{\rm{C n}}\Delta\mu(\epsilon^{ik} C_{k}{}^l Q_l{}^j+\epsilon^{jk} C_k{}^l Q_l{}^i)  \nonumber \\[.2cm]
\overline{t}^{ij}_{\rm{PC}}=& \nu_{\rm PC} \epsilon^{i}{}_{k} H_{d}^{kj}+\nu_{\rm PC}' \epsilon^{kl} Q_{km}H_d^{ml} g^{ij}\nonumber\\
&+\zeta_{\rm PC{\rm n}}\Delta \mu \epsilon^{ik} Q_k{}^j ~,
	\label{ConstitutiveEquationtij_nematic}
\end{align}
where we have used that $[Q_{ik} H_d^k{}_j]_s=\frac{1}{2} Q_{kl} H_d^{kl} g_{ij}$, that $[\boldsymbol{\epsilon} \mathbf{Q}\mathbf{H}_d]_s=[ \mathbf{Q}\mathbf{H}_d \boldsymbol{\epsilon}]_s=[ \mathbf{H}_d \boldsymbol{\epsilon}\mathbf{Q}]_s=-[ \boldsymbol{\epsilon}\mathbf{H}_d \mathbf{Q}]_s =-[ \mathbf{H}_d \mathbf{Q}\boldsymbol{\epsilon}]_s=-[ \mathbf{Q} \boldsymbol{\epsilon}\mathbf{H}_d]_s$ and that $[\epsilon^{ik}Q_{d,kl}H_d^{lj}]_s=\frac{1}{2}\epsilon^{kl}Q_d^{lm}  H_{d, mk} g^{ij}$, which follows from $\mathbf{H}_d$ and $\mathbf{Q}$ being tangent, symmetric and traceless (Eqs. \ref{eq:appendix:product_Q_Q_identity}-\ref{eq:appendix:product_epsilon_Q_Q_identity}), to avoid redundant couplings. We also note that the tensor $\epsilon^{ik} Q_{k}{}^j$ is symmetric. We find only two independent active terms in the contribution $\bar{t}_{\rm C}^{ij}$ proportional to $[\boldsymbol{\epsilon} \mathbf{Q} \mathbf{C}]_s$ and $[\boldsymbol{\epsilon} \mathbf{C} \mathbf{Q}]_s$; following the same arguments as given for the polar case after Eq. \ref{ConstitutiveEquationtij}.

The contributions to the moment tensor, in the decomposition indicated in Eq. \ref{GeneralTensorDecompositionSymmetry}, which depend on the nematic tensor, read
\begin{align}
\overline{m}^{ij}_0=&\tilde{\zeta}_{c{\rm n}}  \Delta \mu  Q^{kl} C_{kl} g^{ij}+\tilde{\zeta}'_{c\rm n}  \Delta \mu Q^{ij}C_{k}{}^k 
 \nonumber\\
\overline{m}^{ij}_{\rm{UD}}=&\beta H_d^{ij}+\beta' Q_{kl} H_d^{kl} g^{ij}+\zeta_{c{\rm n}} \Delta \mu Q^{ij} \nonumber\\
\overline{m}^{ij}_{\rm{C}}=&\beta_{\rm C}  \epsilon^i{}_k H_d^{kj}
+\beta_{\rm C}' \epsilon^{kl} Q_{km}H_d^{ml} g^{ij}+\zeta_{c{\rm C}{\rm n}} \Delta\mu\epsilon^{ik} Q_k{}^j \nonumber\\
\overline{m}^{ij}_{\rm{PC}}=&\zeta_{c\rm{PC n}}\Delta\mu(\epsilon^{ik} Q_{k}{}^l C_l{}^j+\epsilon^{jk} Q_k{}^l C_l{}^i)\nonumber\\
&+\tilde{\zeta}_{c\rm{PC n}}\Delta\mu(\epsilon^{ik} C_{k}{}^l Q_l{}^j+\epsilon^{jk} C_k{}^{l} Q_l{}^i) ~.
	\label{ConstitutiveEquationmij_nematic}
\end{align}
In Eq. \ref{ConstitutiveEquationmij_nematic}, we have only introduced symmetric contributions to the bending moment tensor.  With our simplifying assumptions, there is no contribution to the normal moment $\mathbf{m}_n$, and to the flux of component $\alpha$, $\mathbf{j}^{\alpha}$, involving the nematic order parameter $\mathbf{Q}$ or the molecular field $\mathbf{H}$.

The co-moving, co-rotating derivative of the nematic order parameter field reads:
\begin{align}
D_t Q^{ij}=\mathcal{Q}^{ij}_0+\epsilon_{\rm UD} \mathcal{Q}^{ij}_{\rm UD}+\epsilon_{\rm C} \mathcal{Q}^{ij}_{\rm C}+\epsilon_{\rm PC}\mathcal{Q}^{ij}_{\rm PC}
\end{align}
with
\begin{align}
	\mathcal{Q}_0^{ij}=&\frac{1}{\gamma}H_d^{ij}-\nu \tilde{v}^{ij}	-\nu' v_k{}^k Q^{ij}+\lambda \Delta\mu \,Q^{ij}
\nonumber\\
	\mathcal{Q}_{\rm  UD}^{ij}=&
	-\beta  D_t \tilde{C}^{ij}-\beta' D_tC_k{}^k Q^{ij}\nonumber\\&
	+\lambda_{\rm UD}\Delta\mu \,\tilde{C}^{ij}
	+\lambda_{\rm UD}'\Delta\mu C_k{}^k Q^{ij}
	\nonumber\\[.2cm]
	\mathcal{Q}_{\rm  C}^{ij}=& \beta_{\rm C} \epsilon^i{}_k D_t \tilde{C}^{kj}+\beta_{\rm C}' \epsilon^{i}{}_{k}Q^{kj} D_t C_{l}{}^l \nonumber\\&+
	\lambda_{\rm C}\Delta\mu (\epsilon^i{}_k C^{kj}+\epsilon^j{}_k C^{ki})
+\lambda_{\rm{C n}}\Delta\mu C_k{}^k \epsilon^{ik} Q_k{}^j\nonumber\\
	\mathcal{Q}_{\rm  PC}^{ij}=&\frac{1}{\gamma_{\rm PC}} \epsilon^{i}{}_{k} H_d^{kj}+\nu_{\rm PC} \epsilon^{ik} \tilde{v}_k{}^j\nonumber\\
	&+ \nu'_{{\rm PC}}\epsilon^{i}{}_{k}Q^{kj} v_{l}{}^l +\lambda_{\rm PC}\Delta\mu \epsilon^{ik} Q_{k}{}^{j}	\label{constitutiveequation_dQdt}~.
\end{align}
where we have used that $[C^{i}{}_{k}Q^{kj}]_s - \frac{1}{2} C_{kl}Q^{kl} g^{ij} =\frac{1}{2}C_k{}^k Q^{ij}$, following Eq. \ref{eq:appendix_product_S_Q_identity}, to avoid redundant terms. We have also used the identities \ref{eq:epsilon_C_q_identity}-\ref{eq:epsilon_q_C_identity} as well as the tensor identities $\mathbf{Q}\boldsymbol{\epsilon} \mathbf{C} =-\boldsymbol{\epsilon} \mathbf{Q} \mathbf{C} $, $\mathbf{C} \boldsymbol{\epsilon} \mathbf{Q} = -\mathbf{C} \mathbf{Q}\boldsymbol{\epsilon}$, $[\mathbf{C}\mathbf{Q}\boldsymbol{\epsilon}]_s=-[\boldsymbol{\epsilon} \mathbf{Q}\mathbf{C}]_s$ and $[\mathbf{Q}\mathbf{C}\boldsymbol{\epsilon}]_s=-[\boldsymbol{\epsilon} \mathbf{C}\mathbf{Q}]_s$, and that $D_t Q^{ij}$ is traceless, to obtain the only term proportional to $\boldsymbol{\epsilon}$, $\mathbf{C}$ and $\mathbf{Q}$. With a similar reasoning we obtained one term proportional to $\boldsymbol{\epsilon}$, $v^{ij}$ and $\mathbf{Q}$ and one term proportional to $\boldsymbol{\epsilon}$, $D_t C^{ij}$ and $\mathbf{Q}$. As for the polar case, the odd inverse rotational viscosity $1/\gamma_{\rm PC}$ can be nonzero in systems close to equilibrium only if time-reversal symmetry is broken, 
for example in the presence of a magnetic field. If time-reversal symmetry holds, Onsager symmetry require this
odd inverse rotational viscosity to vanish.

Finally, the rate of fuel consumption has the following contributions which depend on the nematic tensor:
\begin{align}
r_0=&-\zeta_{\rm n} v^{ij} Q_{ij}-(\tilde{\zeta}_{c\rm n} Q^{kl} C_{kl} g^{ij} + \tilde{\zeta}'_{c\rm n} C_k{}^k Q^{ij}  )D_t C_{ij}\nonumber\\
&+\lambda Q_{ij} H_d^{ij}\\
r_{\rm UD}=&-(\tilde{\zeta}_n Q^{kl} C_{kl} g^{ij} +\tilde{\zeta}_n' Q^{ij} C_k{}^k  )v_{ij} -\zeta_{c\rm n} Q^{ij} D_t C_{ij} \nonumber\\
&+(\lambda_{\rm UD} \tilde{C}^{ij}  +\lambda'_{\rm UD} C_k{}^k Q^{ij}   ) {H_d}_{ij}\\
r_{\rm C}=&-( 2 \zeta_{\rm Cn} \epsilon^{ik} Q_{kl} C_l{}^j   + 2 \tilde{\zeta}_{\rm Cn} \epsilon^{ik} C_k{}^l Q_l{}^j  )v_{ij} \nonumber\\
& -  \zeta_{c\rm Cn}\epsilon^{ik} Q_k{}^j   D_t C_{ij}+2 \lambda_{\rm C} \epsilon_{jk} C^{k}{}_{i}  {H_d}^{ij}\nonumber\\
&+\lambda_{\rm{C n}} C_k{}^k \epsilon^{ik} Q_k{}^j {H_d}_{ij} \nonumber\\
r_{\rm PC}=&- \zeta_{\rm PC n} \epsilon^{ik} Q_k{}^j v_{ij} +\lambda_{\rm PC} \epsilon_{ik} Q^k{}_j H_d^{ij}\nonumber\\
&-(2 \zeta_{\rm cPCn} \epsilon^{ik} Q_k{}^l C_l{}^j  +2 \tilde{\zeta}_{\rm cPCn} \epsilon^{ik} C_k{}^l Q_l{}^j ) D_t C_{ij}~.
\end{align}

In the equations above, coefficients which contain the letters $\nu$, $\beta$ or $\zeta$ are all reactive. Coefficients containing the letters $\gamma$, $\lambda$ are dissipative.
\subsubsection{Pseudonematic surfaces}

Here we consider the two types of pseudonematic surfaces and assume that their order is described by, respectively, the pseudotensor $\mathbf{Q}_{\rm UD}$ and the combination of the pseudoscalar and pseudovector $\{\epsilon_{\rm PC},\mathbf{Q}_{\rm UD} \}$.
With this restriction, the contributions to the constitutive equations for the pseudonematic surfaces which depend on the pseudonematic tensor read:
\begin{align}
\overline{t}^{ij}_d=&\nu' Q_{{\rm UD},kl} H_{{\rm UD}d}^{kl} g^{ij}+\tilde{\zeta}_{\rm n}  \Delta \mu  Q_{\rm UD}^{kl} C_{kl} g^{ij} \nonumber\\
&+\tilde{\zeta}'_{\rm n} \Delta \mu Q_{\rm UD}^{ij} C_{k}{}^k +\epsilon_{\rm PC}\left[\nu_{\rm PC}' \epsilon^{kl} Q_{{\rm UD},km}H_{{\rm UD}d}^{ml} g^{ij}\right. \nonumber\\
&\left.+\zeta_{\rm{C n}}\Delta\mu(\epsilon^{i}{}_{k} Q_{\rm UD}^{kl} C_l{}^j+\epsilon^{j}{}_{k} Q_{\rm UD}^{kl} C_l{}^i)\right.\nonumber\\
&\left.+\tilde{\zeta}_{\rm{C n}}\Delta\mu(\epsilon^{ik} C_{kl} Q_{\rm UD}^{lj}+\epsilon^{jk} C_{kl}Q_{\rm UD}^{li}) \right]\\
\overline{m}^{ij}_d=&\beta H_{{\rm UD}d}^{ij}+\zeta_{c{\rm n}} \Delta \mu Q_{\rm UD}^{ij} \nonumber\\
&+\epsilon_{\rm PC}\left[\beta_{\rm C}  \epsilon^i{}_k H_{{\rm UD}d}^{kj} +\zeta_{c{\rm C}{\rm n}}\Delta\mu \epsilon^{i}{}_{k} Q_{\rm UD}^{kj}\right]\\
D_t Q^{ij}_{\rm UD}=&\frac{1}{\gamma}H_{{\rm UD}d}^{ij}+\lambda \Delta\mu Q_{\rm UD}^{ij}	+\lambda_{\rm UD}\Delta\mu \,\tilde{C}^{ij}\nonumber\\
&	-\nu' v_k{}^k Q_{\rm UD}^{ij}-\beta  D_t \tilde{C}^{ij}\nonumber\\
	&
	+\epsilon_{\rm PC}\left[\frac{1}{\gamma_{\rm PC}} \epsilon^{i}{}_{k} H_{{\rm UD}d}^{kj}+ \nu'_{{\rm PC}}\epsilon^{i}{}_{k}Q_{\rm UD}^{kj} v_{l}{}^l\right.\nonumber\\
	&\left.+\frac{1}{2} \beta_{\rm C}\,\left(\epsilon^i{}_k D_t C^{kj}+\epsilon^j{}_k D_t C^{ki}\right) \right.\nonumber\\
	&\left.+	\lambda_{\rm C}\Delta\mu (\epsilon^i{}_k C^{kj}+\epsilon^j{}_k C^{ki})+\lambda_{\rm PC}\Delta\mu \epsilon^{i}{}_{k} Q_{\rm UD}^{kj}	\right]\\
r=&\lambda Q_{\rm UD}^{ij} H_{{\rm UD}d,ij}+\lambda_{\rm UD} \tilde{C}_{ij}      H_{{\rm UD}d}^{ij}-\zeta_{c\rm n} Q_{\rm UD}^{ij} D_t C_{ij}   \nonumber\\
& -(\tilde{\zeta}_{\rm n} Q_{\rm UD}^{kl} C_{kl} g^{ij} +\tilde{\zeta}_{\rm n}' Q_{\rm UD}^{ij} C_k{}^k  )v_{ij}\nonumber\\
&+\epsilon_{\rm PC}\left[-( 2 \zeta_{\rm Cn} \epsilon^{i}{}_{k} Q_{{\rm UD}}^{kl} C_{l}{}^{j}   + 2 \tilde{\zeta}_{\rm Cn} \epsilon^{ik} C_{kl} Q_{\rm UD}^{lj}  )v_{ij}\right.\nonumber\\
&\left.-   \zeta_{\rm cCn}\epsilon^{i}{}_{k} Q_{\rm UD}^{kj}   D_t C_{ij} +2 \lambda_{\rm C} \epsilon^{j}{}_{k} C^{ki} H_{{\rm UD}d,ij}\right.\nonumber\\
&\left.+\lambda_{\rm PC} \epsilon_{ik} {Q_{\rm UD}}^{k}{}_{j} H_{{\rm UD}d}^{ij}\right]~.
\end{align}

\section{Active polar film in a confined geometry}
\label{sec:active_polar_film_confined}

\begin{figure}
 \centering
 \includegraphics[width=1.0\columnwidth]{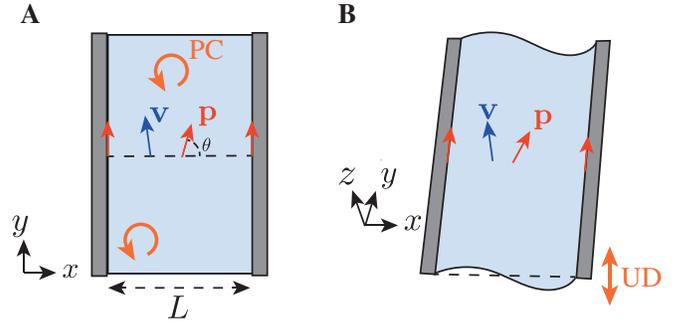}
 \caption{Schematic for a confined polar active film, with planar anchoring at boundaries. A. Flat planar-chiral membrane. B. Deformed active polar surface with broken up-down symmetry.
 \label{fig:application_schematic}
 } 
\end{figure}

We now discuss a one-component polar active film confined along its $x$ axis in between $x=0$ and $x=L$, invariant by translation in the $y$ direction (Fig. \ref{fig:application_schematic}). We consider the limit of low Reynolds number where inertial terms can be neglected. For a flat, polar, achiral surface, and planar anchoring of the polarity at the wall, an instability arises at a threshold value of nematic active stress, leading to a distorted polarity profile and spontaneous flow in the active film \cite{voituriez2005spontaneous}. Here we consider a similar geometry and explore a more general situation. We first investigate the case of a planar-chiral flat surface, 
and then consider an active film which can deform weakly in the third dimension and has broken up-down symmetry (Fig. \ref{fig:application_schematic}).

We describe a polar active surface with a tangent polar vector $\mathbf{p}$. We assume that the equilibrium behaviour of the surface is described by the free energy
\begin{align}
\label{eq:free_energy_application}
F=\int_{\mathcal{S}} dS\left[ \gamma+ \frac{\kappa}{2} (C_i{}^i )^2+\frac{K}{2} (\partial_i \mathbf{p})^2  +\frac{\lambda_p}{2}(\mathbf{p}^2-1)\right]~,
\end{align}
where $\lambda_p$ is a Lagrange multiplier enforcing the norm of $\mathbf{p}$, $\gamma$ is the passive surface tension, $\kappa>0$ the surface bending modulus. $K>0$ is the polarity distortion modulus in the single-constant approximation, and we denote $(\partial_i \mathbf{p})^2=(\partial_i \mathbf{p})\cdot (\partial^i\mathbf{p})$. Here we assume that the passive surface has no spontaneous curvature. We note that the polarity distortion term in $ (\partial_i \mathbf{p})^2=(\nabla_i p_j )(\nabla^i p^j) + p_i C^{ij} p^k C_{kj}$ introduces a coupling between polarity and curvature. For simplicity we did not introduce additional coupling terms between the curvature tensor and polarity in Eq. \ref{eq:free_energy_application}. The surface is assumed to be free from external force and is not subjected to an external potential.

Differentation of the free energy density $f_0=  \gamma+ \frac{\kappa}{2} (C_i{}^i )^2 +K(\partial_i \mathbf{p})^2/2+\frac{\lambda_p}{2}(\mathbf{p}^2-1) $ with respect to $\mathbf{p}$ and the curvature tensor results in the following expressions for thermodynamic fields conjugate to $\mathbf{C}$ and $\mathbf{p}$, as defined in Eq. \ref{FluidMembraneFreeEnergyDensity}:
\begin{align}
K^{ij} = &\kappa C_k{}^k g^{ij}~,\\
\mathbf{h}_0 =& - \lambda_p \mathbf{p}~,\\
\boldsymbol{\pi} =& K \mathbf{e}^i \otimes \partial_i \mathbf{p}~,
\end{align}
and following Eq. \ref{eq:def_total_molecular_field_h}, the total molecular field is then given by
\begin{align}
\label{eq:application_total_molecular_field}
\mathbf{h} = -\lambda_p \mathbf{p} + K\nabla_i (\partial^i \mathbf{p})~.
\end{align}
The corresponding equilibrium tensors are (Eqs. \ref{eq:equilibriumtangentialtension}, \ref{FluidEquilibriumNormalTorqueTensor}):
\begin{align}
t_e^{ij}=&\left[\gamma+\frac{\kappa}{2} (C_k{}^k)^2+\frac{K}{2} (\partial_i \mathbf{p})^2   \right] g^{ij}-\kappa C_k{}^k C^{ij} \nonumber\\
& - K (\partial^i\mathbf{p}) \cdot (\partial^j\mathbf{p})~,\label{eq:application:equilibrium_t}\\
\mathbf{m}_e^{i}=&\kappa C_k{}^k \epsilon^{ij} \mathbf{e}_j + K \mathbf{p} \times \partial^i \mathbf{p}~.\label{eq:application:equilibrium_m}
\end{align}

Deviatoric, non equilibrium contributions to the tension and moment tensors, and the polarity dynamic equation, are determined from the general constitutive equations \ref{Appendix:ConstitutiveEquationtij}-\ref{Appendix: constitutiveequation_dpdt} with the following simplifications: we set $\eta_{\rm PC}=\eta_{c{\rm PC}}=1/\gamma_{\rm PC}=0$ and we do not include cross-coupling viscosities; we also do not include non-equilibrium contributions to $m_n^i$ nor coupling to the vorticity gradient $\omega_{in}$. We also consider the surface to be incompressible, corresponding to the limit $\eta_b\rightarrow \infty$ and $v_k{}^k =0$. In the discussion below, we assume that $\Delta \mu>0$.

\subsection{Flat planar-chiral membrane}

\begin{figure}
 \centering
 \includegraphics[width=1.0\columnwidth]{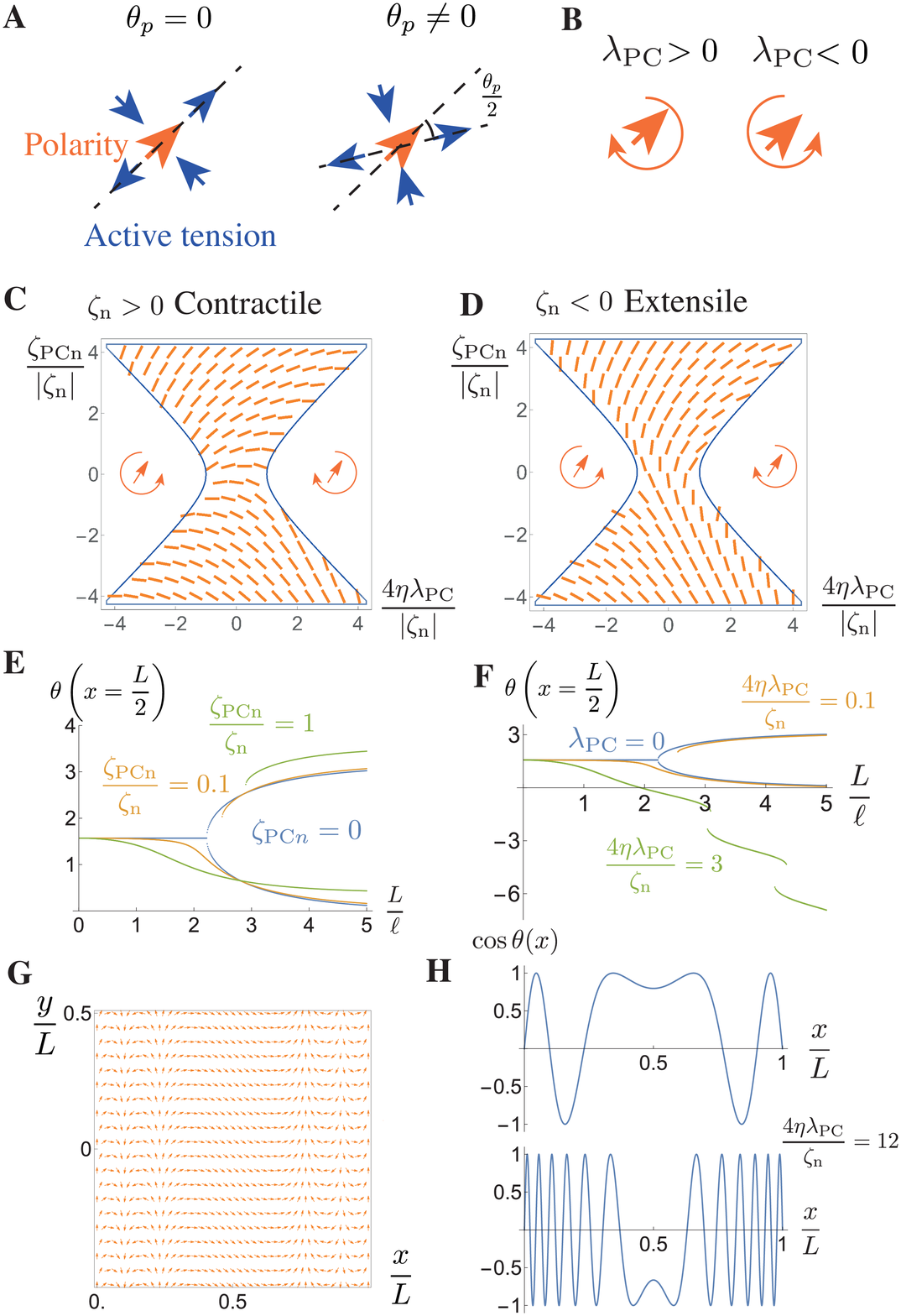}
 \caption{ Phase diagram for a quasi one-dimensional, confined, flat polar active surface with broken planar-chiral symmetry. {\bf A}. Depending on the active tension parameters $\zeta_{{\rm PC n}}$ and $\zeta_{\rm n}$, a principal axis of the active tension tensor deviates by an angle $-\theta_p/2$ from the polarity axis, for a surface with broken planar-chiral symmetry. Here the active tension is represented in the contractile case, $\zeta_{\rm n}>0$. {\bf B}. The parameter $\lambda_{\rm PC}$ triggers an active rotation of the polar vector. {\bf C}, {\bf D}: Phase diagram of homogeneous polarity orientation as a function of the two planar-chirality coefficients, for the contractile and extensile case, for $\nu=\nu_{\rm {PC}}=0$. Outside of the ``hourglass'' shape, the polarity is rotating with a time-varying angular velocity. {\bf E}. Bifurcation plot showing the angle at the center of the confined active film as a function of system size $L$ ($\lambda_{\rm PC}=0$, $\zeta_{\rm n}>0$, $\nu=\nu_{\rm PC}=0$, $\ell=\sqrt{K(1+4\eta/\gamma)/(\zeta_{\rm n} \Delta\mu)}$). For small values of $L$ and $\zeta_{{\rm PC n}}=0$, the polar angle is homogeneous and equal to $\frac{\pi}{2}$; for larger values of $L$ a pitchfork bifurcation appears. The pitchfork bifurcation becomes imperfect for $\zeta_{{\rm PC n}}\neq 0$. {\bf F}. Bifurcation plot as a function of system size $L$, for different values of $\lambda_{\rm PC}$ ($\zeta_{{\rm PC n}}=0$, $\zeta_{\rm n}>0$, $\nu=\nu_{\rm PC}=0$, $\ell=\sqrt{K(1+4\eta/\gamma)/(\zeta_{\rm n}\Delta\mu) }$). For large enough values of $\lambda_{\rm PC}$, the magnitude of $\theta(x=\frac{L}{2})$ keeps increasing as $L$ increases, corresponding to the formation of repeated polarity turns in the system. ({\bf G}, {\bf H}). Steady-state polarity profiles for $\nu=\nu_{\rm PC}=\zeta_{{\rm PCn}}=0$, $\zeta_{\rm n}>0$, $L/\ell=5$, $4\eta \lambda_{\rm PC}/\zeta_{\rm n} = 3$ or as indicated.
 \label{fig:application_PC}
 } 
\end{figure}

We first consider a flat surface, such that $C_{ij}=0$, confined at its boundaries, such that $v_x(x=0)=v_x(x=L)=0$, and we also assume a vanishing shear stress at the boundaries, resulting in $t_{xy}=0$ (Appendix \ref{appendix:application:PC}). The constitutive equations for the tension and the polarity dynamics are:
\begin{align}
t^{ij}=&-P g^{ij} + 2\eta \tilde{v}^{ij} +\frac{\nu}{2}(p^i h^j+p^j h^i - p^k h_k g^{ij})\nonumber\\
&+\frac{\nu_{{\rm PC}}}{2}(\epsilon^{i}{}_{k}h^k p^j+\epsilon^{j}{}_{k} h^k p^i - \epsilon_{kl} h^k p^l g^{ij}) \nonumber\\
&-\frac{1}{2}\epsilon_{kl}p^k h^l  \epsilon^{ij}+\zeta_{\rm n} \Delta \mu q^{ij}+\zeta_{\rm PC{\rm n}}\Delta \mu \epsilon^{ik} q_k{}^j \nonumber\\
&-K ((\nabla^i p_k )(\nabla^j p^k) - \frac{1}{2} (\nabla_l p_k )(\nabla^l p^k )g^{ij})~,\label{eq:tension_tensor_application_flat_PC}\\
D_t p^i&=\frac{1}{\gamma} h^i-\nu \tilde{v}^{ij} p_j+\nu_{\rm{PC}} \epsilon^i{}_k \tilde{v}^{kj}p_j+\lambda_{\rm{PC}}\Delta\mu \epsilon^{ij} p_j \label{eq:polarity_dynamics_application_flat_PC} \,,
\end{align}
where we have introduced a two dimensional pressure $P$ which enforces the incompressibility condition. We take here $\lambda=0$, as the corresponding term acts in the polarity equation to change the norm of the polarity, which is constrained here. In general this term also renormalizes tensions by modifying the molecular field $\mathbf{h}$. In the equations above, $\gamma$ is an inverse rotational viscosity, $\nu$ is a flow-coupling alignment whose role have been discussed extensively \cite{degennesprost, voituriez2005spontaneous}. Constitutive equations related to Eqs. \ref{eq:tension_tensor_application_flat_PC} and \ref{eq:polarity_dynamics_application_flat_PC} have been considered in Refs. \cite{duclos2018spontaneous, maitra2019spontaneous, li2020pattern}. Ref. \cite{maitra2019spontaneous} pointed out that solutions with uniformly rotating polarity can emerge due to the coupling term in $\lambda_{\rm{PC}}$.

Since we consider $|\mathbf{p}|=1$, the polarity vector can be denoted $\mathbf{p}=\cos\theta \mathbf{e}_x + \sin\theta \mathbf{e}_y$. Starting from Eqs. \ref{eq:tension_tensor_application_flat_PC}- \ref{eq:polarity_dynamics_application_flat_PC}, we obtain a dynamic equation for the angle $\theta$ (Appendix \ref{appendix:application:PC}):
\begin{align}
\label{eq:application:dttheta:PC}
\partial_t \theta=&K\left(\frac{1}{\gamma}+\frac{(1-\nu\cos2\theta-\nu_{\rm PC}\sin2\theta)^2}{4 \eta + \gamma (\nu\sin 2\theta-\nu_{\rm {PC}} \cos2\theta)^2}\right)\partial_x^2 \theta\nonumber\\
& -\lambda_{\rm{PC}}\Delta\mu\nonumber\\
&-\frac{(1- \nu  \cos2\theta -\nu_{\rm PC}\sin2\theta) }{4 \eta + \gamma (\nu\sin 2\theta-\nu_{\rm {PC}} \cos2\theta)^2}\times\nonumber\\
&(\zeta_{\rm n} \Delta\mu\sin2\theta -\zeta_{{\rm PC n}} \Delta\mu \cos 2\theta)~.
\end{align}
In this expression the angle $\theta_p$, which depends on active tension parameters, and is defined by 
\begin{align}
\label{eq:def_theta_p}
\tan \theta_p=\frac{\zeta_{{\rm PCn}}}{\zeta_{\rm n} }, ~-\frac{\pi}{2}<\theta_p<\frac{\pi}{2}~,
\end{align}
plays a special role. Indeed the active tension contribution to Eq. \ref{eq:tension_tensor_application_flat_PC} (contribution proportional to $\Delta\mu$) can be rewritten:
\begin{align}
t^{ij}_a=\frac{\zeta_{\rm n} \Delta\mu}{2\cos\theta_p} \left(\begin{array}{cc}\cos(2\theta-\theta_p) & \sin(2\theta-\theta_p) \\\sin(2\theta-\theta_p) & - \cos(2\theta-\theta_p)\end{array}\right)~,
\end{align}
such that $-\theta_p/2$ corresponds to the angle, introduced by the planar-chiral order parameter, between one of the principal directions of the active tension tensor and the polarity vector direction (Fig. \ref{fig:application_PC}A). In addition, in Eq. \ref{eq:application:dttheta:PC}, the active coupling $\lambda_{\rm PC}$ induces a rotation of the polar order parameter on the surface (Fig. \ref{fig:application_PC}B).

We now assume that the flow-aligning parameters $\nu$ and $\nu_{\rm PC}$ can be neglected.  We first consider the situation where the polarity field can be considered uniform (limit of $L\rightarrow \infty$). Steady-states solution for the polarity angle $\theta$ then exists for
\begin{align}
\label{eq:application_PC_threshold_rotation}
\left | \frac{\zeta_{{\rm PCn}}}{\zeta_{\rm n}}\right |>\sqrt{\frac{16 \eta^2 \lambda_{\rm{PC}}^2 }{\zeta_{\rm n}^2}- 1}~,
\end{align}
and the corresponding steady-state solutions are
\begin{align}
\label{eq:PC_solution_theta_s_1}
\theta_s^1 = &\frac{1}{2} \left(\pi+\arcsin\left(\frac{4\eta\lambda_{\rm PC}}{\zeta_{\rm n}}\cos\theta_p\right)+\theta_p \right) + k \pi~,\\
\label{eq:PC_solution_theta_s_2}
\theta_s^2 = &\frac{1}{2} \left(-\arcsin\left(\frac{4\eta\lambda_{\rm PC}}{\zeta_{\rm n}}\cos\theta_p\right)+\theta_p \right) + k \pi~,
\end{align}
with $k$ an integer. A stability analysis (Appendix \ref{appendix:application:PC}) shows that $\theta_s^1$ is stable for $\zeta_{\rm n}<0$ (extensile active stress), while $\theta_s^2$ is stable for $\zeta_{\rm n}>0$ (contractile active stress). For a simple surface, not planar-chiral, the coefficients $\zeta_{{\rm PC}{\rm n}}=\lambda_{\rm PC}=0$ and the angle $\theta_p=0$: in that case $\theta=0$, $\pi$ are the stable solutions for $\zeta_{\rm n}>0$ and $\theta=\pi/2$, $3\pi/2$ are stable solutions for $\zeta_{\rm n}<0$. When $\theta_p \neq 0$ or $\lambda_{\rm PC}\neq0$, the steady-state angles deviate from this solution  as $|\zeta_{\rm PC n}|$ or $|\lambda_{\rm PC}|$ are increased (Fig. \ref{fig:application_PC}C-D). For sufficiently large $|\lambda_{\rm PC}|$, above the threshold defined by Eq. \ref{eq:application_PC_threshold_rotation}, the steady-state solutions are lost and the polarity rotates with the average angular velocity (Fig. \ref{fig:application_PC}C):
\begin{align}
\omega = |\lambda_{\rm PC}|\Delta\mu\sqrt{1-\left(\frac{\zeta_{\rm n}}{4\eta \lambda_{\rm PC}\cos\theta_p}\right)^2}~,
\end{align}
which approaches a constant velocity,  $|\lambda_{\rm PC}|\Delta\mu$ for $|\lambda_{\rm PC}|\rightarrow \infty$, and vanishes at the transition line given by Eq. \ref{eq:application_PC_threshold_rotation}. The direction of rotation is determined by the sign of $\lambda_{\rm PC}$.

For finite values of $L$ and planar anchoring ($\theta(x=0)=\theta(x=L)=\frac{\pi}{2}$), a competition arises between the distortion term proportional to $K$ in Eq. \ref{eq:application:dttheta:PC}, favoring uniform orientation $\theta=\frac{\pi}{2}$, and other physical effects promoting a different polarity orientation. For a simple surface, $\zeta_{{\rm PCn}}=\lambda_{\rm PC}=0$, and for contractile stress $\zeta_{\rm n}>0$, this competition gives rise to a critical length, below which the polarity orientation stays uniform and oriented along the $y$ axis, and above which a spontaneous flow and polarity distortion emerges \cite{voituriez2005spontaneous}. The transition is a pitchfork bifurcation which gives rise to two symmetric configurations, with the polarity tilted to the left or the right. For small, but non-zero values of $\zeta_{{\rm PCn}}$ and $\lambda_{\rm PC}$, the pitchfork bifurcation becomes imperfect (Fig. \ref{fig:application_PC}E-F); as the planar-chiral terms break the symmetry between the two polarity orientations away from $\theta=\frac{\pi}{2}$. For sufficiently large $\lambda_{\rm PC}$, a new regime emerges where the polarity transiently rotates and sets up a number of spatial polarity turns in the system, until the distortion energy prevents the polarity from rotating further and a high-distortion steady-state is reached (Fig. \ref{fig:application_PC}F-G). At steady-state and for large $\lambda_{\rm PC}$, $n\sim L^2 |\lambda_{\rm PC}|\Delta\mu \eta\gamma/(2\pi K(\gamma+4 \eta))$ spatial turns can form in the confined system: as $\lambda_{\rm PC}$ or $L$ are increased, more and more bands of full $2\pi$ polarity rotations emerge (Fig. \ref{fig:application_PC}G-H).
\subsection{Weakly deformed active polar surface with broken up-down symmetry}

\begin{figure}
 \centering
 \includegraphics[width=0.8\columnwidth]{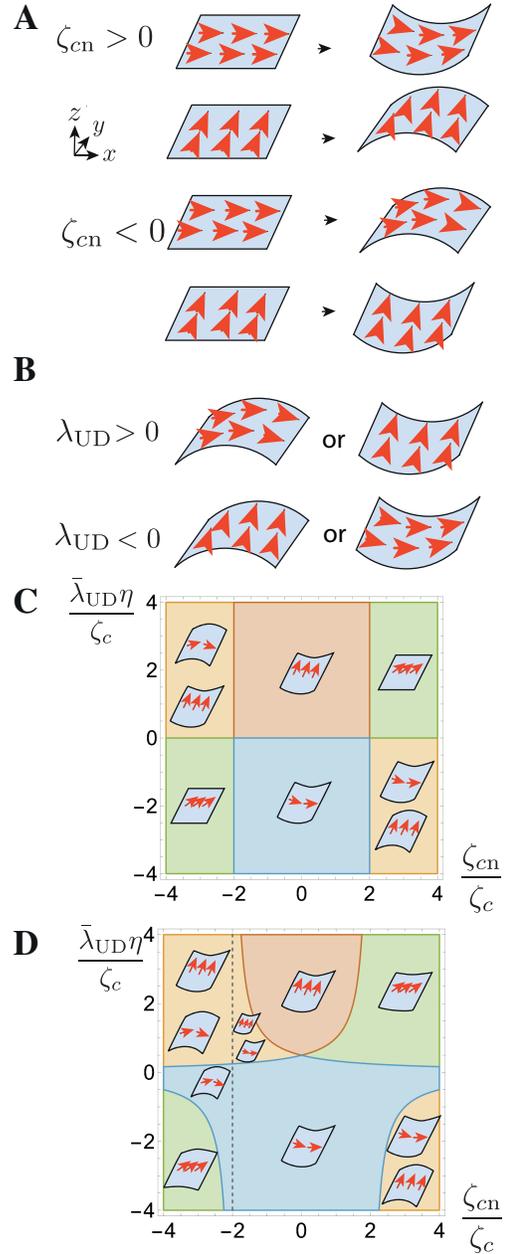}
 \caption{ Phase diagram for a quasi one-dimensional, confined, polar active surface with broken up-down symmetry. { \bf A}. The coefficient $\zeta_{c \rm{n}}\Delta\mu$, an active bending moment, results in internal torques and a preferred curvature direction, depending on the polarity axis orientation. We assume here $\Delta\mu>0$. The schematic on the right indicates the curvature induced in a free surface by the active bending moment $\sim\zeta_{c{\rm n}}$, assuming that the surface deforms only along the $x$ direction. {\bf B}. The coefficient $\lambda_{\rm UD}\Delta\mu$, with dimension of a velocity, is an active response of the polarity field to curvature. The schematic on the right indicates the stationary orientations induced by the coupling $\sim \lambda_{\rm UD}$. {\bf C}, {\bf D}. Phase diagrams of possible stable steady homogeneous states for $\zeta_c >0$ and $\zeta_{\rm n}=0$ (C), or $\zeta_{\rm n} \Delta\mu\kappa/(\zeta_c\Delta\mu)^2=1$ (D). Green: $\theta=\theta_0=\frac{1}{2}\arccos\left(- \frac{2 \zeta_c  }{\zeta_{c\rm n}}+ \frac{\zeta_n \kappa}{\eta \bar{\lambda}_{\rm UD} \Delta\mu  \zeta_{c{\rm n}}  } \right)$ or $\theta=\pi - \theta_0, \pi+\theta_0, 2\pi-\theta_0$; Brown region: $\theta=0, \pi$; Blue region: $\theta=\pi/2, 3\pi/2$; orange region: $\theta=0, \pi/2, \pi, 3\pi/2$. The dashed line indicates a change of curvature sign of the solutions $\theta=0, \pi$, which does not correspond to a bifurcation. One assumes here $\beta=\nu=0$. 
 \label{fig:application}
 } 
\end{figure}

We now consider an active polar surface which does not have a chiral or planar-chiral broken symmetry but has broken up-down symmetry. We consider a situation when the boundaries of the active surface at $x=0$ and $x=L$ are forced to be straight, but the surface can otherwise move without external forces at its boundaries (no tension, no torque). In that case $t_{xx}=0$, $\bar{m}_{xx}=0$ and $t_{xy}=0$. We assume that active terms contributing isotropic and anisotropic bending moduli ($\zeta'_c$, $\tilde{\zeta}_c$, $\tilde{\zeta}_{c\rm n}$, $\tilde{\zeta}'_{c\rm n}$) are vanishing. We also take $\beta'=\lambda=\lambda_{\rm UD}'=0$, as these terms act in the polarity equation to change the norm of the polarity which is constrained here, so that they only renormalize couplings in the constitutive equations for tensions and torques by modifying the molecular field $\mathbf{h}$. With these simplifications, the constitutive equations for the tangential tension tensor, the tangential bending moment and the polarity dynamics are (Eqs. \ref{Appendix:ConstitutiveEquationtij}-\ref{Appendix: constitutiveequation_dpdt}, \ref{appendix_application_eq_t_bar_t}-\ref{eq:appendix_application_bartije}, \ref{eq:application:equilibrium_t}-\ref{eq:application:equilibrium_m}):
\begin{align}
t^{ij}=&-P g^{ij}+2\eta \tilde{v}^{ij}  +(\zeta_{\rm n} + \tilde{\zeta}'_{\rm n} C_{k}{}^k ) \Delta \mu q^{ij} +2 \tilde{\zeta}\Delta \mu\tilde{C}^{ij} \nonumber\\
&+\frac{\nu}{2}(p^i h^j+p^j h^i - p_k h^k g_{ij}) \nonumber\\
&-\frac{1}{2}\left[\epsilon_{kl}p^k h^l -K \epsilon^{km} p_k p_l C^{nl} C_{nm}\right] \epsilon^{ij}\nonumber\\
&- K ((\nabla^i p^k)( \nabla^j p_k) - \frac{1}{2} (\nabla^l p^k)( \nabla_l p_k) g^{ij})\nonumber\\
& - \frac{1}{2}(\bar{m}^{ik} C_k{}^j + \bar{m}^{jk} C_k{}^i- \bar{m}^{lk} C_{kl} g^{ij} \nonumber\\
&- \bar{m}^{lk} C_{lm} \epsilon_k{}^m \epsilon^{ij})\label{eq:application_UD_tij} \\
\overline{m}^{ij}=&(\kappa C_k{}^k+  \eta_{cb} D_t C_{k}{}^{k} )g^{ij}+ K C^{i}{}_k p^k p^j \nonumber\\
& +2 \eta_c D_t \tilde{C}^{ij}+ \zeta_c  \Delta \mu g^{ij} +\zeta_{c\rm n} \Delta \mu q^{ij}\nonumber\\
&+\frac{\beta}{2} (p^i h^j+p^j h^i -p^k h_k g^{ij}) \label{eq:application_UD_mij}\\
D_t p^i =&\frac{1}{\gamma} h^i-\nu \tilde{v}^{ij} p_j+\lambda_{\rm UD}\Delta\mu \,C^{ij}p_j -\beta p^j D_t \tilde{C}^{i}{}_j~.\label{eq:application_UD_dpdt}
\end{align}
In the equations above, $\eta_c$ and $\eta_{cb}$ are shear and bulk bending viscosities. The term in $\zeta_c$ corresponds to an active bending moment which results in a spontaneous curvature of the film $C_k{}^k=-\zeta_c\Delta\mu/\kappa$ in the absence of other effects. Without loss of generality, we assume that the surface is oriented such that $\zeta_c>0$. The term in $\zeta_{c{\rm n}}$ corresponds to an active anisotropic bending moment whose orientation is set by the polarity field (Fig. \ref{fig:application}A).  $\lambda_{\rm UD}$ is an active orientation coupling which orients the polarity along the directions of principal curvatures of the surface (Fig. \ref{fig:application}B).

We perform calculations in the Monge gauge, at first order in the height gradient $\partial_x h$, the velocity field and the polar angle gradient $\partial_x \theta$. The surface is considered incompressible, $v_{xx}=\partial_x v_x +C_{xx} v_n=0$. The only non-zero component of the curvature tensor is $C_{xx}\simeq - \partial_x^2 h$.  For simplicity, we take here the limit where flow-alignment and curvature-alignment couplings are small, such that $\beta=\nu\simeq 0$; and where the bending modulus is large compared to the polarity distortion modulus, $\kappa\gg K$.  We then obtain (Appendix \ref{Appendix:weakly_deformed_active_film}):
\begin{align}
\partial_t C_{xx} =&-\frac{\kappa C_{xx}+\zeta_c\Delta\mu +\zeta_{c{\rm n}}\Delta\mu \frac{1}{2}\cos 2\theta}{ \eta_c +\eta_{cb} }~,\label{eq:weakly_deformed_dtCxx}\\
\partial_t \theta =&\frac{K(4\eta+\gamma)}{4\eta\gamma} \partial_x^2 \theta-   \frac{ \Delta\mu (\zeta_{\rm n} + 2\eta \bar{\lambda}_{\rm UD}    C_{xx} )}{4\eta}\sin2\theta ~.\label{eq:weakly_deformed_dttheta}
\end{align}
where in the last line, we have introduced the parameter $\bar{\lambda}_{\rm UD} = \lambda_{\rm UD} +( \tilde{\zeta}'_{\rm n} - \zeta_{c{\rm n}})/(2\eta)$.
Uniform steady-states correspond either to:
\begin{align}
\label{eq:application_UD_first_steady_state_solution}
 &\cos 2\theta = - \frac{2 \zeta_c  }{\zeta_{c\rm n}}+ \frac{\zeta_n \kappa}{\eta \bar{\lambda}_{\rm UD} \Delta\mu  \zeta_{c{\rm n}}  }  , C_{xx}=-\frac{\zeta_{\rm n}}{2\eta \bar{\lambda}_{\rm UD}    }~,\\
\text{or }&\theta=k \pi, C_{xx} =- \frac{( \zeta_{c\rm n}+2 \zeta_c)\Delta\mu  }{2\kappa }~,\\
 \text{or } &\theta=\frac{\pi}{2} + k \pi , C_{xx} = \frac{(\zeta_{c\rm n}-2 \zeta_c)\Delta\mu }{2\kappa }~,
\end{align}
with $k$ an integer. As we require weak deformations, these solutions apply provided that $C_{xx} L \ll 1$. We determine the stability thresholds of these solutions analytically  for fast curvature relaxation ($\eta_c\rightarrow 0$, $\eta_{cb}\rightarrow 0$) (Appendix \ref{appendix:quasi_1D_active_film}). In the limit of an infinite system, $L\rightarrow \infty$, one obtains the phase diagram of Fig. \ref{fig:application}C . The phase diagram exhibits two regions of solution coexistence where the active film can have opposite curvatures, depending on the orientation of the polarity (orange regions in Fig. \ref{fig:application}C). Remarkably, in the absence of a contractile or extensile anisotropic active tension ($\zeta_{\rm n}=0$), two regions of the diagram have flat solutions despite the existence of an active bending moment and therefore a film spontaneous curvature (green regions in Fig. \ref{fig:application}C). In these parameter regions, the polarity orients so as to cancel the spontaneous curvature in the $x$ direction. Therefore, varying $\bar{\lambda}_{\rm UD}$ or $\zeta_{c{\rm n}}$ allows to switch between a flat shape with tilted polarity, and a curved shape with the polarity aligned along, or orthogonal to, the curvature principal axis. For $\zeta_{\rm n}\neq0$, the flat solution is replaced by a curved layer, whose curvature depends on the active film contractile or extensile anisotropic active tension, $\zeta_{\rm n}\Delta\mu$ (Fig. \ref{fig:application}D).

For $\bar{\lambda}_{\rm UD}>0$, $\zeta_{c\rm n}<2\bar{\zeta}_c$ or $\bar{\lambda}_{\rm UD}<0$, $\zeta_{c\rm n}>2\bar{\zeta}_c$, the homogeneous solution $\theta=\frac{\pi}{2}$ and $C_{xx} = (\zeta_{c\rm n}-2 \bar{\zeta}_c)\Delta\mu  /(2\kappa )$ is always stable for planar anchoring of the polarity ($\theta(0)=\frac{\pi}{2}$ and $\theta(L)=\frac{\pi}{2}$). Away from this region, this solution is stable only for a sufficient small system size $L$, and the stability condition is
\begin{align}
L< \pi   \sqrt{\frac{\kappa  K (4\eta+\gamma)\pi^2}{2\eta\gamma \bar{\lambda}_{\rm UD} ( 2 \bar{\zeta}_c-\zeta_{c\rm n} )(\Delta\mu)^2}}~.
\end{align}
This situation is reminiscent of the flowing flat film discussed in the previous section, but the physics at play is different. Here a distortion introduced in the polarity field away from the homogeneous steady-state not only couples to the emergence of a flow field, but also responds to the curvature of the surface, which can favour further polarity rotation (Fig. \ref{fig:application}A-B).

\section{Discussion}

In this work we have considered polar and nematic surfaces and categorized different phases according to their symmetries. We find 7 possible polar surfaces, 7 possible nematic surfaces, which add to the 5 types of in-plane isotropic surfaces we have previously discussed \cite{salbreux2017activesurfaces}. Two polar surfaces and two nematic surfaces actually involve pseudovectors and pseudotensors which do not transform as vectors and tensors under all symmetries; for this reason we call them pseudopolar and pseudonematic surfaces. Some of these 19 phases have experimental realisations, notably with lipid membranes and epithelial layers, which we discuss in appendix \ref{appendix:surface_glossary}: the pseudopolar phase for instance is simply realised with a phospholipid bilayer with molecules tilted relative to the normal. In some cases, for instance in the case of the pseudonematic surface, we do not know of a physical realisation. Remarkably however the case of the supracellular actin fiber arrangement in {\it Hydra}, which provides a remarkable example of a complex arrangement of orthogonal fibers in two parallel layers in a living organism, may be seen as an approximate realisation of such a pseudonematic surface \cite{Maroudas-Sacks:2021aa}. Surfaces with order parameters with higher symmetries, such as hexatic surfaces, could in principle be categorized following the reasoning outlined in section \ref{subsection_symmetries}. We note that an hexatic phase has recently been found in a model of self-propelled active tissues \cite{pasupalak2020hexatic}.

We have introduced differential operators $\nabla$ and $D$ which are helpful in maintaining notations compact and distinguishing various physical effects. For instance, the corotational variation of the polar order parameter $\mathbf{p}$ enters the expression for the infinitesimal variation of free energy \ref{eq:equilibrium_variations_W}, as it describes variations of the order parameter which do not arise from the surface rotation. 

We have derived constitutive equations for polar and active surfaces that depend on their broken symmetries. Constitutive equations are obtained by identifying thermodynamics forces and fluxes based on a derivation of entropy production, generalizing to active polar and nematic surfaces previous results obtained for multicomponent isotropic passive or active surfaces \cite{lomholt2005general,salbreux2017activesurfaces} or for passive nematic surfaces \cite{napoli2010equilibrium, napoli2016hydrodynamic, santiago2018stresses}. Constitutive equations for the tension tensor, bending moment tensor, have passive, equilibrium contributions which follow from the surface free-energy, and deviatoric contributions, notably viscous and active effects, whose full expressions are given in Appendix \ref{appendix:full_constitutive_equations}. Some of the active terms we identified are well-known in the active matter literature, such as the nematic active tension $\zeta_{\rm n}\Delta\mu$ \cite{Marchetti_RMP, julicher2018hydrodynamic}; however we also find a large number of additional active couplings, involving notably the curvature tensor $C_{ij}$ or, with appropriate broken symmetries, the Levi-Civita tensor $\epsilon_{ij}$. 

To identify some of the physical effects associated to polar and nematic active couplings in surfaces, we have considered here two examples of application of the framework of active polar and nematic surfaces by considering a quasi one-dimensional confined surface with planar anchoring at its boundaries. We have allowed first for planar-chiral effect in a flat surface and then for out-of-plane deformation of a surface with broken up-down symmetry. For a flat surface with a broken planar-chiral symmetry, the system undergoes a transition from a distorted but steady polarity field, to a permanently rotating polarity field \cite{maitra2019spontaneous}, when the confinement distance is large. We find that boundary conditions can oppose this rotation, instead setting up a spatial pattern consisting of repeated full turns of the polarity vector. In a weakly-deformed surface with a broken up-down symmetry, we find that depending on the magnitude of active couplings, the active surface can switch between a state with tilted polarity and a curvature that is controlled by the nematic active tension, and a state with polarity aligned along or orthogonal to the direction of curvature, and curvature controlled by active bending moments. Here, we have restricted our analysis to quasi-one dimensional patterns of flows and polarities; we note that further instabilities could in general emerge in the longitudinal direction along the confinement walls \cite{bell2022active}.

In this work we have considered that the active surface is subjected to external forces and torques, but we have not treated explicitly the surface surroundings; notably exchange of chemical species which would modify the mass and concentration balance equations \ref{MassBalanceEquation}-\ref{ConcentrationBalanceEquation}. It would be interesting to consider these effects in future work. Here we have also considered active effects giving rise to internal tension and torques, but we have not considered active external forces giving rise to a net surface velocity \cite{ramaswamy2000nonequilibrium, maitra2014activating}.

The role of intrinsic or spin angular momentum density has been considered in  three-dimensional passive and active fluids \cite{furthauer2012active, julicher2018hydrodynamic}. It would be interesting to study active surfaces taking account such a degree of freedom, which would require modifying the torque balance equation \ref{TorqueBalanceEquation} by retaining an inertial term that has so far been neglected.

The examples treated here already show that interactions of polar or nematic order and curvature changes in active surfaces can exhibit a range of behaviours which is only beginning to be explored. We expect that further applications of our framework will reveal other physical effects, and will be useful notably to describe biological morphogenesis \cite{khoromskaia2021active}.

\section*{Acknowledgements}
GS acknowledges support from the Francis Crick Institute, which receives its core funding from Cancer Research UK (FC001317), the UK Medical Research Council (FC001317), and the Wellcome Trust (FC001317). JP's research was supported by Seed fund of Mechanobiology Institute and Singapore Ministry of Education Tier 3 grant, MOET32020-0001. GS thanks Quentin Vagne and Nicolas Cuny for comments on the manuscript.

\appendix

\section{Notations and relations of differential geometry}
\label{appendix_diff_geom}
Here we summarize our notations and give relations of differential geometry for the surface. 

The surface is embedded in a 3D cartesian basis whose coordinates are denoted by greek indices $\alpha,\beta...$. $\tilde{\epsilon}_{ijk}$ is the fully antisymmetric 3D unit tensor (Levi-Civita tensor) and $\tilde{\mathbf{e}}_{\alpha}$ the tangent vectors of the 3D cartesian basis. The cross product is then denoted for two vectors $\mathbf{a}$, $\mathbf{b}$:
\begin{align}
\mathbf{a}\times\mathbf{b}=\tilde{\epsilon}_{\alpha\beta\gamma} a_{\beta} b_{\gamma} \tilde{\mathbf{e}}_{\alpha}~,
\end{align}
and for a vector $\mathbf{a}$ and second rank tensor $\mathbf{B}$, we use the notation:
\begin{align}
\label{eq:cross_product_tensor_notation}
\mathbf{a} \times_1 \mathbf{B}&=- \mathbf{B}  \times_1 \mathbf{a}= \tilde{\epsilon}_{\alpha\gamma\delta} a_{\gamma} B_{\delta\beta} \tilde{\mathbf{e}}_{\alpha} \otimes \tilde{\mathbf{e}}_{\beta}~,\nonumber\\
\mathbf{a} \times_2 \mathbf{B}&=-\mathbf{B} \times_2 \mathbf{a} =\tilde{\epsilon}_{\alpha\gamma\delta} a_{\gamma} B_{\beta\delta} \tilde{\mathbf{e}}_{\beta} \otimes \tilde{\mathbf{e}}_{\alpha}~,
\end{align}
which can generalized for a vector $\mathbf{a}$ and a tensor of rank $n$, $\mathbf{B}=B_{i_1...i_n} \tilde{\mathbf{e}}_{i_1}\otimes...\otimes\tilde{\mathbf{e}}_{i_n}$:
\begin{align}
\label{eq:cross_product_tensor_notation_generalized}
\mathbf{a} \times_k \mathbf{B}=&\tilde{\epsilon}_{\alpha\gamma\delta} a_{\gamma} B_{i_1... i_{k-1}\delta i_{k+1}... i_n}\nonumber\\
& \tilde{\mathbf{e}}_{i_1}\otimes...\otimes\tilde{\mathbf{e}}_{i_{k-1}} \otimes \tilde{\mathbf{e}}_{\alpha}  \otimes \tilde{\mathbf{e}}_{i_{k+1}} \otimes...\otimes \tilde{\mathbf{e}}_{i_n} ~.
\end{align}
We also find it convenient to introduce a cross-product notation, returning a vector from the product of two second-rank tensors $\mathbf{A}$ and $\mathbf{B}$:
\begin{align}
\mathbf{A}\times_1\mathbf{B}=&\tilde{\epsilon}_{\alpha\beta\gamma} A_{\beta\delta} B_{\gamma\delta}\tilde{\mathbf{e}}_{\alpha}\nonumber\\
\mathbf{A}\times_2\mathbf{B}=&\tilde{\epsilon}_{\alpha\beta\gamma} A_{\delta\beta} B_{\delta\gamma}\tilde{\mathbf{e}}_{\alpha}~. \label{eq:appendix_cross_product_tensor}
\end{align}
This allows to generalize the circular shifts relations of the triple product for three vectors $\mathbf{a}$, $\mathbf{b}$, $\mathbf{c}$:
\begin{align}
\mathbf{a}\cdot(\mathbf{b}\times\mathbf{c}) = \mathbf{b}\cdot(\mathbf{c}\times\mathbf{a})=  \mathbf{c}\cdot(\mathbf{a}\times\mathbf{b})~\label{eq:appendix:triple_product_vectors}
\end{align}
to a triple product for a vector and two second-rank tensors:
\begin{align}
\mathbf{A}:(\mathbf{b}\times_1\mathbf{C}) = \mathbf{b}\cdot(\mathbf{C}\times_1\mathbf{A})=  \mathbf{C}:(\mathbf{A}\times_1\mathbf{b})~\nonumber,\\
\mathbf{A}:(\mathbf{b}\times_2\mathbf{C}) = \mathbf{b}\cdot(\mathbf{C}\times_2\mathbf{A})=  \mathbf{C}:(\mathbf{A}\times_2\mathbf{b})~\label{eq:appendix:triple_product_tensors},
\end{align}
where we use the contraction notation $\mathbf{A}:\mathbf{B} = A_{\alpha\beta}B_{\alpha\beta}$.

We consider a curved surface $\mathbf{X}(s^1,s^2)$ parametrised by two generalised coordinates $s^1$, $s^2$ (Fig. \ref{fig:schematic}C). We use latin indices to refer to surface coordinates.  The tangent vectors are given by 
\begin{align}
\label{eq:appendix:defcovariant_basis}
\mathbf{e}_i=\partial_i \mathbf{X}~,
\end{align}
 with $\partial_i=\partial/\partial s^i$ and the unit normal vector by 
 \begin{align}
 \label{eq:appendix:defnormalvector}
 \mathbf{n}=\frac{\mathbf{e}_1\times\mathbf{e}_2}{|\mathbf{e}_1\times\mathbf{e}_2|}~.
 \end{align}

The metric $g_{ij}$ and curvature tensor $C_{ij}$ associated to $\mathbf{X}$ are defined by
\begin{eqnarray}
\label{MetricAndCurvature}
g_{ij}={ \mathbf e_i}\cdot{ \mathbf e_j}\mbox{ , } C_{ij}=-(\partial_i\partial_j \mathbf{X})\cdot\mathbf{n}~.
\end{eqnarray}

The vectors $\mathbf{e}^1$, $\mathbf{e}^2$ of the dual basis are defined by
\begin{align}
\label{eq:appendix:defcontravariantbasis}
\mathbf{e}_i \cdot \mathbf{e}^j =\delta_i^j~.
\end{align}
The inverse of the metric is denoted $g^{ij}$, such that $g_{ij}g^{jk} =\delta_i^k$.
The components of a vector $\mathbf{a}$ or a tensor $\mathbf{B}$ can be written:
\begin{align}
\mathbf{a}=&a^i \mathbf{e}_i = a_i \mathbf{e}^i\\
\mathbf{B}=&B^{ij} \mathbf{e}_i \otimes\mathbf{e}_j=B_{ij} \mathbf{e}^i \otimes\mathbf{e}^j~,
\end{align}
and the metric $g^{ij}$ can be used to raise or lower indices, i.e. $a^i=g^{ij} a_j$ and $a_i=g_{ij}a^j$. We use the notation $\partial^i=g^{ij}\partial_j$.

We denote $dl$ with $dl^2=g_{ij}ds^i ds^j$ a line element on the surface, and $dS=\sqrt{g} ds^1 ds^2$  a surface element, where $g=\det g_{ij}$ is the determinant of the metric tensor.

The derivatives of the basis and normal vectors are given by the Gauss-Weingarten equations
\begin{eqnarray}
\label{eq:appendix:Gauss_Weingarten_1}
\partial_i { \mathbf n}&=&C_i{}^j  { \mathbf e_j},\\
\partial_i \mathbf{e}_j&=&-C_{ij}\mathbf{n}+\Gamma_{ij}^k \mathbf{e}_k,
\end{eqnarray}
where $\Gamma_{ij}^k$  are the Christoffel symbols.

We introduce the Riemann tensor as
\begin{align}
\label{eq:def_Riemann_tensor}
R^l{}_{kji} &= \partial_j \Gamma_{ik}^l-\partial_i \Gamma_{jk}^l+ \Gamma_{ik}^m\Gamma_{jm}^l- \Gamma_{jk}^m\Gamma_{im}^l~,
\end{align}
which can be expressed in terms of the curvature tensor:
\begin{align}
\label{eq:riemann_tensor_curvature_tensor}
R_{ijkl}=&C_{ik} C_{jl}-C_{il}C_{jk}\\
=&{\rm det} (C_k{}^l) (g_{ik} g_{jl}-g_{il}g_{jk}) ~,
\end{align}
where the first identity follows from the definition \ref{eq:def_Riemann_tensor} and the relation $(\partial_i \partial_j \mathbf{e}_k)\cdot\mathbf{e}^l=(\partial_j \partial_i \mathbf{e}_k)\cdot\mathbf{e}^l$.

The Levi-Civita tensor on the curved surface is defined as:
\begin{equation}
\label{DefinitionEpsilonij}
\epsilon_{ij}=\mathbf{n}\cdot(\mathbf{e}_i\times\mathbf{e}_j)~,
\end{equation}
and it satisfies the identity
\begin{equation}
\epsilon_{ij}\epsilon^{jk}=-\delta_i^k~.
\end{equation}
The Levi-Civita tensor can be used to express vectorial products of the basis vectors:
\begin{eqnarray}
\label{CrossProductIdentity1}
\textbf{n}\times\textbf{e}_i=\epsilon_i{}^{j} \textbf{e}_j ~,\\
\label{CrossProductIdentity2}
\mathbf{e}_i\times\mathbf{e}_j=\epsilon_{ij}\mathbf{n}~.
\end{eqnarray}

A tensor with two indices can generally be decomposed into a symmetric and antisymmetric part:
\begin{eqnarray}
A^{ij}&=&A^{ij}_s+A^{ij}_a\label{TensorSymmetricAntisymmetricDecomposition}\\
&=&A^{ij}_s+\frac{1}{2} A^{kl}\epsilon_{kl} \epsilon^{ij}~.
\end{eqnarray}
We also use the notation $[A^{ij}]_s=A^{ij}_s$. The traceless part of a tensor $\mathbf{A}$ is denoted $\tilde{\mathbf{A}}$ with:
\begin{align}
\tilde{A}^{ij}=A^{ij}-\frac{1}{2}A_k{}^k g^{ij}~.
\end{align}

In Eqs. \ref{def_covariant_diff_p} and \ref{def_covariant_diff_Q} we introduce the generic covariant differentiation operator defined by $\nabla a^i = (d\mathbf{a})\cdot\mathbf{e}^i$ and $\nabla B^{ij} = (d \mathbf{B})\cdot(\mathbf{e}^i\otimes \mathbf{e}^j)$. From the differentiation rule $d(\mathbf{a}\otimes\mathbf{b})=(d\mathbf{a})\otimes \mathbf{b} + \mathbf{a}\otimes (d\mathbf{b} )$ one obtains directly the product rule $\nabla(a^i b^j) =(\nabla a^i) b^j + a^i (\nabla b^j)$. 
Denoting the metric tensor $\mathbf{G}= g_{ij} \mathbf{e}^i\otimes \mathbf{e}^j$, 
one obtains the identity using Eqs. \ref{MetricAndCurvature} and \ref{eq:appendix:defcontravariantbasis}:
\begin{align}
\nabla g_{ij}=0~,
\end{align}
from which it also follows that $\nabla_i g_{jk}=0$ and $\nabla_t g_{ij}=0$. The corotational variation and time derivative of the metric also vanishes, $D g_{ij}=0$ and $D_t g_{ij}=0$, by direct application of Eq. \ref{eq:appendix:corotation_q_explicit_expression}. These identities can be used to lower or raise indices inside the covariant and corotational derivatives.

Denoting the Levi-Civita antisymmetric tensor $\boldsymbol{\epsilon}=\epsilon_{ij} \mathbf{e}^i\otimes\mathbf{e}_j$, one obtains using Eq. \ref{CrossProductIdentity2} that its covariant variation vanishes:
\begin{eqnarray}
\nabla \epsilon_{ij}=0~,
\end{eqnarray}
which also implies $\nabla_i \epsilon_{jk}=0$ and $\nabla_t \epsilon_{ij}=0$. By direct application of Eq. \ref{eq:appendix:corotation_q_explicit_expression}, one also has $D\epsilon_{ij}=0$ and $D_t \epsilon_{ij}=0$.

The curvature tensor satisfies the Mainardi-Codazzi equation \cite{kreyszig1968introduction}:
\begin{equation}
\label{MainardiCoddazi}
\nabla_i C_{jk}=\nabla_j C_{ik}.
\end{equation}
The curvature tensor also satisfies the identity
\begin{eqnarray}
\nabla_i\left(C^i{}_j-C_k{}^k \delta^i{}_j\right)=0\label{DivergenceCurvatureTensor},
\end{eqnarray}
as well as the relation
\begin{equation}
\label{DesernoGaussEquation}
C_{ik} C^{k}{}_j =C_k{}^k C_{ij}- g_{ij} \det(C_{k}{}^{l}).
\end{equation}

The divergence theorem on a curved surface reads \cite{capovilla2002stresses}:
\begin{equation}
\label{DivergenceTheorem}
\int_{\mathcal{S}} dS \nabla_i f^i=\int_{\mathcal{C}} dl \nu_i f^i,
\end{equation}
where $\mathcal{S}$ is the surface enclosed by $\mathcal{C}$, $\boldsymbol{\nu}$ is a unit vector tangent to $\mathcal{S}$, outward-pointing and normal to the contour $\mathcal{C}$, and $dl$ is an infinitesimal line element going along the contour $\mathcal{C}$. 

The two-components tensor obtained from the gradient of a tangent vector $\mathbf{p}$ is denoted
\begin{align}
\label{def:appendix:vector_gradient}
\boldsymbol{\nabla} \otimes \mathbf{p} = &\mathbf{e}^i \otimes \partial_i \mathbf{p}\\
=&(\nabla_i p^j) \mathbf{e}^i \otimes \mathbf{e}_j -C_{ij} p^j \mathbf{e}^i \otimes\mathbf{n}~,
\end{align}
and it contains a component which is not tangent. Similarly the gradient of a two-components tangent tensor $\mathbf{Q}$ is a tensor of rank 3:
\begin{align}
\label{def:appendix:tensor_gradient}
\boldsymbol{\nabla} \otimes \mathbf{Q}=&\mathbf{e}^i \otimes \partial_i \mathbf{Q}\\\
=&(\nabla_i Q^{jk}) \mathbf{e}^i \otimes \mathbf{e}_j\otimes \mathbf{e}_k - C_{ij} Q^{jk} \mathbf{e}^i\otimes\mathbf{n}\otimes\mathbf{e}_k \nonumber\\
&- C_{ik} Q^{jk} \mathbf{e}^i \otimes\mathbf{e}_j\otimes \mathbf{n}~.
\end{align}
Covariant derivatives are not commutative and verify
\begin{align}
[\nabla_i, \nabla_j ] p^k=& \nabla_i (\nabla_j p^k) - \nabla_j (\nabla_i p^k)   \nonumber\\
=&R^k{}_{lij} p^l~,\\
[\nabla_i, \nabla_j ] Q^{kl} = &\nabla_i (\nabla_j Q^{kl}) - \nabla_j (\nabla_i Q^{kl}) 
 \nonumber\\
 =&  R^k{}_{mij} Q^{ml}+R^l{}_{mij} Q^{km}~,
\end{align}
where $\left[\cdot,\cdot\right]$ is the commutation operator, and $R_{ijkl}$ is the Riemann tensor introduced in Eq. \ref{eq:def_Riemann_tensor}. The commutation relation between covariant derivatives can be rewritten using the Riemann tensor expression in terms of the curvature tensor, Eq. \ref{eq:riemann_tensor_curvature_tensor}:
\begin{align}
\label{eq:covariant_derivative_commutation}
[\nabla_i, \nabla_j ] p^k=&C_i{}^{k}C_{lj}p^l -C_{il} C_{j}{}^{k} p^l~,\\
\label{eq:covariant_derivative_tensor_commutation}
[\nabla_i, \nabla_j ] Q^{kl} =& C_{i}{}^{k}C_{mj}Q^{ml} - C_{im} C^{k}{}_{j} Q^{ml}\nonumber\\
&+ C_{i}{}^{l}C_{mj}Q^{km} - C_{im} C_{j}{}^{l} Q^{km}~.
\end{align}

We further note that for a scalar field $f$, a vector field $\mathbf{a}$ or a tensor field $\mathbf{B}$,
\begin{align}
\nabla_i (\partial_j f)&=\nabla_j (\partial_i f)\label{eq:commutation_derivative_scalar}\\
\nabla_i(\partial_j \mathbf{a} ) &= \nabla_j (\partial_i \mathbf{a})\label{eq:commutation_derivative_vector}\\
\nabla_i(\partial_j \mathbf{B} ) &= \nabla_j (\partial_i \mathbf{B})~,\label{eq:commutation_derivative_tensor}
\end{align}
which follow from the symmetry relations of Christoffel coefficients $\Gamma_{ij}^k = \Gamma_{ji}^k$.

Denoting $\mathbf{S}$ a tangent symmetric tensor and $\mathbf{Q}$ a tangent traceless symmetric tensor, we obtain the following product identities:
\begin{align}
\label{eq:appendix_product_S_Q_identity}
S_{ik}Q^k{}_j =& \frac{1}{2} S_k{}^k Q_{ij} + \frac{1}{2} S_{kl} Q^{kl} g_{ij}+\frac{1}{2} \epsilon_{kl} Q^{lm} S_m{}^k \epsilon_{ij}~,\\
\label{eq:appendix_product_Q_S_identity}
Q_{ik}S^k{}_j =& \frac{1}{2} S_k{}^k Q_{ij} + \frac{1}{2} S_{kl} Q^{kl} g_{ij}-\frac{1}{2} \epsilon_{kl} Q^{lm} S_m{}^k \epsilon_{ij}~,\\
\label{eq:appendix_product_epsilon_S_identity}
\epsilon_{ik}S^k{}_j =& \epsilon_j{}^k S_{ki} + S_k{}^k \epsilon_{ij}~, \\
\label{eq:appendix_product_epsilon_S_identity}
S_{ik} \epsilon^{k}{}_j=&S_{jk} \epsilon^{k}{}_i+S_k{}^k \epsilon_{ij}~,\\
\label{eq:appendix_product_epsilon_Q_identity}
\epsilon_{ik} Q^{k}{}_{j}=&-Q_{ik} \epsilon^k{}_j~.
\end{align}
The two first relations imply, for two traceless symmetric tensor $\mathbf{Q}$ and $\mathbf{Q}'$,
\begin{align}
\label{eq:appendix:product_Q_Q_identity}
[Q_{ik} {Q'}^k{}_j]_s =&[Q_{ik}' Q^k{}_j]_s= \frac{1}{2} Q_{kl} {Q'}^{kl} g_{ij}~,\\
\label{eq:appendix:product_epsilon_Q_Q_identity}
[\epsilon_{ik} Q^{kl}Q'_{lj}]_s =& [ Q_{ik}{Q'}^{kl}\epsilon_{lj} ]_s
=[ Q'_{ik}\epsilon^{kl}Q_{lj} ]_s\nonumber\\
 =& \frac{1}{2} \epsilon_{mk} Q^{kl} Q'_{lm} g_{ij}~.
\end{align}

\section{Variation of vectorial and tensorial surface quantities}
\label{VarSurfTensors}

Here we discuss infinitesimal variations of surface quantities and expressions for the corotational and covariant variations introduced in the manuscript. 

\subsection{Infinitesimal variations of fundamental vectors and tensors}

We consider here an infinitesimal surface displacement $d \mathbf{X}$ of the surface. Using the definitions \ref{eq:appendix:defcovariant_basis}, \ref{eq:appendix:defnormalvector}, \ref{MetricAndCurvature}, \ref{eq:appendix:defcontravariantbasis}, we obtain the following expressions for the infinitesimal variations of the tangent and normal vector, and metric and curvature tensors:
\begin{align}
d\mathbf{e}_i=&(\nabla_i d X^j + C_i{}^j d X_n ) \mathbf{e}_j+(\partial_i  d X_n- C_{ij} dX^j  ) \mathbf{n},\label{VariationTangentVector}\\
d\mathbf{e}^i=&-(\nabla^j d X^i+C^{ij} d X_n )\mathbf{e}_j+(\partial^i d X_n-C^i{}_jd X^j )\mathbf{n},\label{VariationTangentVectorContravariant}\\
d\mathbf{n}=&(-\partial_i  dX_n+C_{ij}d X^j  ) \mathbf{e}^i,\label{VariationNormalVector}\\
d g_{ij}=&\nabla_i d X_j+\nabla_j d X_i+2C_{ij} d X_n \label{VariationMetricCovariant}~,\\
d \sqrt{g} =& \frac{1}{2}\sqrt{g} g^{ij} dg_{ij}\label{Variation_sqrt_metric_determinant} ~,\\
d C_{ij}=& - \nabla_i (\partial_j dX_n) +C_{ik} C_j{}^k dX_n + (\nabla_k C_{ij}) dX^k\nonumber\\
& + C_{ik} \nabla_j dX^k + C_{jk} \nabla_i dX^k\label{VariationCurvatureTensor}~.
\end{align}

We write an arbitrary change in surface scalar fields, for instance a concentration field $c^\alpha$, as:
\begin{align}
	\label{eq:variation_c_alpha_appendix}
	d c^{\alpha}=- c^{\alpha}g^{ij}\frac{d g_{ij}}{2}+\dbar c^{\alpha}
\end{align}
with $\dbar c^{\alpha} =d(\sqrt{g} c^{\alpha})/\sqrt{g}$ a change of concentration that does not arise by ``geometric'' dilution, whose effect is captured by the first term in the right-hand side of Eq. \ref{eq:variation_c_alpha_appendix}.

\subsection{Infinitesimal surface rotation and rotation gradient}
The surface infinitesimal rotation $d\boldsymbol{\vartheta}$ associated with the surface infinitesimal displacement $d \mathbf{X}$ has explicit expression:
\begin{align}
d\boldsymbol{\vartheta}=&\frac{1}{2}\nabla\times d\mathbf{X}\nonumber\\
=&\epsilon^{ij}(\partial_j d X_n-C_{jk} d X^k) \mathbf{e}_i+\frac{1}{2} (\nabla_i d X_j) \epsilon^{ij}\mathbf{n}\label{def_rotation_rate_appendix}~.
\end{align}
To obtain this expression, we have assumed that the derivative of the deformation in the direction normal to the surface reads $\partial_n d\mathbf{X}=-((\partial_i d\mathbf{X})\cdot\mathbf{n} )\mathbf{e}^i$ \cite{salbreux2017activesurfaces}. 

We note the following useful identities involving the differential surface rotation:
\begin{align}
\label{eq:rotational_variation_tangent_vector}
d\mathbf{e}_i =& \frac{d g_{ij}}{2} \mathbf{e}^j+d\boldsymbol{\vartheta}\times\mathbf{e}_i~,\\
d\mathbf{e}^i=&- g^{ik}  \frac{dg_{kj}}{2}\mathbf{e}^j+d\boldsymbol{\vartheta}\times\mathbf{e}^i~,\\
\label{eq:rotational_variation_normal_vector}
d\mathbf{n} =& d\boldsymbol{\vartheta}\times\mathbf{n}~.
\end{align}

In Appendix \ref{sec_Appendix_virtual_work_polar_surface} we require an expression for in-plane gradients of the local rotation variation, $(\partial_id \boldsymbol{\vartheta})\cdot \mathbf{e}_j$, in terms of the curvature and the metric.  Writing the curvature tensor $\mathbf{C}= \mathbf{e}^i\otimes(\partial_i \mathbf{n})$ which follows from the Gauss-Weingarten equation \ref{eq:appendix:Gauss_Weingarten_1}, and calculating its corotational variation (Eq. \ref{def_corot_variation_Q}) using the relations \ref{eq:rotational_variation_tangent_vector} -\ref{eq:rotational_variation_normal_vector}, we note that we the surface gradient of rotation can be rewritten:
\begin{align}
	(\partial_id\boldsymbol{\vartheta})\cdot \mathbf{e}_j
	&=\epsilon_{kj} \left(D C_i{}^k+\frac{dg_{il}}{2}C^{kl} \right)\, .
	\label{inplanegradtheta}
\end{align}

The normal part of the gradient of rotation is coupled to the normal part of the moment tensor $m_n^i$ (Eq. \ref{eq:VirtualWork}) and can be written
\begin{align}
 (\partial_i d\boldsymbol{\vartheta})\cdot\mathbf{n}=\partial_id\vartheta_n-C_i{}^jd\vartheta_j~.
\end{align}

\subsection{Corotational infinitesimal variations}
\label{appendix:corot_inf_var}

In Eqs. \ref{def_corot_variation_p} and \ref{def_corot_variation_Q}, we introduce the corotational infinitesimal variation $D$, and its components in Eqs. \ref{Eq:corotational_variation_p} and \ref{Eq:corotational_variation_Q}.  If $\mathbf{p}$ and $\mathbf{Q}$ are tangent vectors and second-rank tensors, $D\mathbf{p}$ and $D\mathbf{Q}$ are also tangent vectors and second-rank tensors. Indeed:
\begin{align}
\label{eq:appendix:normal_part_corotational_variation}
(D\mathbf{p})\cdot\mathbf{n} =& (d\mathbf{p})\cdot\mathbf{n} - (d\boldsymbol{\vartheta}\times\mathbf{p})\cdot\mathbf{n}\nonumber\\
=&p^i \left[d\mathbf{e}_i\cdot\mathbf{n}  - (d\boldsymbol{\vartheta}\times\mathbf{e}_i)\cdot\mathbf{n}\right]=0~,
\end{align}
where we have used the identity \ref{eq:rotational_variation_tangent_vector} in the last line. A similar calculation leads to $D\mathbf{Q}$ also a tangent tensor.
Explicit expressions in components of the infinitesimal corotational variation of a tangent vector $\mathbf{p}$ and a tangent tensor $\mathbf{Q}$ are:
\begin{align}
\label{eq:appendix:corotation_p_explicit_expression}
D p^i =& \nabla p^i +d \vartheta_n \epsilon^{ij} p_j~,\\
\label{eq:appendix:corotation_q_explicit_expression}
D Q^{ij} =& \nabla Q^{ij} +d \vartheta_n  \epsilon^{ik}Q_k{}^j +d \vartheta_n  \epsilon^{jk}Q^i{}_k~,
\end{align}
with $d\vartheta_n = \frac{1}{2} \epsilon^{ij} \nabla_i dX_j$ (Eq. \ref{def_rotation_rate_appendix}).

We now give explicit expressions for the infinitesimal corotational variation of the normal vector, metric tensor, curvature tensor, of the gradient of a tangent vector and of the gradient of a second-rank tangent tensor.  Using the definitions \ref{def_corot_variation_p} and \ref{def_corot_variation_Q}, Eqs. \ref{eq:rotational_variation_normal_vector} and \ref{eq:appendix:corotation_q_explicit_expression} and the variations \ref{VariationTangentVectorContravariant}, \ref{VariationCurvatureTensor}, we find the following relations:
\begin{align}
	D\mathbf{n}=&\mathbf{0}	\, ,
	\label{Dnormal}\\
	D g_{ij}=&0~,\\
	DC_{ij}=&- \nabla_i (\partial_j dX_n) -C_{ik} C_j{}^k dX_n + (\nabla_k C_{ij}) dX^k\nonumber\\
	&+d \vartheta_n  \epsilon_{ik}C^k{}_j +d \vartheta_n  \epsilon_{jk}C_i{}^k\label{Dcurvature_components} ~.
\end{align} 
This last expression also implies $DC_{ij}=dC_{ij} - C^k{}_j \frac{dg_{ki}}{2} - C_i{}^l \frac{dg_{lj}}{2}$.

Next, we obtain an expression for the co-rotational variation of the components of the gradient $ \boldsymbol{\nabla}\otimes \mathbf{p}$ of a tangent vector $\mathbf{p}$, defined in Eq. \ref{def:appendix:vector_gradient}:
\begin{align}
	D( \boldsymbol{\nabla}\otimes \mathbf{p})=&\mathbf{e}^i \otimes \left[( \partial_i D\mathbf{p}) -\frac{d g_{ij}}{2} \partial^j \mathbf{p}+   (\partial_i d\boldsymbol{\vartheta} )\times\mathbf{p} \right]~.
	\label{Dgradacomponents}
\end{align}
Similarly, the components of the co-rotational variation of the gradient of a tangent second-rank tensor $D(\boldsymbol{\nabla}\otimes \mathbf{Q})$, defined in Eq. \ref{def:appendix:tensor_gradient}, read:
\begin{align}
	D(\boldsymbol{\nabla}\otimes\mathbf{Q})=&\mathbf{e}^i\otimes\left[
	\partial_i (D\mathbf{Q})-\frac{d g_{ij}}{2} \partial^j \mathbf{Q}
+ (\partial_id\boldsymbol{\vartheta})\times_1 \mathbf{Q}\right.\nonumber\\
&\left.+ (\partial_id\boldsymbol{\vartheta})\times_2 \mathbf{Q} \right]~.
	\label{DgradBcomponents}
\end{align}
In Appendix \ref{Appendix_entropy_production_rate}, we use Eqs.~\ref{Dgradacomponents} and \ref{DgradBcomponents} in calculating the variation of a free energy with gradient of polarity or gradient of nematic tensor term. Eq. \ref{DgradBcomponents} also implies the following relations:
\begin{align}
D(\boldsymbol{\nabla}\otimes\mathbf{Q})\cdot(\mathbf{e}_i\otimes\mathbf{e}_j\otimes\mathbf{e}_k) =& D(\boldsymbol{\nabla}\otimes\mathbf{Q})\cdot(\mathbf{e}_i\otimes\mathbf{e}_k\otimes\mathbf{e}_j)\nonumber\\
 D(\boldsymbol{\nabla}\otimes\mathbf{Q})\cdot(\mathbf{e}_i\otimes\mathbf{n}\otimes\mathbf{e}_j) =& D(\boldsymbol{\nabla}\otimes\mathbf{Q})\cdot(\mathbf{e}_i\otimes\mathbf{e}_j\otimes\mathbf{n})\nonumber\\
 D(\boldsymbol{\nabla}\otimes\mathbf{Q})\cdot(\mathbf{e}_i\otimes\mathbf{n}\otimes\mathbf{n}_j)=&0.
 \label{eq:appendix_properties_DgradQ}
\end{align}
If the tangent tensor $\mathbf{Q}$ is symmetric traceless, $D\mathbf{Q}$ is also symmetric traceless. This follows from Eq. \ref{eq:appendix:corotation_q_explicit_expression} and Eq. \ref{eq:appendix_product_epsilon_Q_identity} if $\nabla Q_{ij}$ is symmetric traceless. This is the case as 
\begin{align}
\nabla Q_{ij}=&(d\mathbf{Q})\cdot(\mathbf{e}_i\otimes\mathbf{e}_j)\nonumber\\
=&dQ_{ij} + Q_{kj} d\mathbf{e}^k\cdot\mathbf{e}_i + Q_{ik} d\mathbf{e}^k\cdot\mathbf{e}_j~,
\end{align}
such that $\nabla Q_{ij}$ is symmetric if $Q_{ij}$ is symmetric; and
\begin{align}
\nabla Q_k{}^k=&(d\mathbf{Q})\cdot(\mathbf{e}_k\otimes\mathbf{e}^k)\nonumber\\
=&d(\mathbf{Q} : \mathbf{e}_k\otimes\mathbf{e}^k)-\mathbf{Q} : d(\mathbf{e}_k\otimes\mathbf{e}^k)\nonumber\\
=&d(Q_k{}^k)~,
\end{align}
such $\nabla Q_{ij}$ is traceless if $Q_{ij}$ is traceless.

\subsection{Corotational time derivatives}
\label{appendix:subsec:corotational_time_derivatives}
In Eqs. \ref{eq:advected_time_derivative_p}-\ref{eq:advected_time_derivative_Q}, we have introduced the Lagrangian time derivative denoted ${\rm d}/{\rm d}t$. We first clarify here this definition. For a field $f(s^1, s^2,t)$ on the surface changing with time, one can calculate the change of $f$ along the trajectory of the flow, i.e. the time derivative of $f(\mathbf{s}(t) , t )$ where $\mathbf{s}(t)$ are the coordinates labelling a point moving with the flow.  We use an Eulerian description with respect to tangential flows, and Lagrangian description with respect to normal flows, such that points with given coordinates follow the normal component of the flow. Then
\begin{align}
\label{eq:appendix:convected_time_derivative}
\frac{{\rm d} f}{{\rm d} t} = \partial_t\left[f(\mathbf{s}(t),t)\right] = \partial_t f + v^i \partial_i f~,
\end{align}
where we have used that moving along a line of flow with $\mathbf{s}(t)$, one has $(\mathbf{X}+d\mathbf{X})(\mathbf{s}+d\mathbf{s})=\mathbf{X}(\mathbf{s}) + \mathbf{v} dt$; therefore expanding in $d\mathbf{s}$, $dt$, and using that $d\mathbf{X}$ is along the normal to the surface, one obtains $ds^i = v^i dt$. The definition \ref{eq:appendix:convected_time_derivative} can be extended directly to vector and tensorial fields on the surface.

In Eqs. \ref{def_corot_derivative_p} and \ref{def_corot_derivative_Q}, we have introduced the corotational derivative operators $D_t$. The explicit expressions in components of the corotational derivatives of a tangent vector $\mathbf{p}$ and a tangent tensor $\mathbf{Q}$ are, following Eqs. \ref{eq:appendix:corotation_p_explicit_expression}-\ref{eq:appendix:corotation_q_explicit_expression}:
\begin{align}
D_t p^i =& \nabla_t p^i+\omega_n \epsilon^{ij} p_j\label{eq:appendix:corotational_time_derivative_p} \\
D_t Q^{ij} =& \nabla_t Q^{ij} + \omega_n \epsilon^{ik}Q_k{}^j + \omega_n \epsilon^{jk}Q^i{}_k\label{eq:appendix:corotational_time_derivative_Q}~,
\end{align}
with $\omega_n =\frac{1}{2}\epsilon^{ij} \nabla_i v_j$.

 Following the definition for the covariant time derivative, Eq. \ref{def_time_covariant_diff_p}, the covariant time derivative of a tangent polar vector $\mathbf{p}$ reads:
\begin{align}
\nabla_t p^i=& (\partial_t \mathbf{p}+v^j\partial_j \mathbf{p})\cdot\mathbf{e}^i\nonumber \\
=& \partial_t p^i + v^j \nabla_j p^i+ v_n  C^i{}_j p^j  \label{eq:appendix:advected_time_derivative_p}~,
\end{align}
where we have used Eq. \ref{VariationTangentVector}, replacing $d X_j$ by 0 and $dX_n$ by $v_n dt$, to obtain the relation $(\partial_t \mathbf{e}_i)\cdot \mathbf{e}_j=C_{ij} v_n$.

Similarly for a tangent second-rank tensor field $\mathbf{Q}$ one has from Eq. \ref{def_time_covariant_diff_Q}:
\begin{align}
\nabla_t Q^{ij}&=(\partial_t \mathbf{Q}+ v^k \partial_k \mathbf{Q}):(\mathbf{e}^i\otimes\mathbf{e}^j) \nonumber\\
&=\partial_t Q^{ij} + v^k \nabla_k Q^{ij}+ v_n C^j{}_k   Q^{ik} +v_n  C^i{}_k  Q^{kj} \label{eq:appendix:advected_time_derivative_Q}~.
\end{align}

Combining Eqs. \ref{eq:appendix:corotational_time_derivative_p} and \ref{eq:appendix:advected_time_derivative_p} for a tangent vector, and Eqs. \ref{eq:appendix:corotational_time_derivative_Q} and \ref{eq:appendix:advected_time_derivative_Q} for a tangent second-rank tensor, the full explicit expression for the corotational time-differential operator $D_t$ is, for a tangent vector $\mathbf{p}$ and a tangent second-rank tensor $\mathbf{Q}$:
\begin{align}
 D_t p^i =& \partial_t p^i + v^j \nabla_j p^i  +v_n  C^i{}_j p^j  +\omega_n \epsilon^{i}{}_{j} p^j~, \\
D_t Q^{ij} =& \partial_t Q^{ij} + v^k \nabla_k Q^{ij}+ v_n C^j{}_k Q^{ik}   + v_n  C^i{}_k Q^{kj}  \nonumber\\
& +\omega_n \epsilon^{ik} Q_k{}^j+ \omega_n \epsilon^{j}{}_{k} Q^{ik} \,.
\label{DQij}
\end{align}
In the expressions above, the partial time derivatives are taken for components chosen to be expressed in a covariant or contravariant basis, so that in general $\partial_t p_i \neq g_{ij} \partial_t p^j$,$ \partial_t Q^{ij} \neq g^{ik} g^{jl}\partial_t Q_{kl}$. For instance, the tensor $D_tQ_{ij}$ reads, using the time derivative of the contravariant components $Q_{ij}$:
\begin{align}
D_t Q_{ij} =& \partial_t Q_{ij} + v^k \nabla_k Q_{ij}- v_n C^j{}_k Q^{ik}   - v_n  C^i{}_k Q^{kj}  \nonumber\\
& +\omega_n \epsilon_{ik} Q^k{}_j + \omega_n \epsilon_{jk} Q_{i}{}^{k}\,.
\end{align}

\section{Rotation of order parameters}
\label{appendix_order_parameters}
In this Appendix we give the table of conserved and broken symmetries for order parameters which result from the transformation $\mathbf{p}\rightarrow\hat{\mathbf{p}}$ and $\mathbf{Q}\rightarrow\doublehat{\mathbf{Q}}$ of one of the phases discussed in the main text. Here $\mathbf{p}$ denotes a tangent vector or pseudovector, and $\mathbf{Q}$ a tangent tensor or pseudotensor. In this Appendix, for a given transformation $T$, we denote $\mathbf{T}$ the associated tensor; and we use the notations $T[\mathbf{p}]=\epsilon_{\mathbf{p}}^T  \mathbf{T}\mathbf{p}$ and $T[\mathbf{Q}]=\epsilon_{\mathbf{Q}}^T  \mathbf{T}\mathbf{Q} \mathbf{T}^{-1}$ for the application of this transformation; where the matrix products are understood here in a cartesian basis. We note that $\mathbf{p}$ and $\mathbf{Q}$ have signature $+1$ with respect to rotations around the normal.

We denote $R_n^{\frac{1}{2}}$ the rotation by $\pi/2$ around the normal vector. For polar phases we find, with the definitions of $\mathbf{p}$ and $\hat{\mathbf{p}}=R_n^{\frac{1}{2}}[ \mathbf{p}]$ introduced in the main text:
\begin{center}
	\begin{tabular}{|c|c|c|c|c|c|c|c|c|c|}\hline
		\makecell{Order\\ parameters }&$M_n$ & $M_{\mathbf{p}}$ & $R_{\mathbf{p}}$ & $M_{\hat{\mathbf{p}}}$& $R_{\hat{\mathbf{p}}}$ &$I$&$R_n$\\\hline
		 	 $\hat{\mathbf{p}}$, $\mathbf{p}_{\rm PC}$& 1 & X & X &1&1&X&X\\\hline
			 $\hat{\mathbf{p}}_{\rm C}$, $\mathbf{p}_{\rm UD}$& X&1 & X &X&1&1&X\\\hline
	$\{\epsilon_{\rm UD}$, $ \hat{\mathbf{p}}\}$   & X & X & X &1&X&X&X\\\hline
		$\{\epsilon_{\rm C}$, $\hat{\mathbf{p}}\}$& X&X & X &X&1&X&X\\\hline
	\end{tabular} 
\end{center}
These relations follow from Table \ref{polar_surfaces_table} and the identities which apply for a tangent vector or pseudovector $\mathbf{p}$:
\begin{align}
\label{eq:appendix_order_parameter_RpP}
R_{\mathbf{p}} [\mathbf{p}] = \mathbf{p}& \iff R_{\hat{\mathbf{p}}} [\hat{\mathbf{p}}] = \hat{\mathbf{p}}\\
R_{\hat{\mathbf{p}}} [\mathbf{p}] = \mathbf{p} &\iff R_{\mathbf{p}} [\hat{\mathbf{p}}] = \hat{\mathbf{p}}\\
M_{\mathbf{p}} [\mathbf{p}] = \mathbf{p} &\iff M_{\hat{\mathbf{p}}} [\hat{\mathbf{p}}] = \hat{\mathbf{p}}\\
\label{eq:appendix_order_parameter_MhatpP}
M_{\hat{\mathbf{p}}} [\mathbf{p}] = \mathbf{p}& \iff M_{\mathbf{p}} [\hat{\mathbf{p}}] = \hat{\mathbf{p}}~,
\end{align}
which imply that the transformation $\mathbf{p}\rightarrow \hat{\mathbf{p}}$ leads to an exchange of the columns $M_{\mathbf{p}} \leftrightarrow M_{\hat{\mathbf{p}}}$ and  $R_{\mathbf{p}} \leftrightarrow R_{\hat{\mathbf{p}}}$ in the table of conserved and broken symmetries. The relations \ref{eq:appendix_order_parameter_RpP}-\ref{eq:appendix_order_parameter_MhatpP} follow from 
\begin{align}
R_{\hat{\mathbf{p}}} [\hat{\mathbf{p}}]&=R_n^{\frac{1}{2}}[R_{\mathbf{p}}[ \mathbf{p}]]\\
 R_{\mathbf{p}} [\hat{\mathbf{p}}] &=R_n^{\frac{1}{2}} [R_{\hat{\mathbf{p}}} [\mathbf{p}]]\\
M_{\hat{\mathbf{p}}} [\hat{\mathbf{p}}]&=R_n^{\frac{1}{2}}[M_{\mathbf{p}}[ \mathbf{p}]]\\
M_{\mathbf{p}} [\hat{\mathbf{p}}] &=R_n^{\frac{1}{2}} [M_{\hat{\mathbf{p}}} [\mathbf{p}]]~,
\end{align}
which follow from the relations $R_{\hat{\mathbf{p}}} R_n^{\frac{1}{2}} = R_n^{\frac{1}{2}}R_{\mathbf{p}}$, $ R_{\mathbf{p}} R_n^{\frac{1}{2}} = R_n^{\frac{1}{2}}R_{\hat{\mathbf{p}}}$, $M_{\hat{\mathbf{p}}}  R_n^{\frac{1}{2}}  = R_n^{\frac{1}{2}} M_{\mathbf{p}}$ and  $M_{\mathbf{p}}  R_n^{\frac{1}{2}}  = R_n^{\frac{1}{2}} M_{\hat{\mathbf{p}}} $.

For nematic phases we find:
\begin{center}
	\begin{tabular}{|c|c|c|c|c|c|c|c|c|c|c|c|}\hline
\makecell{Order\\ parameters }&$M_n$ &\makecell{ $M_{\mathbf{q}}$\\or \\$M_{\hat{\mathbf{q}}}$} &\makecell{$R_{\mathbf{q}}$\\or\\ $R_{\hat{\mathbf{q}}}$}  &$I$&$R_n$  & $R_n^{\pm\frac{1}{2}}$& $R_n^{\pm\frac{1}{2}} M_n$ & $M_{\doublehat{\mathbf{q}}}$ & $ R_{\doublehat{\mathbf{q}}}$\\\hline
             $ \doublehat{\mathbf{Q}}$, $\mathbf{Q}_{\rm PC}$  &1&X&X&1&1&X&X&1&1\\\hline
                        $\doublehat{\mathbf{Q}}_{\rm UD}$, $\mathbf{Q}_{\rm C}$        &X&X&1&X&1&X&1&1&X\\\hline
                      	$\{\epsilon_{\rm UD}$, $\doublehat{\mathbf{Q}}\}$                    &X&X&X&X&1&X&X&1&X\\\hline
                                        $\{\epsilon_{\rm C}$, $\doublehat{\mathbf{Q}}\}$      &X&X&X&X&1&X&X&X&1\\\hline
	\end{tabular} 
	\end{center}
These relations follow from Table \ref{nematic_surfaces_table} and the identities which apply for a tangent tensor or pseudotensor $\mathbf{Q}$:
\begin{align}
\label{eq:appendix_order_parameter_RqQ}
R_{\mathbf{q}} [\mathbf{Q}] = \mathbf{Q} &\iff R_{\doublehat{\mathbf{q}}} [\doublehat{\mathbf{Q}}] = \doublehat{\mathbf{Q}}\\
R_{\doublehat{\mathbf{q}}} [\mathbf{Q}] = \mathbf{Q} &\iff R_{\mathbf{q}} [\doublehat{\mathbf{Q}}] = \doublehat{\mathbf{Q}}\\
M_{\mathbf{q}} [\mathbf{Q}] = \mathbf{Q}& \iff M_{\doublehat{\mathbf{q}}} [\doublehat{\mathbf{Q}}] = \doublehat{\mathbf{Q}}\\
\label{eq:appendix_order_parameter_MhatqQ}
M_{\doublehat{\mathbf{q}}} [\mathbf{Q}] = \mathbf{Q}& \iff M_{\mathbf{q}} [\doublehat{\mathbf{Q}}] = \doublehat{\mathbf{Q}}~,
\end{align}
which imply that the transformation $\mathbf{Q}\rightarrow \doublehat{\mathbf{Q}}=R_n^{\frac{1}{4}} [\mathbf{Q}]$ leads to an exchange of the columns $M_{\mathbf{q}} \leftrightarrow M_{\doublehat{\mathbf{q}}}$ and  $R_{\mathbf{q}} \leftrightarrow R_{\doublehat{\mathbf{q}}}$ in the table of conserved and broken symmetries. Here $R_n^{\pm \frac{1}{4}}$ is the rotation by $\pm\pi/4$ around the normal. The relations \ref{eq:appendix_order_parameter_RqQ}-\ref{eq:appendix_order_parameter_MhatqQ} follow from 
\begin{align}
R_{\doublehat{\mathbf{q}}} [\doublehat{\mathbf{Q}}]&=R_n^{\frac{1}{4}}[R_{\mathbf{q}}[ \mathbf{Q}]]\\
 R_{\mathbf{q}} [\doublehat{\mathbf{Q}}] &=R_n^{\frac{1}{4}} [R_{\doublehat{\mathbf{q}}} [\mathbf{Q}]]\\
M_{\doublehat{\mathbf{q}}} [\doublehat{\mathbf{Q}}]&=R_n^{\frac{1}{4}}[M_{\mathbf{q}}[ \mathbf{Q}]]\\
M_{\mathbf{q}} [\doublehat{\mathbf{Q}}] &=R_n^{\frac{1}{4}} [M_{\doublehat{\mathbf{q}}} [\mathbf{Q}]]~,
\end{align}
which follow from the relations $R_{\doublehat{\mathbf{q}}} R_n^{\frac{1}{4}} = R_n^{\frac{1}{4}}R_{\mathbf{q}}$, $ R_{\mathbf{q}} R_n^{\frac{1}{4}} = R_n R_n^{\frac{1}{4}}R_{\doublehat{\mathbf{q}}}$, $M_{\doublehat{\mathbf{q}}} R_n^{\frac{1}{4}}  = R_n^{\frac{1}{4}} M_{\mathbf{q}}$ and  $M_{\mathbf{q}}  R_n^{\frac{1}{4}}  = R_n R_n^{\frac{1}{4}} M_{\doublehat{\mathbf{q}}}$, and that all tensors and pseudotensors $\mathbf{Q}$ are invariant by $R_n$.

\section{Tension and moment tensors and conservation equations on a curved surface}
\label{appendix_ForceBalance_conservation}
\subsection{Force and torques on the surface}

The force $\textbf{f}$ and torque $\boldsymbol{\Gamma}$ on a line of length $dl$ with {\it unit} vector $\boldsymbol{\nu}=\nu^i \mathbf{e}_i$, tangential to the surface and normal to the line 
can be expressed as
\begin{eqnarray}
\label{DefinitionsTension}
\mathbf{f}&=&dl\; \nu^i \textbf{t}_i=dl\; \nu_i \textbf{t}^i,\\
\label{DefinitionsMoment}
\boldsymbol{\Gamma}&=&dl\; \nu^i \textbf{m}_i=dl\; \nu_i \textbf{m}^i \quad ,
\end{eqnarray}
where we have introduced the tension ${\bf t}_i$ and moment ${\bf m}_i$ per unit length (Fig. \ref{fig:schematic}B, D). Decomposing $\textbf{t}^i$ and $\textbf{m}^i$  in tangential and normal components as 
\begin{eqnarray}
\textbf{t}^i&=&t^{ij} \textbf{e}_j+t_{n}^i \textbf{n},\label{TensionProjection}\\
\textbf{m}^i&=&m^{ij}\textbf{e}_j+m^i_{n} \mathbf{n} \quad , \label{MomentProjection}
\end{eqnarray}
defines the tension and moment per unit length tensors $t^{ij}$, $t_n^i$, $m^{ij}$ and $m_n^i$.

\subsection{Conservation equations}
\label{subsec:appendix:conservation_equations}

Here we recall conservation equations for the surface mass, concentration of chemical species, energy, entropy and free energy given for a Eulerian representation in Ref. \cite{salbreux2017activesurfaces}. Mass balance reads
\begin{eqnarray}
\label{MassBalanceEquation}
\partial_t \rho+\nabla_i(\rho v^i ) +  v_n C_i{}^i \rho=0,
\end{eqnarray}
with $\rho$ the mass density on the surface and 
$\mathbf{v}=v^i \mathbf{e}_i+v_n \mathbf{n}$ is the center-of-mass velocity.

The concentrations $c^{\alpha}$ obey the balance equation
\begin{eqnarray}
\label{ConcentrationBalanceEquation}
\partial_t c^{\alpha}+\nabla_i J^{\alpha, i}+v_n C_i{}^i c^{\alpha}=r^{\alpha},
\end{eqnarray}
where $J^{\alpha, i}=c^{\alpha} v^i+{j^{\alpha,i}}$ is the tangential flux in the surface of molecule $\alpha$, $j^{\alpha,i}$ is the flux relative to the center of mass, and $r^{\alpha}$ denote source and sink terms corresponding to chemical reactions in the surface. In this work we do not consider exchanges of chemical species between the surface and its surrounding environment, which would contribute additional source terms in Eq. \ref{ConcentrationBalanceEquation}. Mass conservation implies the following relation between fluxes of molecules and chemical rates
\begin{eqnarray}
\sum_{\alpha} m^{\alpha} j^{\alpha,i}&=&0,\label{BalanceTangentialFluxes}\\
\sum_{\alpha} m^{\alpha} r^{\alpha}&=&0.\label{BalanceChemicalReactions}
\end{eqnarray}
The rate $r^\alpha$ at which species $\alpha$ is generated by chemical reactions can be rewritten as a sum over contributions from different chemical reactions \cite{julicher2018hydrodynamic}
\begin{align}
\label{eq:appendix_def_reaction_rate}
r^\alpha=-\sum_{I} a^{\alpha, I} r^I,
\end{align}
where the index $I$ labels chemical reactions, and the numbers $a^{\alpha, I}$ are the stoichiometric coefficients for reaction $I$:
\begin{align}
\label{eq:chemical_reactions}
\sum_{\alpha} a^{\alpha, I} A^{\alpha} \rightleftharpoons  0~.
\end{align}
In Eq. \ref{eq:chemical_reactions}, $A^{\alpha}$ is the chemical symbol for species $\alpha$. Mass conservation implies that $\sum_{\alpha} a^{\alpha, I} m_{\alpha}=0$.

The conservation of energy and the balance of entropy and free energy density have the form:
\begin{align}
\partial_t e+\nabla_i [v^i e + j^{e,i}]+v_n C_i{}^i e =&J^e_n\label{BalanceInternalEnergy},\\
\partial_t s+\nabla_i [v^i s + j^{s,i}]+v_n C_i{}^i s =&J^s_n+\theta \label{BalanceEntropy},\\
\partial_t f+\nabla_i [v^i f + j^{f,i}]+v_n C_i{}^i f=&J^f_n-T\theta-J_s^i \nabla_i T\nonumber\\
&-(\partial_t T )s\label{BalanceFreeEnergy},
\end{align}
where $e$ and $s$ are the energy and entropy density respectively, $J^e_n$ and $J^s_n$ are energy and entropy fluxes entering the surface from the adjacent bulk,  $J^{e,i}=v^i e + j^{e,i}$ and $J^{s,i}=v^i s + j^{s,i}$ are the tangential energy and entropy fluxes within the surface, and $J^f_n=J^e_n-TJ^s_n$ and $J^{f,i}=J^{e,i}-TJ^{s,i}=v^i f + j^{f,i}$ are the normal and tangential fluxes of free energy. The entropy production rate within the surface is denoted $\theta$. Eq. \ref{BalanceFreeEnergy} is obtained from the relation $f=e-Ts$ and Eqs. \ref{BalanceInternalEnergy} and \ref{BalanceEntropy}. In this manuscript we consider for simplicity the isothermal case.

Considering a surface $\mathcal{S}$ enclosed by a contour $\mathcal{C}$, moving with the velocity field $\mathbf{v}$ and such that the contour $\mathcal{C}$ follows the flow, the rate of change of free energy $F=\int_{\mathcal{S}} dS f$ is then
\begin{align}
\label{eq:appendix_free_energy_variation}
\frac{dF}{dt} = \int_{\mathcal{S}} dS[J_n^f - T \theta] -\int_{\mathcal{C}} dl \nu_i j_f^i~.
\end{align}

\section{Virtual work}
\label{sec_Appendix_virtual_work_polar_surface}
In this Appendix we derive an alternative expression for the infinitesimal virtual work defined in Eq. \ref{eq:VirtualWorkDef} \cite{salbreux2017activesurfaces}. Using the force and torque balance equations \ref{ForceBalanceEquation} and \ref{TorqueBalanceEquation}, the virtual work defined in Eq. \ref{eq:VirtualWorkDef} can be re-written 
\begin{align}
d W=&\int_{\mathcal{S}} dS\,\left[\mathbf{t}^i \cdot d \mathbf{e}_i
+\mathbf{m}^i\cdot \partial_i d\boldsymbol{\vartheta}+(\mathbf{t}^i\times \mathbf{e}_i)\cdot d\boldsymbol{\vartheta}\right]	\nonumber\\
=&\int_{\mathcal{S}} dS\,\left[t^{ij}\frac{d g_{ij}}{2}+\mathbf{m}^i\cdot\partial_i d\boldsymbol{\vartheta}\right]~,	\label{eq:AppendixVirtualWork_first_expression}
\end{align}
where Eq.~\ref{eq:rotational_variation_tangent_vector} has been used to replace d$\mathbf{e}_i$ in the first line.
By re-expressing $(\partial_id\boldsymbol{\vartheta})\cdot\mathbf{e}_j$ with the aid of Eq.~\ref{inplanegradtheta} one then finds:
\begin{align}
	\label{eq:AppendixVirtualWork}
\delta W
=&\int_\mathcal{S} dS\left[\overline{t}^{ij}\frac{d g_{ij}}{2}  + \overline{m}^{ij}
D C_{ij}+m^{i}_{n} (\partial_i d\boldsymbol{\vartheta})\cdot\mathbf{n}\right]\,,
\end{align}
where we use the modified tensors introduced in Eqs. \ref{eq:tbardef} and \ref{eq:mbardef}.

We note that the definition of the virtual work differential \ref{eq:VirtualWorkDef} allows for a redefinition of the tension and moment tensors $\mathbf{t}^i\rightarrow \mathbf{t}^i+\epsilon^{ik} \partial_k (A\mathbf{n})$, $\mathbf{m}^i \rightarrow \mathbf{m}^i+A g^{ij} \mathbf{e}_j$ with an arbitrary scalar field $A$, which leaves the virtual work differential invariant. Indeed for any scalar field $A$:
\begin{align}
\oint_{\mathcal{C}} dl \nu_i\left(\mathbf{t}^i\cdot d \mathbf{X}+\mathbf{m}^i\cdot d\boldsymbol{\vartheta}\right)=\hspace{4cm}\nonumber\\
\oint_{\mathcal{C}} dl \nu_i\left([\mathbf{t}^i+\epsilon^{ik} \partial_k (A\mathbf{n})]\cdot d \mathbf{X}+(\mathbf{m}^i+A g^{ij} \mathbf{e}_j)\cdot d\boldsymbol{\vartheta}\right)~,
\end{align}
which can be proven using using that the difference $\Delta$ between the right-hand side and left-hand side in the equation above is:
\begin{align}
\Delta&=\oint_{\mathcal{C}} dl \nu_i[ \partial_k (A \mathbf{n})\cdot d\mathbf{X} \epsilon^{ik} +  A d\theta^i] \nonumber\\
&=\oint_{\mathcal{C}} dl \nu_i[ \partial_k (A \mathbf{n})\cdot d\mathbf{X} +  A \mathbf{n}\cdot \partial_k d\mathbf{X}] \epsilon^{ik}\nonumber\\
&=\oint_{\mathcal{C}} dl \nu_i \partial_k (A dX_n)\epsilon^{ik} \nonumber\\
&=\oint_{\mathcal{S}} dS \nabla_i( \partial_k (A dX_n))\epsilon^{ik} \nonumber\\
&= 0~,
\end{align}
where we have used Eq. \ref{def_rotation_rate_appendix}, the divergence theorem \ref{DivergenceTheorem}, and the commutation relation \ref{eq:commutation_derivative_scalar}.

The transformation $\mathbf{m}^i \rightarrow \mathbf{m}^i+A g^{ij} \mathbf{e}_j$ corresponds to $\bar{m}^{ij}\rightarrow \bar{m}^{ij} -A \epsilon^{ij}$, using the definition for the modified moment tensor, Eq. \ref{eq:mbardef}. Choosing $A$ to be equal to the antisymmetric part of the tensor $\bar{m}^{ij}$, $A=\frac{1}{2}\bar{m}^{ij}\epsilon_{ij}$, then implies that $\bar{m}^{ij}$ can always be symmetrised at the cost of a redefinition of the tension tensor $\mathbf{t}^i$.
\section{Translation and rotation invariance}
\label{AppendixGibbsDuhem}
In this Appendix we obtain Gibbs-Duhem relation for polar and nematic surfaces, which result from the invariance of the surface free energy under solid translations and rotations.
\subsection{Polar surface}
\subsubsection{Invariance by translation}
To express invariance by translation for a polar surface, we consider the free energy density$f_0(c^{\alpha}, \mathbf{C}, \mathbf{p}, \boldsymbol{\nabla}\otimes\mathbf{p})$ introduced in Eq. \ref{TotalFreeEnergyDensity} with its differential given in Eq. \ref{FluidMembraneFreeEnergyDensity}. Because the free energy density $f_0$ does not depend on the surface position $\mathbf{X}$, we obtain:
\begin{align}
\partial_i f_0 =& \mu^{\alpha} \partial_i c^{\alpha} + \mathbf{K} :\partial_i \mathbf{C}  - \mathbf{h}_0\cdot \partial_i \mathbf{p} + \boldsymbol{\pi}: \partial_i (\boldsymbol{\nabla}\otimes\mathbf{p})\nonumber\\
=& \mu^{\alpha} \partial_i c^{\alpha} + \mathbf{K} :\partial_i \mathbf{C}  - \mathbf{h} \cdot \partial_i \mathbf{p} +(\nabla_j \boldsymbol{\pi}^j) \cdot \partial_i \mathbf{p} \nonumber\\&+ \boldsymbol{\pi}^j\cdot \nabla_i (\partial_j \mathbf{p})\nonumber\\
=& \mu^{\alpha} \partial_i c^{\alpha} + \mathbf{K} :\partial_i \mathbf{C}  - \mathbf{h} \cdot \partial_i \mathbf{p} +\nabla_j (\boldsymbol{\pi}^j \cdot \partial_i \mathbf{p} )\nonumber\\
&+ \boldsymbol{\pi}^j\cdot \left[\nabla_i (\partial_j \mathbf{p} )- \nabla_j (\partial_i \mathbf{p})\right]\nonumber\\
=& \mu^{\alpha} \partial_i c^{\alpha} + \mathbf{K} :\partial_i \mathbf{C}  - \mathbf{h} \cdot \partial_i \mathbf{p} +\nabla_j (\boldsymbol{\pi}^j \cdot \partial_i \mathbf{p} )\label{appendix:eq_invariance_translation_polar}~,
\end{align}
where one has used that $\boldsymbol{\pi} = \mathbf{e}_i\otimes \boldsymbol{\pi}^i$, the definition of the total molecular field $\mathbf{h}$ (Eq. \ref{eq:def_total_molecular_field_h}), the definition of the gradient of polarity \ref{def:appendix:vector_gradient} and the commutation relation Eq. \ref{eq:commutation_derivative_vector}.
This relation reflects the property that the free energy density $f_0$ is a function only of the surface internal variables, and leads to the relation given in Eq. \ref{GibbsDuhemTranslation}.

\subsubsection{Invariance by rotation}
\label{appendix_polar_invariance_rotation}
We now discuss the invariance of free energy under an infinitesimal rotation $d\boldsymbol{\vartheta}$. Under this rotation the polar vector, curvature tensor, and gradient of polarity tensors transform as:
\begin{align}
d \mathbf{p} &=d\boldsymbol{\vartheta}\times\mathbf{p}~,\label{appendix_polar_invariance_rotation_deltap}\\
d \mathbf{C}&= d\boldsymbol{\vartheta}\times_1\mathbf{C} + d\boldsymbol{\vartheta}\times_2\mathbf{C} ~,\label{appendix_polar_invariance_rotation_deltaC}\\
d (\boldsymbol{\nabla}\otimes\mathbf{p}) &=  d\boldsymbol{\vartheta}\times_1(\boldsymbol{\nabla}\otimes\mathbf{p}) + d\boldsymbol{\vartheta}\times_2(\boldsymbol{\nabla}\otimes\mathbf{p})~,
\end{align}
which correspond to the requirements of a vanishing corotational variation $D\mathbf{p}=0$, $D\mathbf{C}=0$, $D(\boldsymbol{\nabla}\otimes\mathbf{p})=0$. The concentration fields do not change, $dc^{\alpha}=0$. One then obtains for the change of free energy density under this rotation:
\begin{align}
d f_0 =& \mathbf{K}:(d\boldsymbol{\vartheta}\times_1\mathbf{C}) +\mathbf{K}:(d\boldsymbol{\vartheta}\times_2\mathbf{C}) - \mathbf{h}_0 \cdot (d\boldsymbol{\vartheta}\times\mathbf{p})\nonumber\\
& + \boldsymbol{\pi}: (d\boldsymbol{\vartheta}\times_1(\boldsymbol{\nabla}\otimes\mathbf{p}) ) + \boldsymbol{\pi}: (d\boldsymbol{\vartheta}\times_2(\boldsymbol{\nabla}\otimes\mathbf{p}) ) \nonumber\\
 =& d\boldsymbol{\vartheta} \cdot \left(\mathbf{C}\times_1 \mathbf{K}  +\mathbf{C}\times_2 \mathbf{K} + \mathbf{h}_0\times \mathbf{p} - \boldsymbol{\pi} \times_1( \boldsymbol{\nabla}\otimes\mathbf{p})   \right.  \nonumber\\
&\left.- \boldsymbol{\pi} \times_2 (\boldsymbol{\nabla}\otimes\mathbf{p})\right)\nonumber\\
 =& d\boldsymbol{\vartheta} \cdot (\mathbf{C}\times_1 \mathbf{K}  +\mathbf{C}\times_2 \mathbf{K} + \mathbf{h}_0\times \mathbf{p} - \epsilon_{i}{}^{j}( \boldsymbol{\pi}^i \cdot \partial_j \mathbf{p} ) \mathbf{n}\nonumber\\
&  - \boldsymbol{\pi}^i\times\partial_i \mathbf{p} )~.
\end{align}
where we have
used that $\boldsymbol{\pi} =\mathbf{e}_i \otimes \boldsymbol{\pi}^i$, and used the notation for cross-products of tensors introduced in Eq. \ref{eq:appendix_cross_product_tensor} and the related circular shift relation of triple products, Eqs. \ref{eq:appendix:triple_product_vectors} and \ref{eq:appendix:triple_product_tensors}. Invariance of the free energy density under a local rotation then leads to the following condition:
\begin{align}
\label{eq:appendix:polar_invariance_rotation}
\boldsymbol{\pi}^i \times\partial_i \mathbf{p}+\epsilon_{i}{}^{j}( \boldsymbol{\pi}^i\cdot\partial_j \mathbf{p}) \mathbf{n} =\mathbf{h}_0\times\mathbf{p}+\mathbf{C}\times_1 \mathbf{K}  +\mathbf{C}\times_2 \mathbf{K}~.
\end{align}
which is reported in the main text in Eq. \ref{GibbsDuhemRotation}. 
This equation can be projected on the normal and tangent directions to give the conditions:
\begin{align}
2 C_i{}^k \epsilon_k{}^j K^i{}_j+(\mathbf{h}_0\times\mathbf{p})\cdot\mathbf{n}-\epsilon_{i}{}^{j}  \boldsymbol{\pi}^i\cdot\partial_j \mathbf{p}  &\nonumber\\
 - (\boldsymbol{\pi}^i\times\partial_i \mathbf{p})\cdot\mathbf{n}&=0\\
(\boldsymbol{\pi}^i \times\partial_i \mathbf{p})\cdot\mathbf{e}_j&=0
\end{align}
where one has used the symmetry of the tensor $K_{ij}$, and that $\mathbf{K}$ and $\mathbf{h}_0$ are tangent to the surface.

\subsection{Nematic surface}
\subsubsection{Invariance by translation}
We now consider a nematic surface with free energy density $f_0(c^{\alpha}, \mathbf{C}, \mathbf{Q}, \boldsymbol{\nabla}\otimes\mathbf{Q})$ with $\mathbf{Q}$ the nematic tensor on the surface. One then obtains the gradient of the free energy density, using that it does not depend on the surface position:
\begin{align}
\partial_i f_0 =& \mu^{\alpha} \partial_i c^{\alpha} + \mathbf{K} :\partial_i \mathbf{C}  - \mathbf{H}_0 :\partial_i \mathbf{Q} + \boldsymbol{\Pi}\triplecontract \partial_i (\boldsymbol{\nabla}\otimes\mathbf{Q})\nonumber\\
 =& \mu^{\alpha} \partial_i c^{\alpha} + \mathbf{K} :\partial_i \mathbf{C}  - \mathbf{H} : \partial_i \mathbf{Q}+(\nabla_j \boldsymbol{\Pi}^j ):\partial_i \mathbf{Q}\nonumber\\
 &+ \boldsymbol{\Pi}^j: \nabla_i \partial_j\mathbf{Q}\nonumber\\
  =& \mu^{\alpha} \partial_i c^{\alpha} + \mathbf{K} :\partial_i \mathbf{C}  - \mathbf{H} : \partial_i \mathbf{Q}+\nabla_j( \boldsymbol{\Pi}^j : \partial_i \mathbf{Q})\nonumber\\
 &+ \boldsymbol{\Pi}^j: (\nabla_i \partial_j\mathbf{Q}-\nabla_j \partial_i\mathbf{Q})\nonumber\\
   =& \mu^{\alpha} \partial_i c^{\alpha} + \mathbf{K} :\partial_i \mathbf{C}  - \mathbf{H} : \partial_i \mathbf{Q}+\nabla_j( \boldsymbol{\Pi}^j : \partial_i \mathbf{Q})~,
   \label{appendix:eq_invariance_translation_nematic}
\end{align}
where one has used that $\boldsymbol{\Pi}=\mathbf{e}_i\otimes \boldsymbol{\Pi}^i$, the definition of the total molecular field $\mathbf{H}$ (Eq. \ref{eq:def_total_molecular_field_nematic}) and the commutation relation Eq. \ref{eq:commutation_derivative_tensor}. Rearranging then gives the Gibbs-Duhem relation,  Eq. \ref{eq:Gibbs_Duhem_translation_nematic}.

\subsubsection{Invariance by rotation}
\label{sec:appendix_invariance_rotation_nematic}
We proceed as in Appendix \ref{appendix_polar_invariance_rotation} and consider the effect of a local infinitesimal rotation $d\boldsymbol{\vartheta}$ on the free energy density $f_0$.  The curvature tensor transforms as in Eq. \ref{appendix_polar_invariance_rotation_deltaC} under a rotation, and the nematic tensor and its gradient transform as
\begin{align}
d \mathbf{C}=& d\boldsymbol{\vartheta}\times_1\mathbf{C} + d\boldsymbol{\vartheta}\times_2\mathbf{C} ~,\label{appendix_nematic_invariance_rotation_deltaC}\\
d \mathbf{Q}=& d\boldsymbol{\vartheta}\times_1\mathbf{Q} + d\boldsymbol{\vartheta}\times_2\mathbf{Q} ~,\label{appendix_nematic_invariance_rotation_deltaQ}\\
d (\boldsymbol{\nabla}\otimes\mathbf{Q}) =&  d\boldsymbol{\vartheta}\times_1(\boldsymbol{\nabla}\otimes\mathbf{Q}) + d\boldsymbol{\vartheta}\times_2(\boldsymbol{\nabla}\otimes\mathbf{Q})\nonumber\\&+ d\boldsymbol{\vartheta}\times_3(\boldsymbol{\nabla}\otimes\mathbf{Q})~,
\end{align}
which correspond to $D\mathbf{C}=0$, $D \mathbf{Q}=0$ and $D (\boldsymbol{\nabla}\otimes\mathbf{Q}) =0$. Here we use the notation for cross-product of tensors introduced in Eq. \ref{eq:cross_product_tensor_notation_generalized}. Under the infinitesimal rotation one also has $dc^{\alpha}=0$.
This then leads to the change of free energy density under the rotation:
\begin{align}
d f_0 =& \mathbf{K}:(d\boldsymbol{\vartheta}\times_1\mathbf{C}) +\mathbf{K}:(d\boldsymbol{\vartheta}\times_2\mathbf{C}) - \mathbf{H}_0 :(d\boldsymbol{\vartheta}\times_1\mathbf{Q})\nonumber\\
&  - \mathbf{H}_0: (d\boldsymbol{\vartheta}\times_2\mathbf{Q})+ \boldsymbol{\Pi}\triplecontract (d\boldsymbol{\vartheta}\times_1(\boldsymbol{\nabla}\otimes\mathbf{Q}) )\nonumber\\
& + \boldsymbol{\Pi}\triplecontract  (d\boldsymbol{\vartheta}\times_2(\boldsymbol{\nabla}\otimes\mathbf{Q}) ) + \boldsymbol{\Pi}\triplecontract  (d\boldsymbol{\vartheta}\times_3(\boldsymbol{\nabla}\otimes\mathbf{Q}) ) \nonumber\\
=& d\boldsymbol{\vartheta}\cdot( \mathbf{C}\times_1\mathbf{K}+\mathbf{C}\times_2\mathbf{K} + \mathbf{H}_0 \times_1\mathbf{Q}+ \mathbf{H}_0 \times_2\mathbf{Q} \nonumber\\
&- \epsilon_i{}^j (\boldsymbol{\Pi}^i  :\partial_j\mathbf{Q}) \mathbf{n}- \boldsymbol{\Pi}^k\times_1 \partial_k \mathbf{Q} - \boldsymbol{\Pi}^k\times_2  \partial_k \mathbf{Q})~,
\end{align}
where we have 
used that $\boldsymbol{\Pi} =\mathbf{e}_i \otimes \boldsymbol{\Pi}^i $, and used the notation for cross-products of tensors introduced in Eq. \ref{eq:appendix_cross_product_tensor} and the related circular shift relation of triple products, Eqs. \ref{eq:appendix:triple_product_vectors} and \ref{eq:appendix:triple_product_tensors}.  Invariance of the free energy density under this rotation then leads to the relation:
\begin{align} 
\label{eq:appendix_invariance_rotation_nematic}
 \boldsymbol{\Pi}^k\times_1 \partial_k \mathbf{Q} + \boldsymbol{\Pi}^k\times_2  \partial_k \mathbf{Q}+\epsilon_i{}^j (\boldsymbol{\Pi}^i  :\partial_j\mathbf{Q}) \mathbf{n}=\nonumber\\
 \mathbf{H}_0 \times_1\mathbf{Q}+ \mathbf{H}_0 \times_2\mathbf{Q}+\mathbf{C}\times_1\mathbf{K}+\mathbf{C}\times_2\mathbf{K} ~,
\end{align}
which is reported in Eq. \ref{eq:Gibbs_Duhem_rotation_nematic}.

\section{Equilibrium forces}
\subsection{Generic equilibrium polar fluid membrane}
 \label{AppendixEquilibriumPolar}

\subsubsection{Internal free energy}
\label{AppendixEquilibriumPolar:internal_free_energy}
Here we calculate the free energy change associated to an infinitesimal surface shape change, using the infinitesimal variation of the surface free energy density given in Eq. \ref{FluidMembraneFreeEnergyDensity}. We consider a shape change of the surface, such that a point at $\mathbf{X}(s^1,s^2)$ is displaced to $\mathbf{X}(s^1,s^2)+d\mathbf{X}(s^1,s^2)$. We allow for an arbitrary change in $\mathbf{p}$ and in the concentration $ c^{\alpha}$. Calculating the variation of the free energy about equilibrium, we have
\begin{align}
	d F_0&=\int_\mathcal{S} dS\,\left(f_0\, g^{ij}\frac{d g_{ij}}{2}+ d f_0\right)~,
\end{align}
where we have used the variation of the square root of the metric determinant, Eq. \ref{Variation_sqrt_metric_determinant}.
The first term arises from area variation and the second from variation of $f_0$.  Using the expression for the free energy density differential $df_0$, this can be rewritten
\begin{align}
\label{eq:deltaF0:polar:intermediatestep}
	d F_0&=\int_{\mathcal{S}} dS\,\left[(f_0-\mu^\alpha c^\alpha)g^{ij}\frac{d g_{ij}}{2}-\mathbf{h}_0\cdot  d\mathbf{p}+\mathbf{K}:d\mathbf{C} \right.\nonumber\\
&	\left.+\boldsymbol{\pi}: d(\boldsymbol{\nabla}\otimes \mathbf{p} )
+\mu^{\alpha} \dbar c^{\alpha}\right]~,
\end{align}
where the variation of concentration that does not arise from dilution, $\dbar c^{\alpha}$, has been defined in Eq. \ref{eq:variation_c_alpha_appendix}. This expression can be further rewritten,
\begin{align}
	dF_0
	=&	\int_{\mathcal{S}} dS\,\left[t_e^{ij} \frac{d g_{ij}}{2}		+ \mathbf{m}_e^i\cdot\partial_i d\boldsymbol{\vartheta}-\mathbf{h}\cdot D\mathbf{p}+\mu^{\alpha} \dbar c^{\alpha}\right]	\nonumber\\
&+\oint_\mathcal{C} dl \nu_i\,\boldsymbol{\pi}^i\cdot D\mathbf{p}~,
\label{eq:deltaF0:polar:finalstep}
\end{align}

with 
\begin{align}
\mathbf{t}_e^i =&\left[ (f_0-\mu^{\alpha} c^{\alpha}) g^{ij} - K^{ik} C^j{}_k - \boldsymbol{\pi}^i\cdot\partial^j \mathbf{p}\right]\mathbf{e}_j\nonumber\\
&+t_{e,n}^i \mathbf{n}~,\label{eq:ti_equilibrium_definition}\\
 \mathbf{m}_e^i=&K^{ij}\epsilon_j{}^k \mathbf{e}_k+\mathbf{p}\times\boldsymbol{\pi}^i \label{eq:mi_equilibrium_definition}~.
\end{align}
We postpone to the end of this section the calculation showing the equality between Eqs. \ref{eq:deltaF0:polar:intermediatestep} and \ref{eq:deltaF0:polar:finalstep}.  We first note that using the expression for the variation of the free energy, eq. \ref{eq:equilibrium_variations_W}, and the virtual work expression Eq.~\ref{eq:AppendixVirtualWork_first_expression}, $\mathbf{t}_e^i $ and $ \mathbf{m}_e^i$ are indeed the equilibrium tension and moment tensors.

To obtain an explicit expression for the normal part of the equilibrium tension tensor $t^i_{e,n}$, we use the tangential torque balance condition Eq. \ref{TorqueBalanceTangential}. We find:
\begin{align}
t_{e,n}^i =& \nabla_j \bar{m}_e^{ji} +C_j{}^k m_{n,e}^j \epsilon^i {}_{k}+\Gamma^{{\rm ext}, j}\epsilon^i{}_j \nonumber\\
=&\nabla_j K^{ji}-\epsilon^{ik} \nabla_j[\left(\boldsymbol{\pi}^j\times\mathbf{p})\cdot\mathbf{e}_k \right]\nonumber\\
&- \epsilon^i {}_{k}C_j{}^k (\boldsymbol{\pi}^j\times\mathbf{p} )\cdot\mathbf{n}+\Gamma^{{\rm ext}, j}\epsilon^i{}_j \nonumber\\
=&\nabla_j K^{ji}-\epsilon^{ik} [\left( \nabla_j\boldsymbol{\pi}^j )\times\mathbf{p})\cdot\mathbf{e}_k \right] +\Gamma^{{\rm ext}, j}\epsilon^i{}_j \nonumber\\
=&\nabla_j K^{ji}+\epsilon^i{}_j\left[\boldsymbol{\Gamma}^{{\rm ext}} - \mathbf{h}\times\mathbf{p}\right]\cdot\mathbf{e}^j\label{eq:appendix_equilibrium_normal_tension_polar}~,
\end{align}
where we have used the relation coming from invariance by rotation, Eq. \ref{eq:appendix:polar_invariance_rotation}, and in the last line that $\mathbf{h}_0\times\mathbf{p}$ is normal to the surface.

We now establish the equality between Eqs. \ref{eq:deltaF0:polar:intermediatestep} and \ref{eq:deltaF0:polar:finalstep}. The infinitesimal variation of vectors $\mathbf{a}$ and second-rank tensors $\mathbf{B}$ can be decomposed in a corotational part and a infinitesimal rotation arising from the surface rotation $d\boldsymbol{\vartheta}$, $d \mathbf{a}=D\mathbf{a}+ d\boldsymbol{\vartheta}\times \mathbf{a}$ and  $d\mathbf{B}=D\mathbf{B}+ d\boldsymbol{\vartheta}\times_1 \mathbf{B}+ d\boldsymbol{\vartheta}\times_2 \mathbf{B}$ (Eqs \ref{def_corot_variation_p} and \ref{def_corot_variation_Q}). A local infinitesimal rotation of the surface and its associated vector and tensor fields does not change the free energy density of the surface. This property has been used in Appendix \ref{AppendixGibbsDuhem} to obtain a Gibbs-Duhem relation.  Such a local rotation is equivalent to performing a rotation of the vectors $\mathbf{a}$ and tensors $\mathbf{B}$ together with the shape rotation, such that $D\mathbf{a}=0$ and $D\mathbf{B}=0$ . Therefore, only corotational variations of vectors and tensor lead to a change in the surface free energy, and one can rewrite Eq. \ref{eq:deltaF0:polar:intermediatestep}:
\begin{align}
	dF_0=&\int_{\mathcal{S}} dS\,\left[(f_0-\mu^\alpha c^\alpha)g^{ij}\frac{d g_{ij}}{2}-\mathbf{h}_0\cdot D\mathbf{p}+\mathbf{K}:D\mathbf{C} \right.\nonumber\\
&\left.	+\boldsymbol{\pi}: D(\boldsymbol{\nabla}\otimes \mathbf{p} )
+\mu^{\alpha} \dbar c^{\alpha}\right]\nonumber\\[0.2cm]
	=&	\int_{\mathcal{S}} dS\,\left[((f_0-\mu^\alpha c^\alpha)g^{ij}-\boldsymbol{\pi}^i \cdot \partial^j \mathbf{p})\frac{d g_{ij}}{2}\right.\nonumber\\
	&\left.-\mathbf{h}_0\cdot D\mathbf{p}+K^{ij} D C_{ij}+
	\boldsymbol{\pi}^i\cdot\partial_i D\mathbf{p}\right.\nonumber\\
	&\left.+\boldsymbol{\pi}^i\cdot (\partial_i d\boldsymbol{\vartheta}\times\mathbf{p})+\mu^{\alpha} \dbar c^{\alpha}\right]\nonumber \\[.2cm]
	=&	\int_{\mathcal{S}} dS\,\left[((f_0-\mu^\alpha c^\alpha)g^{ij}-\boldsymbol{\pi}^i \cdot \partial^j \mathbf{p})\frac{d g_{ij}}{2}\right.\nonumber\\
&\left.	-\mathbf{h}\cdot D\mathbf{p}+K^{ij} \left(-\frac{d g_{ik}}{2}C^k{}_j +\epsilon_{j}{}^{k} \partial_i d\boldsymbol{\theta}\cdot\mathbf{e}_k\right)\right.\nonumber\\
	&\left.+(\partial_i d\boldsymbol{\vartheta})\cdot(\mathbf{p}\times \boldsymbol{\pi}^i)+\mu^{\alpha} \dbar c^{\alpha}\right]	\nonumber \\
&+\oint_\mathcal{C} dl \nu_i\,\boldsymbol{\pi}^i\cdot D\mathbf{p}\nonumber\\
	=&	\int_{\mathcal{S}} dS\,\left[((f_0-\mu^\alpha c^\alpha)g^{ij}-\boldsymbol{\pi}^i \cdot \partial^j \mathbf{p} -K^{ik} C_k{}^j)\frac{d g_{ij}}{2}\right.\nonumber\\
&\left.	-\mathbf{h}\cdot D\mathbf{p}	+( K^{ik} \epsilon_k{}^j \mathbf{e}_j +\mathbf{p} \times \boldsymbol{\pi}^i  )\cdot\partial_i d\boldsymbol{\vartheta}+\mu^{\alpha} \dbar c^{\alpha}\right]	\nonumber\\
&+\oint_\mathcal{C} dl \nu_i\,\boldsymbol{\pi}^i\cdot D\mathbf{p}~.
\label{deltaF0polar}
\end{align}

To go from the first to the second line above, we have used 
Eq.~\ref{Dgradacomponents}.  To go from the second to the third line, we have performed an integration by parts and have used the definition
 $\mathbf{h}=\mathbf{h}_0+\nabla_i\boldsymbol{\pi}^i$.  We have also used Eq. \ref{inplanegradtheta} to reexpress the differential of the curvature tensor in terms of $(\partial_i d\boldsymbol{\theta})\cdot\mathbf{e}_j$ and the variations of the metric tensor.
 
\subsubsection{External forces and torques arising from an external potential}
\label{appendix_force_torque_external_potential}
We consider here forces arising from an external potential $U=\int_{\mathcal{S}} dS c^{\alpha} u^{\alpha}(\mathbf{X}, \mathbf{n}, \mathbf{p})$, where $u^{\alpha}$ acts on component $\alpha$. Variation of the surface shape $d \mathbf{X}$ and of the polarity vector $\mathbf{p}$ leads to
\begin{align}
d U = &\int_{\mathcal{S}} dS \left[ c^{\alpha}\left[\frac{ \partial u^{\alpha}}{\partial \mathbf{X}} \cdot d \mathbf{X} + \frac{\partial u^{\alpha}}{\partial \mathbf{n}} \cdot (d\boldsymbol{\vartheta}\times\mathbf{n}) \right.\right.\nonumber\\
&\left.\left.+\frac{\partial u^{\alpha}}{\partial \mathbf{p}}\cdot  \left(d\boldsymbol{\vartheta}\times\mathbf{p}\right) +\frac{\partial u^{\alpha}}{\partial \mathbf{p}} \cdot D \mathbf{p}\right]+u^{\alpha}\dbar c^{\alpha} \right]\,,
\end{align} 
where we have used Eq. \ref{eq:variation_c_alpha_appendix} for $\dbar c^{\alpha}$, a change of concentration not arising from dilution; $d \mathbf{n}=d \boldsymbol{\vartheta}\times \mathbf{n}$;  and Eq. \ref{def_corot_variation_p},  giving here $d\mathbf{p}=d\boldsymbol{\vartheta}\times \mathbf{p}+D\mathbf{p}$. 
 The expression for $d U$ can then be rewritten
\begin{align}
d U =& \int_{\mathcal{S}} dS \left[c^{\alpha}\left[\frac{ \partial u^{\alpha}}{\partial \mathbf{X}} \cdot d \mathbf{X}- \left(\frac{\partial u^{\alpha}}{\partial \mathbf{n}}\times \mathbf{n}
+\frac{\partial u^{\alpha}}{\partial \mathbf{p}}\times \mathbf{p}\right)\cdot d\boldsymbol{\vartheta}\right.\right.\nonumber\\
&\left.\left.+\frac{\partial u^{\alpha}}{\partial \mathbf{p}} \cdot D \mathbf{p}\right]+u^{\alpha} \dbar c^{\alpha}  \right]\,.
\end{align}

Comparing this expression to the external virtual work $d W_{\rm ext}$ defined in Eq. \ref{eq:VirtualWorkextDef}, and using the expression for the variation of the external potential \ref{eq:equilibrium_variations_Wext}, one obtains expressions for the equilibrium external force density, torque density, molecular field, and external chemical potentials:
\begin{align}
\mathbf{f}^{\rm ext}_e &= -c^{\alpha} \frac{ \partial u^{\alpha}}{\partial \mathbf{X}} ~,\label{eq:external_force_density_polar_appendix}\\
\boldsymbol{\Gamma}^{\rm ext}_e&=c^{\alpha}\left(\frac{\partial u^{\alpha}}{\partial \mathbf{n}}\times \mathbf{n}
+\frac{\partial u^{\alpha}}{\partial \mathbf{p}}\times \mathbf{p}\right)~,
\label{eq:external_torque_density_polar_appendix}\\
\mathbf{h}^{\rm ext} &=c^{\alpha} \frac{\partial  u^{\alpha}}{\partial\mathbf{p}}~,\label{eq:external_molecular_field_polar_appendix}\\
\mu^{\rm ext,\alpha}&=-u^{\alpha}~.
\end{align}
With these definitions, the variation of external potential can be written
\begin{align}
\label{eq:variation_dU_polar}
dU =& \int_{\mathcal{S}} dS \left[-\mathbf{f}^{\rm ext}_e  \cdot d\mathbf{X}-\boldsymbol{\Gamma}^{\rm ext}\cdot d\boldsymbol{\vartheta}+\mathbf{h}^{\rm ext} \cdot D\mathbf{p} \right.\nonumber\\
&\left.- \mu^{\rm ext,\alpha} \dbar c^{\alpha}\right]~.
\end{align}
As a result of these equilibrium equations, the normal torque can be written:
\begin{align}
\label{appendix:equilibrium_normal_torque_polar}
\Gamma_{e,n}^{\rm ext} = (\mathbf{h}^{\rm ext}\times\mathbf{p})\cdot\mathbf{n}~,
\end{align}
and we also note the following identity:
\begin{align}
\label{appendix:gradient_external_chemical_potential_polar}
c^{\alpha} \partial_i \mu^{{\rm ext},\alpha} =& -c^{\alpha} \partial_i u^{\alpha}\nonumber\\
=& -c^{\alpha}\left[ \frac{ \partial u^{\alpha}}{\partial \mathbf{X}}  \cdot\mathbf{e}_i  +\frac{\partial u^{\alpha}}{\partial \mathbf{n}} \cdot\partial_i \mathbf{n}  + \frac{\partial  u^{\alpha}}{\partial\mathbf{p}} \cdot\partial_i \mathbf{p}\right]\nonumber\\
=& f_{e,i}^{\rm ext} + C_{ij}\epsilon^{jk} \left[\boldsymbol{\Gamma}_e^{\rm ext} - \mathbf{h}^{\rm ext}\times\mathbf{p}\right]\cdot\mathbf{e}_k\nonumber\\
&- \mathbf{h}^{\rm ext} \cdot\partial_i \mathbf{p}~.
\end{align}

\subsection{Generic equilibrium nematic fluid membrane}
\label{sec:appendix_eq_nematic_fluid_membrane}
\subsubsection{Internal free energy}
\label{AppendixEquilibriumNematic:internal_free_energy}
We now consider a nematic fluid membrane with free energy for a region of surface $\mathcal{S}$  given by $F_0=\int_{\mathcal{S}} dS f_0$ and free energy differential given by Eq. \ref{eq:free_energy_differential_nematic}. The surface is also subjected to an external potential $U=\int_{\mathcal{S}} dS c^{\alpha} u^{\alpha}(\mathbf{X}, \mathbf{n}, \mathbf{Q})$, where $u^{\alpha}$ is an external potential density acting on species $\alpha$. We consider a shape change of the surface, displacing $\mathbf{X}(s^1,s^2)$  to $\mathbf{X}(s^1,s^2)+d\mathbf{X}(s^1,s^2)$. 

Subjected to a variation of the surface shape, of the concentration field $c^{\alpha}$ and of $\mathbf{Q}$, the variation of the free energy about equilibrium  is 
\begin{align}
\label{eq:deltaF0:nematic:intermediatestep}
	d F_0=&\int_\mathcal{S} dS\,\left(f_0\, g^{ij}\frac{d g_{ij}}{2}+ d f_0\right)\nonumber\\
=&\int_{\mathcal{S}} dS\,\left[(f_0-\mu^\alpha c^\alpha)g^{ij}\frac{d g_{ij}}{2}-\mathbf{H}_0 : d\mathbf{Q}+\mathbf{K}:d\mathbf{C} \right.\nonumber\\
&	\left.+\boldsymbol{\Pi}\triplecontract d(\boldsymbol{\nabla}\otimes \mathbf{Q} )
+\mu^{\alpha} \dbar c^{\alpha}\right]~,
\end{align}
where the variation of concentration that does not arise from dilution, $\dbar c^{\alpha}$, has been defined in Eq. \ref{eq:variation_c_alpha_appendix}. This expression can be rewritten:
\begin{align}
	dF_0
=&	\int_{\mathcal{S}} dS \left[ t_e^{ij}\frac{d g_{ij}}{2}+\mathbf{m}_e^i\cdot \partial_id\boldsymbol{\vartheta}-\mathbf{H} : D\mathbf{Q}+\mu^{\alpha} \dbar c^{\alpha} \right]\nonumber \\
&+\oint_\mathcal{C} dl \,\nu_i \boldsymbol{\Pi}^i:D\mathbf{Q}~,\label{eq:deltaF0:nematic:finalstep}
\end{align}
with
\begin{align}
	\mathbf{t}_{e}^{i}&=\left[(f_0-\mu^\alpha c^\alpha) g^{ij}-K^{ik}C_k{}^j	-\boldsymbol{\Pi}^i :\partial^j\mathbf{Q}\right]\mathbf{e}_j
\nonumber\\
&+t_{e,n}^i \mathbf{n} ~, \label{eq:teqnematic}\\	 
	\mathbf{m}_e^i&=K^{ik} \epsilon_{k}{}^{j} \mathbf{e}_j- \boldsymbol{\Pi}^i\times_1\mathbf{Q}-\boldsymbol{\Pi}^i\times_2\mathbf{Q}~.\label{eq:meqnematic} 
\end{align}
As for the polar case, we show the equality of Eq. \ref{eq:deltaF0:nematic:intermediatestep} and \ref{eq:deltaF0:nematic:finalstep} at the end of this section. We first note that using the expression for the variation of the free energy \ref{eq:equilibrium_variations_W_nematic} and the virtual work expression Eq.~\ref{eq:AppendixVirtualWork_first_expression}, $\mathbf{t}_e^i $ and $ \mathbf{m}_e^i$ are indeed the equilibrium tension and moment tensors. 

To obtain an explicit expression for the normal part of the equilibrium tension tensor $t^i_{e,n}$, we use the tangential torque balance condition Eq. \ref{TorqueBalanceTangential}. We find:
\begin{align}
t_{e,n}^i =& \nabla_j \bar{m}_e^{ji} +C_j{}^k m_{n,e}^j \epsilon^i {}_{k}+\Gamma^{{\rm ext}, j}\epsilon^i{}_j \nonumber\\
=&\nabla_j K^{ji}-\epsilon^{ik} \nabla_j[\left(\boldsymbol{\Pi}^j\times_1\mathbf{Q}+\boldsymbol{\Pi}^j\times_2\mathbf{Q})\cdot\mathbf{e}_k \right]\nonumber\\
&- \epsilon^i {}_{k}C_j{}^k (\boldsymbol{\Pi}^j\times_1\mathbf{Q}+\boldsymbol{\Pi}^j\times_2\mathbf{Q})\cdot\mathbf{n}+\Gamma^{{\rm ext}, j}\epsilon^i{}_j \nonumber\\
=&\nabla_j K^{ji}-\epsilon^{ik}\left[(( \nabla_j\boldsymbol{\Pi}^j )\times_1\mathbf{Q}+( \nabla_j\boldsymbol{\Pi}^j )\times_2\mathbf{Q})\cdot\mathbf{e}_k \right] \nonumber\\
&+\Gamma^{{\rm ext}, j}\epsilon^i{}_j \nonumber\\
=&\nabla_j K^{ji}+\epsilon^i{}_j\left[\boldsymbol{\Gamma}^{{\rm ext}} - (\mathbf{H}\times_1\mathbf{Q}+\mathbf{H}\times_2\mathbf{Q})\right]\cdot\mathbf{e}^j~,
\label{eq:appendix_equilibrium_normal_tension_nematic}
\end{align}
where we have used the relation coming from invariance by rotation, Eq. \ref{eq:appendix_invariance_rotation_nematic}; and in the last line that the vectors $\mathbf{H}_0\times_1\mathbf{Q}$ and  $\mathbf{H}_0\times_2\mathbf{Q}$ are normal to the surface.

We now establish the equality between Eqs. \ref{eq:deltaF0:nematic:intermediatestep} and \ref{eq:deltaF0:nematic:finalstep}. Starting from Eq. \ref{eq:deltaF0:nematic:intermediatestep}, we obtain:
\begin{align}
	dF_0=&\int_{\mathcal{S}} dS \left[(f_0-\mu^\alpha c^\alpha)g^{ij}\frac{d  g_{ij}}{2}-\mathbf{H}_0 :D\mathbf{Q}+\mathbf{K}:D\mathbf{C} \right.\nonumber\\
	&\left.+\boldsymbol{\Pi}\triplecontract D (\boldsymbol{\nabla}\otimes\mathbf{Q})
	+\mu^{\alpha} \dbar c^{\alpha}\right]\nonumber\\
	=&	\int_{\mathcal{S}} dS \left[\left((f_0-\mu^\alpha c^\alpha)g^{ij}-\boldsymbol{\Pi}^i :\partial^j\mathbf{Q}\right)\frac{d g_{ij}}{2}\right.\nonumber\\
	&\left.-\mathbf{H}_0 :D\mathbf{Q}+K^{ij} D C_{ij}+ \boldsymbol{\Pi}^i:\partial_i D\mathbf{Q} \nonumber \right.\\
&\left.+\boldsymbol{\Pi}^i : (\partial_id\boldsymbol{\vartheta}\times_1\mathbf{Q})+\boldsymbol{\Pi}^i : (\partial_id\boldsymbol{\vartheta}\times_2\mathbf{Q})+\mu^{\alpha} \dbar c^{\alpha}\right]\nonumber \\
=&	\int_{\mathcal{S}} dS \left[\left((f_0-\mu^\alpha c^\alpha)g^{ij}-\boldsymbol{\Pi}^i :\partial^j\mathbf{Q}\right)\frac{d g_{ij}}{2}\right.\nonumber\\
	&\left.-\mathbf{H} : D\mathbf{Q}+K^{ij} \left(-\frac{d g_{ik}}{2}C^k{}_j +\epsilon_{j}{}^{k} \partial_i d\boldsymbol{\theta}\cdot\mathbf{e}_k\right)\nonumber \right.\\
&\left.-(\partial_id\boldsymbol{\vartheta})\cdot ( \boldsymbol{\Pi}^i\times_1\mathbf{Q}+\boldsymbol{\Pi}^i\times_2\mathbf{Q})+\mu^{\alpha} \dbar c^{\alpha}\right]\nonumber \\
&+\oint_\mathcal{C} dl \,\nu_i \boldsymbol{\Pi}^i:D\mathbf{Q}\nonumber\\
=&	\int_{\mathcal{S}} dS \left[\left(\left(f_0-\mu^\alpha c^\alpha)g^{ij}-\boldsymbol{\Pi}^i :\partial^j\mathbf{Q}-K^{ik}C_k{}^j\right)\right)\frac{d g_{ij}}{2}\right.\nonumber\\
&\left.+ (K^{ij} \epsilon_{j}{}^{k} \mathbf{e}_k-\boldsymbol{\Pi}^i\times_1\mathbf{Q}-\boldsymbol{\Pi}^i\times_2\mathbf{Q})\cdot \partial_id\boldsymbol{\vartheta}\right.\nonumber \\
	&\left.-\mathbf{H} : D\mathbf{Q}+\mu^{\alpha} \dbar c^{\alpha} \right]+\oint_\mathcal{C} dl \,\nu_i \boldsymbol{\Pi}^i:D\mathbf{Q}\,.\label{deltaF0nematic}
\end{align}
In the first line, we have used the invariance of the free energy density by solid rotation (Appendix \ref{sec:appendix_invariance_rotation_nematic}).
To go from the first to the second line, we have used 
Eq.~\ref{DgradBcomponents}.
 To go from the second to the third line, we have performed an integration by parts, we have used the definition for the total molecular field
$\mathbf{H}=\mathbf{H}_0+\nabla_i\boldsymbol{\Pi}^i$, Eq. \ref{inplanegradtheta} and the triple product relations \ref{eq:appendix:triple_product_tensors}.

\subsubsection{Forces and torques arising from an external potential}
\label{appendix_force_torque_external_potential_nematic}
We consider here forces arising from an external potential $U=\int_{\mathcal{S}} dS c^{\alpha} u^{\alpha}(\mathbf{X}, \mathbf{n}, \mathbf{Q})$, where $u^{\alpha}$ acts on component $\alpha$. Variation of the surface shape $d \mathbf{X}$ and of the second-rank tensor $\mathbf{Q}$ leads to
\begin{align}
	d U = &\int_{\mathcal{S}} dS \left[c^{\alpha}\left[\frac{ \partial u^{\alpha}}{\partial \mathbf{X}} \cdot d \mathbf{X}  + \frac{\partial u^{\alpha}}{\partial \mathbf{n}} \cdot (d \boldsymbol{\vartheta}\times\mathbf{n})\right. \right.\nonumber\\
	&\left.\left.+\frac{\partial u^{\alpha}}{\partial \mathbf{Q}} : \left(d\boldsymbol{\vartheta}\times_1\mathbf{Q}+d\boldsymbol{\vartheta}\times_2\mathbf{Q}\right) +\frac{\partial u^{\alpha}}{\partial \mathbf{Q}} :  D \mathbf{Q}\right]\right.\nonumber\\
	&\left.+u^{\alpha}\dbar c^{\alpha} \right]
\end{align} 
where we have used Eq. \ref{def_corot_variation_Q}, giving here $d\mathbf{Q}=d\boldsymbol{\vartheta}\times_1 \mathbf{Q}+d\boldsymbol{\vartheta}\times_2 \mathbf{Q}+D\mathbf{Q}$, and the variation of the normal vector, Eq. \ref{eq:rotational_variation_normal_vector}. The expression for $d U$ can then be rewritten
\begin{align}
	d U =& \int_{\mathcal{S}} dS \left[c^{\alpha}\left[\frac{ \partial u^{\alpha}}{\partial \mathbf{X}} \cdot d \mathbf{X} +\frac{\partial u^{\alpha}}{\partial \mathbf{Q}} : D \mathbf{Q}\right]+u^{\alpha} \dbar c^{\alpha} \right.\nonumber\\
	&\left.- c^{\alpha}\left(\frac{\partial u^{\alpha}}{\partial \mathbf{n}}\times \mathbf{n}
	+
	\frac{\partial u^{\alpha}}{\partial \mathbf{Q}} \times_1 \mathbf{Q}+
	\frac{\partial u^{\alpha}}{\partial \mathbf{Q}} \times_2 \mathbf{Q}\right)\cdot d\boldsymbol{\vartheta} \right]\,
\end{align}
where we have used the triple product relations \ref{eq:appendix:triple_product_tensors}. 
Then, comparing this expression to the external virtual work $d W_{\rm ext}$ defined in Eq. \ref{eq:VirtualWorkextDef}, and using the equilibrium relation Eq. \ref{eq:equilibrium_variations_Wext_nematic}, one obtains expressions for the equilibrium external force density, torque density, molecular field, and external chemical potentials:
\begin{align}
	\mathbf{f}^{\rm ext}_e &= -c^{\alpha} \frac{ \partial u^{\alpha}}{\partial \mathbf{X}}  \label{eq:externalforcedensitynematicappendix}\\
	\boldsymbol{\Gamma}^{\rm ext}_e&=c^{\alpha}\left(\frac{\partial u^{\alpha}}{\partial \mathbf{n}}\times \mathbf{n}
+\frac{\partial u^{\alpha}}{\partial \mathbf{Q}} \times_1 \mathbf{Q}+
	\frac{\partial u^{\alpha}}{\partial \mathbf{Q}} \times_2 \mathbf{Q}\right)
	\label{eq:externaltorquedensitynematicappendix}\\
	\mathbf{H}^{\rm ext} &=c^{\alpha} \frac{\partial  u^{\alpha}}{\partial\mathbf{Q}}\label{eq:externalmolecularfieldnematicappendix}\\
	\mu^{\rm ext,\alpha}&=-u^{\alpha}~.
\end{align}
With these definitions, the variation of external potential can be written
\begin{align}
\label{eq:variation_dU_nematic}
dU =& \int_{\mathcal{S}} dS \left[-\mathbf{f}^{\rm ext}_e  \cdot d\mathbf{X}-\boldsymbol{\Gamma}^{\rm ext}\cdot d\boldsymbol{\vartheta}+\mathbf{H}^{\rm ext} :D\mathbf{Q} \right.\nonumber\\
&\left.- \mu^{\rm ext,\alpha} \dbar c^{\alpha}\right]~.
\end{align}
We note that since $\mathbf{H}^{\rm ext}$ and $\mathbf{Q}$ are symmetric and $\mathbf{Q}$ is a tangent tensor, the normal torque can be written:
\begin{align}
\label{eq:appendix:external_normal_torque_nematic}
	\Gamma_n^{\rm ext} =&[ \mathbf{H}^{\rm ext}\times_1\mathbf{Q} + \mathbf{H}^{\rm ext}\times_2\mathbf{Q} ]\cdot\mathbf{n} \nonumber\\
	=&2\epsilon_i{}^kH^{{\rm ext}, ij} Q_{kj}\,.
\end{align}
We also note the following identity:
\begin{align}
\label{appendix:gradient_external_chemical_potential_nematic}
c^{\alpha} \partial_i \mu^{{\rm ext},\alpha} =& -c^{\alpha} \partial_i u^{\alpha}\nonumber\\
=& -c^{\alpha}\left[ \frac{ \partial u^{\alpha}}{\partial \mathbf{X}}  \cdot\mathbf{e}_i  +\frac{\partial u^{\alpha}}{\partial \mathbf{n}} \cdot\partial_i \mathbf{n}  + \frac{\partial  u^{\alpha}}{\partial\mathbf{Q}} :\partial_i \mathbf{Q}\right]\nonumber\\
=& f_{e,i}^{\rm ext} + C_{ij}\epsilon^{jk} \left[\boldsymbol{\Gamma}_e^{\rm ext} - \mathbf{H}^{\rm ext}\times_1\mathbf{Q} \right.\nonumber\\
&\left.- \mathbf{H}^{\rm ext}\times_2\mathbf{Q}\right]\cdot\mathbf{e}_k- \mathbf{H}^{\rm ext} :\partial_i \mathbf{Q}~.
\end{align}

\section{Entropy production rate}
\label{Appendix_entropy_production_rate}
\subsection{Entropy production rate for a polar fluid surface}
\label{AppendixEntropyProductionRatePolar}
In this Appendix we calculate the entropy production rate of a polar surface using a Lagrangian approach. That is, we follow a moving fluid surface element labeled by its center of mass ($s^1,s^2$), moving with velocity $\mathbf{v}$. Expressions for the gradient of flow $v_{ij}$, the vorticity $\boldsymbol{\omega}$, the corotational of the curvature tensor $D_t C_{ij}$ can then found from the replacements $d\mathbf{X}\rightarrow \mathbf{v} dt$, $dg_{ij}\rightarrow 2 v_{ij} dt$, $d\boldsymbol{\vartheta}\rightarrow \boldsymbol{\omega} dt$, $DC_{ij}\rightarrow D_t C_{ij} dt$ in Eqs. \ref{VariationMetricCovariant}, \ref{def_rotation_rate_appendix}, and \ref{Dcurvature_components}. Corresponding expressions are reported in Eqs. \ref{eq:expression_vij}-\ref{eq:expression_Dt_Cij}. Including the external potential $U$, the total free energy is 
\begin{align}
F_{\rm tot}=F+U=\int_\mathcal{S} dS\,(f_{\rm kin}+f_0+c^\alpha u^\alpha)~,
\end{align}
where $f_\mathrm{kin}=\rho\,v^2/2$ is the kinetic energy density. The differential of $f_0$ is given in Eq. \ref{FluidMembraneFreeEnergyDensity}:
\begin{equation}
df_0=\mu^{\alpha} d c^{\alpha}+\mathbf{K}:  d\mathbf{C} - \mathbf{h}_0 \cdot d \mathbf{p} +\boldsymbol{\pi}:  d(\boldsymbol{\nabla}\otimes\mathbf{p}) \,.\\
\end{equation}

We consider the time derivative of the free energy, $\dot{F}=\dot{F}_0+\dot{F}_\mathrm{kin}$, and first 
calculate $\dot{F}_\mathrm{kin}$:
\begin{align}
	\dot{F}_\mathrm{kin}&=\frac{d}{dt}\int_{\mathcal{S}} dS\,f_\mathrm{kin}\nonumber\\
	&=\int_\mathcal{S} dS\,\left[\frac{\rho v^2}{2}\,v_i{}^i+\frac{{\rm d}}{{\rm d}t}\left(\frac{\rho\,v^2}{2}\right)\right]\,,
\end{align}
where ${\rm d}/{\rm d}t$ denotes a Lagrangian time derivative (Eq. \ref{eq:appendix:convected_time_derivative}) and where we have used Eq. \ref{Variation_sqrt_metric_determinant}.  In the Lagrangian approach,
\begin{align}
\frac{{\rm d}}{{\rm d}t}\left(\frac{\rho\,v^2}{2}\right)&=\frac{v^2}{2}\frac{{\rm d} \rho}{{\rm d}t}+\rho\mathbf{v}\cdot\frac{{\rm d}\mathbf{v}}{{\rm d}t} =-\frac{v^2}{2}\rho\,v_i{}^i+\mathbf{v}\cdot\rho\mathbf{a}\,,
\end{align}
where we have used the mass conservation equation:
\begin{align}
\frac{{\rm d}\rho}{{\rm d}t} + \rho v_i{}^i=0~,
\end{align}
 to go from the first to the second line. We thus obtain
\begin{align}
	\dot{F}_\mathrm{kin}&=\int_\mathcal{S}dS 
	\rho\mathbf{a}\cdot\mathbf{v} ~.
\end{align}
We can express the first term in the integral above in terms of forces and moments by noting that, since we are taking a Lagrangian approach, we may make the replacements $d\mathbf{X}\to \mathbf{v}\,dt$, $d \boldsymbol{\vartheta}\to \boldsymbol{\omega}\,dt$, $d g_{ij}/2\to v_{ij}dt$, and $D C_{ij} \to D_t C_{ij}dt$ in the equality between Eqs. \ref{eq:VirtualWorkDef} and \ref{eq:VirtualWork}.  We thereby obtain
\begin{align}
	\dot{F}_\mathrm{kin}
	=&\int_\mathcal{S}dS\,\left[-\overline{t}^{ij} v_{ij}-\overline{m}^{ij} D_t C_{ij}
	-m_n^i \omega_{in}\right.\nonumber\\&\left.
	+\mathbf{f}^\mathrm{ext}\cdot\mathbf{v}+\boldsymbol{\Gamma}^\mathrm{ext}\cdot\boldsymbol{\omega}
	\right] +\oint_\mathcal{C} dl\nu_i \left[\mathbf{t}^i\cdot \mathbf{v}+\mathbf{m}^i\cdot\boldsymbol{\omega}\right]\,.
	\label{dFkindt}
\end{align}
with $\omega_{in} = (\partial_i\boldsymbol{\omega})\cdot\mathbf{n}$.

We next calculate $\dot{F}_0=d/dt \int_\mathcal{S}dS\,f_0$.  Using the differential of $f_0$, Eq.~\ref{FluidMembraneFreeEnergyDensity}, we have
\begin{align}
	\dot{F}_0&=\int_\mathcal{S} dS\,\left[f_0 v_i{}^i+
	\sum_{\alpha}\mu^\alpha\partial_t c^\alpha+\mathbf{K} : \frac{{\rm d} \mathbf{C}}{{\rm d}t} -\mathbf{h}_0 \cdot  \frac{{\rm d}\mathbf{p} }{{\rm d}t}\right.\nonumber\\
	&\left.+\boldsymbol{\pi} : \frac{{\rm d}}{{\rm d}t} (\nabla\otimes\mathbf{p})\right]\,.
	\label{dF0dt}
\end{align}

The second term on the right-hand side can be re-written using mass conservation for species $\alpha$, Eq. \ref{ConcentrationBalanceEquation} in a Lagrangian form,  that is,
\begin{equation}
\label{eq:dcalpha_variation_Lagrangian}
\frac{{\rm d} c^\alpha}{{\rm d}t}+c^\alpha v_i{}^i=r^\alpha-\nabla_i j^{\alpha,i}\,,
\end{equation}
where $r^\alpha$ is the chemical reaction rate at which a molecule of species is produced, and $j^{\alpha,i}$ is the relative flux of $\alpha$.  As a result,
\begin{equation}
\mu^\alpha \frac{{\rm d} c^\alpha}{{\rm d}t}=-\mu^\alpha c^\alpha v_i{}^i+r^\alpha \mu^\alpha+j^{\alpha,i}\partial_i\mu^\alpha-\nabla_i(\mu^\alpha j^{\alpha,i})\,,
\label{fluxspeciesalpha}
\end{equation}
Using the balance equations for tangential and normal fluxes, Eq.~\ref{BalanceTangentialFluxes}, and for chemical reactions~\ref{BalanceChemicalReactions}, we then have:
\begin{align}
\sum_{\alpha}\mu^\alpha \frac{{\rm d} c^\alpha}{{\rm d}t} =&
 \sum_{\alpha}[-v_i{}^i \mu^\alpha c^\alpha +r^\alpha\mu^\alpha]
\nonumber\\&
+\sum_{\alpha=2}^N [j^{\alpha,i}\partial_i \overline{\mu}^\alpha
-\nabla_i(\overline{\mu}^\alpha j^{\alpha,i})
]\,,
\end{align}
where
\begin{equation}
\overline{\mu}^\alpha=\mu^\alpha-\frac{m^\alpha}{m^1}\mu^1
\end{equation}
is the relative chemical potential of species $\alpha$.

With implicit summation from $\alpha=1$ to $N$ for terms with $\mu^\alpha$, and from $\alpha=2$ to $N$ for terms with $\overline{\mu}^\alpha$, we therefore have
\begin{align}
	\dot{F}_0=&\int_\mathcal{S} dS\,\left[(f_0-
\mu^\alpha c^\alpha) v_i{}^i+\mathbf{K} : \frac{{\rm d} \mathbf{C}}{{\rm d}t} -\mathbf{h}_0 \cdot\frac{{\rm d} \mathbf{p}}{{\rm d}t}\right.\nonumber\\
&\left.+\boldsymbol{\pi} : \frac{{\rm d} }{{\rm d}t} (\nabla\otimes\mathbf{p})+
j^{\alpha,i}\partial_i \overline{\mu}^\alpha
+r^\alpha\mu^\alpha\right]\nonumber\\&
	-\oint \nu_i dl\, \overline{\mu}^\alpha j^{\alpha,i}\,.
	\label{dF0dt2}
\end{align}
The remaining terms in Eq.~\ref{dF0dt} can be transformed using the results from section \ref{AppendixEquilibriumPolar} by using the equality of Eqs. \ref{eq:deltaF0:polar:intermediatestep} and \ref{eq:deltaF0:polar:finalstep}, with the replacements $d g_{ij}/2\to v_{ij}dt$,  $d \mathbf{p} \to  \frac{{\rm d}\mathbf{p}}{{\rm d}t} dt$, $d \mathbf{C} \to  \frac{{\rm d}\mathbf{C}}{{\rm d}t} dt$, $d(\boldsymbol{\nabla}\otimes\mathbf{p})\rightarrow  \frac{{\rm d} }{{\rm d}t} (\nabla\otimes\mathbf{p}) dt$, $d \boldsymbol{\vartheta}\to \boldsymbol{\omega}\,dt$ and $D\mathbf{p}\rightarrow D_t \mathbf{p} dt$:
\begin{align}
	\dot{F}_0=& \int_\mathcal{S} dS\,\Big[\overline{ t}_e^{ij} v_{ij}
	+\overline{m}_{e}^{ij}D_t C_{ij}-\mathbf{h}\cdot D_t\mathbf{p}+m_{n,e}^i \omega_{in} \nonumber\\
	&+j^{\alpha,i}\partial_i \overline{\mu}^\alpha
+r^\alpha\mu^\alpha\Big]
	\nonumber +\oint_\mathcal{C} dl\nu_i \left(-\overline{\mu}^\alpha j^{\alpha,i}+\boldsymbol{\pi}^i\cdot D_t\mathbf{p}\right)
	\,.
\end{align}
Finally, we calculate the time derivative of the external potential $U$, using Eq. \ref{eq:variation_dU_polar} and Eq. \ref{eq:dcalpha_variation_Lagrangian}:
\begin{align}
\dot{U}= &\int_{\mathcal{S}} dS \left[-\mathbf{f}^{\rm ext}_e \cdot \mathbf{v} - \boldsymbol{\Gamma}^{\rm ext}_e \cdot \boldsymbol{\omega} + \mathbf{h}^{\rm ext} \cdot D_t  \mathbf{p} \right.\nonumber\\
&\left.-  \mu^{{\rm ext}, \alpha} r^{\alpha}+ \overline{\mu}^{{\rm ext}, \alpha} \nabla_i j^{\alpha,i} \right]\,,
\end{align}
where 
\begin{equation}
\label{eq:def_relative_external_chemical_potential_appendix}
\overline{\mu}^{{\rm ext},\alpha}=\mu^{{\rm ext},\alpha}-\frac{m^\alpha}{m^1}\mu^{{\rm ext},1}
\end{equation}
is the external relative chemical potential.
Putting together $\dot{F}_\mathrm{kin}$, $\dot{F}_0$ and $\dot{U}$, we finally obtain Eq. \ref{eq:Fdot}.

\subsection{Entropy production rate for a nematic fluid surface}
\label{AppendixEntropyProductionRateNematic}
We now calculate the entropy production rate for a nematic surface. 
We carry over much of the derivation from the previous section, in particular the calculation of $\dot{F}_{\rm kin}$ is unchanged. In the calculation of $\dot{F}_0$, we use the differential of the free energy introduced in Eq. \ref{eq:free_energy_differential_nematic}.
We also use the equality between Eq. \ref{eq:deltaF0:nematic:intermediatestep} and eq. \ref{eq:deltaF0:nematic:finalstep}. We then obtain:
\begin{align}
	\dot{F}_0=&\int_\mathcal{S} dS\,\Big[ \overline{t}_e^{ij} v_{ij}
	+\overline{m}_{e}^{ij}D_t C_{ij}-\mathbf{H} : D_t\mathbf{Q}+m_{n,e}^i \omega_{in}\nonumber\\
	&+r^\alpha\mu^\alpha+j^{\alpha,i}\partial_i\overline{\mu}^\alpha\Big]+\oint_\mathcal{C} dl\nu_i (-\overline{\mu}^\alpha j^{\alpha,i}+\boldsymbol{\Pi}^i : D_t\mathbf{Q})
	\,.
	\label{dF0dtnematic}
\end{align}
The time derivative of the external potential $U$ reads, using Eq. \ref{eq:variation_dU_nematic}:
\begin{align}
\label{eq:dotU_nematic}
\dot{U}=& \int_{\mathcal{S}} dS \left[-\mathbf{f}^{\rm ext}_e \cdot \mathbf{v} - \boldsymbol{\Gamma}^{\rm ext}_e \cdot \boldsymbol{\omega} + \mathbf{H}^{\rm ext} :  D_t  \mathbf{Q}\right.\nonumber\\
&\left. -\mu^{{\rm ext}, \alpha} r^{\alpha} + \overline{\mu}^{{\rm ext}, \alpha} \nabla_i j^{\alpha,i} \right]\,,
\end{align}
where the relative external chemical potential $\bar{\mu}^{{\rm ext},\alpha}$ is defined in Eq. \ref{eq:def_relative_external_chemical_potential_appendix}.

We can then compute $\dot{F}_{\rm tot}=\dot{F}+\dot{U}=\dot{F}_0+\dot{F}_{\rm kin}+\dot{U}$ with the aid of Eqs.~\ref{dFkindt}, \ref{dF0dtnematic} and \ref{eq:dotU_nematic}, obtaining Eq. \ref{eq:total_rate_free_energy_nematic}.

\section{Full constitutive equations}
\label{appendix:full_constitutive_equations}
Here we give complete constitutive equations, including terms for isotropic surfaces which do not couple to the polar or nematic order parameter on the surface. We include terms which are first-order in the curvature tensor $C_{ij}$, first order in polarity vector $\mathbf{p}$ and its associated nematic tensor $q_{ij}=p_ip_j - \frac{1}{2} p^2 g_{ij}$ for polar surfaces and first order in the nematic tensor $Q_{ij}$ for nematic surfaces. We do not include cross-coupling terms in the gradient of the chemical potential, $\partial_i \bar{\mu}_d^{\alpha}$ or the flux $j^{\alpha,i}$, except for active couplings with the fuel hydrolysis chemical potential, $\Delta \mu$. For simplicity, ``viscous'' coupling coefficients between $\bar{t}^{ij}_d$, $\bar{m}^{ij}$, $m_{n,d}^i$ and $v_{ij}$, $D_tC_{ij}$, $\omega_{in}$ which depend on the curvature, polarity vector or nematic tensor are not considered. The same simplification is applied to ``diagonal" coupling coefficients between a flux and its conjugate force. We do not include couplings between the mechanical tensors and the deviatoric molecular field, and conversely between $D_t \mathbf{p}$ or $D_t \mathbf{Q}$ and $v_{ij}$, $D_tC_{ij}$, $\omega_{in}$, that depend on the curvature tensor. 

\subsection{Polar surface}
The tension tensor has contributions:
\begin{align}
\overline{t}^{ij}_0=&2\eta \tilde{v}^{ij} +\eta_b v_k{}^k g^{ij}+\zeta \Delta \mu g^{ij} +\zeta_{\rm n} \Delta \mu q^{ij}\nonumber\\
&+\frac{\nu}{2}(p^i h_d^j+p^j h_d^i-h_{d,k} p^k g
^{ij}) +\nu' h_{d,k} p^k g
^{ij}\nonumber\\[.2cm]
\overline{t}^{ij}_{\rm{UD}}=&2 \bar{\eta} D_t \tilde{C}^{ij}+\bar{\eta}_b D_t C_k{}^k  g^{ij}+2 \tilde{\zeta}\Delta \mu\tilde{C}^{ij} \nonumber\\
&+\zeta' \Delta\mu C_k{}^k g^{ij}  +\tilde{\zeta}_{\rm n} \Delta \mu  q^{kl} C_{kl} g^{ij}  +\tilde{\zeta}'_{\rm n} \Delta \mu  q^{ij} C_{k}{}^k 
 \nonumber \\[.2cm]
\overline{t}^{ij}_{\rm{C}}=&\eta_{\rm{C}} \left(\epsilon^{ik} D_t C_{k}{}^{j}+\epsilon^{jk} D_t C_{k}{}^{i}\right)\nonumber\\
&+\zeta_{\rm{C}}\Delta\mu \left(\epsilon^i{}_k C^{kj}+\epsilon^j{}_k C^{ki}\right)\nonumber\\
&+\zeta_{\rm{C n}}\Delta\mu(\epsilon^{ik} q_{k}{}^l C_l{}^j+\epsilon^{jk} q_{k}{}^{l} C_l{}^i)\nonumber\\
&+\tilde{\zeta}_{\rm{C n}}\Delta\mu(\epsilon^{ik} C_{k}{}^l q_l{}^j+\epsilon^{jk} C_k{}^l q_l{}^i)  \nonumber \\[.2cm]
\overline{t}^{ij}_{\rm{PC}}=&\eta_{\rm{PC}} \left(\epsilon^i{}_k v^{kj}+\epsilon^j{}_k v^{ki}\right) + \nu_{\rm PC}' \epsilon_{kl} p^k h_d^l g^{ij}\nonumber\\
&+\frac{\nu_{{\rm PC}}}{2}(\epsilon^{i}{}_{k}h_d^k p^j+\epsilon^{j}{}_{k} h_d^k p^i-\epsilon_{lk} h_d^k  p^l g^{ij})\nonumber\\
&+\zeta_{\rm PC{\rm n}}\Delta \mu  \epsilon^{ik} q_k{}^j .
\label{Appendix:ConstitutiveEquationtij}
\end{align}

The bending moment tensor reads
\begin{align}
\overline{m}^{ij}_0=&2 \eta_c D_t \tilde{C}^{ij}+  \eta_{cb} D_t C_{k}{}^{k} g^{ij}+2 \tilde{\zeta_c} \Delta \mu  \tilde{C}^{ij}+\zeta_c'  \Delta\mu C_k{}^k g^{ij}\nonumber\\
&+\tilde{\zeta}_{c\rm n} \Delta \mu  q^{kl} C_{kl} g^{ij} +\tilde{\zeta}'_{c\rm n} \Delta \mu  C_{k}{}^k  q^{ij}  \nonumber\\
\overline{m}^{ij}_{\rm{UD}}=&2\bar{\eta} \tilde{v}^{ij}+\bar{\eta}_b v_k{}^k g^{ij}+\frac{\beta}{2} (p^i h_d^j+p^j h_d^i-p_k h_d^k g^{ij})  \nonumber\\
&+ \beta' p_k h_d^k g^{ij}+ \zeta_c  \Delta \mu g^{ij}+\zeta_{c\rm n} \Delta \mu q^{ij}\nonumber \\
\overline{m}^{ij}_{\rm{C}}=&-\eta_{\rm{C}} \left(\epsilon^i{}_k v^{kj}+\epsilon^j{}_k v^{ki}\right)+\beta_{\rm{C}}'  \epsilon^{k}{}_{l} p_k h_d^l g^{ij} \nonumber\\
&+\frac{\beta_{\rm{C}}}{2}(\epsilon^{i}{}_{k}h_d^k p^j +\epsilon^{j}{}_{k}h_d^k  p^i -\epsilon_{lk} h_d^k  p^l g^{ij})\nonumber\\
&
+\zeta_{c{\rm C}{\rm n}}\Delta \mu \epsilon^{ik} q_k{}^j
\nonumber\\
\overline{m}^{ij}_{\rm{PC}}=&\eta_{\rm{cPC}} \left(\epsilon^{ik} D_t C_{k}{}^{j}+\epsilon^{jk} D_t C_{k}{}^{i}\right)\nonumber\\
&+\zeta_{\rm{PC}}\Delta \mu \left(\epsilon^i{}_k C^{kj}+\epsilon^j{}_k C^{ki}\right)\nonumber\\
&+\zeta_{c\rm{PC n}}\Delta\mu(\epsilon^{ik} q_{k}{}^l C_l{}^j+\epsilon^{jk} q_k{}^l C_l{}^i)\nonumber\\
&+\tilde{\zeta}_{c\rm{PC n}}\Delta\mu(\epsilon^{ik} C_{k}{}^l q_l{}^j+\epsilon^{jk} C_k{}^{l} q_l{}^i) ~.
 \label{Appendix:ConstitutiveEquationmij}
\end{align}
Contributions to the tensor $m_n$ read:
\begin{align}
m^i_{n0}=&\kappa \omega^i_n+\chi_{\rm p} \epsilon^{ij} p_j\Delta \mu
+\psi\, \epsilon^{i}{}_{j} h_d^j \nonumber\\
m^i_{n\rm{UD}}=&(\overline{\chi}_{\rm p} \epsilon^{i}{}_{k} C^{kj} +\overline{\chi}'_{\rm p}  \epsilon^{j}{}_{k} C^{ki} ) p_j \Delta \mu \nonumber\\
m^i_{n\rm{C}}=&(\chi_{\rm{C}p} C^{ij}  +\chi_{\rm{C}p}' C_k{}^k g^{ij})p_j  \Delta \mu\nonumber \\
m^i_{n\rm{PC}}=&\kappa_{\rm{PC}} \epsilon^{ij}\omega_{jn} +\chi_{\rm{PC}\rm{p}} p^i \Delta \mu  + \psi_{\rm PC} h_d^i~.
\label{Appendix:ConstitutiveEquationmn}
\end{align}
with $\omega_{in}=\partial_i \omega_n - C^{ij} \omega_j$.
Terms contributing to the dynamics of the polarity field read:
\begin{align}
P_0^i=&\frac{1}{\gamma} h_d^i-\nu \tilde{v}^{ij} p_j-\nu' v_k{}^k p^i+\lambda \Delta\mu p^i+\psi\,\epsilon^{ij}\omega_{jn}\nonumber\\[.2cm]
P_{\rm  UD}^i=&
-\beta p_j D_t \tilde{C}^{ij}-\beta' D_t C_k{}^k p^i\nonumber\\
&+\lambda_{\rm UD}\Delta\mu \,C^{ij}p_j 
+\lambda'_{\rm UD}\Delta\mu \, C^{j}{}_jp^i\nonumber\\[.2cm]
P_{\rm  C}^i=&\beta_{\rm{C}} \epsilon^i{}_k D_t \tilde{C}^{kj} p_j +\beta_{\rm{C}}' \epsilon^{ij} p_j D_t C_k{}^k\nonumber\\
&+(\lambda_{\rm C} \epsilon^{i}{}_{k} C^{kj} +\lambda_{\rm C}'\epsilon^{j}{}_{k} C^{ki} ) p_j \Delta \mu
\nonumber\\
P_{\rm  PC}^i=&\frac{1}{\gamma_{\rm PC}}\epsilon^{i}{}_{j} h_d^j+\nu_{\rm{PC}} \epsilon^i{}_k \tilde{v}^{kj}p_j+\nu_{\rm{PC}}'v_k{}^k \epsilon^{ij} p_j\nonumber\\
&+\lambda_{\rm{PC}}\Delta\mu \epsilon^{ij} p_j -\psi_{\rm PC} \omega^{i}_{n}\,.
\label{Appendix: constitutiveequation_dpdt}
\end{align}

The flux of species $\alpha=2...N$, relative to the centre of mass has different contributions given by
\begin{align}
j^{\alpha, i}_{0}=&-L^{\alpha\beta} \partial^i \bar{\mu}_d^{\beta}+\kappa_{\rm p}^{\alpha} p^i \Delta \mu  \nonumber\\
j^{\alpha, i}_{\rm{UD}}=&(\kappa_{\rm{UD}p}^{\alpha} C^{ij}  +\kappa_{\rm{UD}p}'^{\alpha} C_k{}^k g^{ij})p_j  \Delta \mu\nonumber \\
j^{\alpha,  i}_{\rm{C}}=&(\kappa_{\rm Cp}^{\alpha} \epsilon^{i}{}_{k} C^{kj} +\kappa_{\rm Cp}'^{\alpha}  \epsilon^{j}{}_{k} C^{ki} ) p_j \Delta \mu \nonumber\\
j^{\alpha, i}_{\rm PC}=&-L_{\rm PC}^{\alpha\beta} \epsilon^{ij} \partial_j \bar{\mu}_d^{\beta}+\kappa_{\rm PCp}^{\alpha} \epsilon^{ij} p_j\Delta \mu
~,
\label{Appendix:ConstitutiveEquationfluxj}
\end{align}
where we have used the relation $\epsilon^i{}_k C^{kj} = \epsilon^j{}_k C^{ki}+C_k{}^k \epsilon^{ij}$ to avoid introducing redundant couplings in the equation for $\mathbf{j}_{C}$. 

Finally, the rate of fuel consumption has the following decomposition:
\begin{align}
r_0=&-\zeta v_{k}{}^k -2\tilde{\zeta}_c \tilde{C}^{ij} D_t \tilde{C}^{ij}- \zeta'_c C_k{}^k D_t C_k{}^k-\zeta_{\rm n} v^{ij} q_{ij}\nonumber\\
&-(\tilde{\zeta}_{c\rm n} q^{kl} C_{kl} g^{ij} + \tilde{\zeta}'_{c\rm n} C_k{}^k q^{ij}  ) D_t C_{ij}\nonumber\\
&-\chi_{\rm p} \epsilon^{ij}  p_j \omega_{in} +\lambda p_i h_d^i- \kappa_{\rm p}^{\alpha} p^i\partial_i \bar{\mu}_d^{\alpha}+\Lambda\Delta\mu\nonumber\\
r_{\rm UD}&=-\zeta' C_k{}^k v_k{}^k-2\tilde{\zeta}\tilde{C}^{ij}v_{ij}-\zeta_c D_t C_k{}^k \nonumber\\
&-(\tilde{\zeta}_{\rm n} q^{kl} C_{kl} g^{ij} +\tilde{\zeta}_{\rm n}' q^{ij} C_k{}^k  )v_{ij} -\zeta_{\rm cn} q^{ij} D_t C_{ij}  \nonumber\\
&-(\overline{\chi}_{\rm p} \epsilon^i{}_k C^{kj}+\overline{\chi}'_{\rm p}\epsilon^j{}_k C^{ki} )p_j \omega_{in}\nonumber\\
&+(\lambda_{\rm UD} C^{ij}  +\lambda'_{\rm UD} C_k{}^k g^{ij}) p_j h_{d,i}\nonumber\\
&-( \kappa_{\rm UDp}^{\alpha}C^{ij}+\kappa_{\rm UDp}'^{\alpha} C_{k}{}^k g^{ij})p_j\partial_i \bar{\mu}_d^{\alpha}\nonumber\\
r_{\rm C}=&-2\zeta_{\rm C} \epsilon_{ik}C^{kj} v^i{}_j-( 2 \zeta_{\rm Cn} \epsilon^{ik} q_{kl} C_l{}^j   + 2 \tilde{\zeta}_{\rm Cn} \epsilon^{ik} C_k{}^l q_l{}^j  )v_{ij} \nonumber\\
&-  \zeta_{\rm cCn}\epsilon^{ik} q_k{}^j    D_t C_{ij}\nonumber\\
&-(\chi_{\rm Cp}C^{ij}+\chi_{\rm Cp}' C_{k}{}^k g^{ij}  )p_j \omega_{in} \nonumber\\
&+(\lambda_{\rm C}\epsilon^{i}{}_{k} C^{kj} +\lambda_{\rm C}' \epsilon^{j}{}_{k} C^{ki}) p_j h_{d,i}\nonumber\\
&-(\kappa_{\rm Cp}^{\alpha} \epsilon^i{}_k C^{kj} +{\kappa'}_{\rm Cp}^{\alpha}\epsilon^j{}_k C^{ki}  )p_j\partial_i \bar{\mu}_d^{\alpha}\nonumber\\
r_{\rm PC}=& -2\zeta_{\rm PC} \epsilon_{ik} C^{kj} D_t C^i{}_j- \zeta_{\rm PC n} \epsilon^{ik} q_k{}^j v_{ij}\nonumber\\
&-(2 \zeta_{\rm cPCn} \epsilon^{ik} q_k{}^l C_l{}^j  +2 \tilde{\zeta}_{\rm cPCn} \epsilon^{ik} C_k{}^l q_l{}^j )D_t C_{ij}\nonumber\\
&- \chi_{\rm PCp} p^i \omega_{in}+ \lambda_{\rm PC}\epsilon^{ij} p_j h_{d,i}-\kappa_{\rm PCp}^{\alpha} \epsilon^{ij} p_j\partial_i \bar{\mu}_d^{\alpha}
\label{Appendix:ConstitutiveEquationr}.
\end{align}
The odd or Hall viscosities $\eta_{\rm PC}$, $\eta_{c{\rm PC}}$, $\kappa_{\rm PC}$, the transport coefficients $L_{\rm PC}^{\alpha\beta}$ and the inverse rotational viscosity $1/\gamma_{\rm PC}$ are odd under time-reversal; they can be non-zero for a system close to equilibrium for instance in the presence of a magnetic field \cite{avron1998odd}. They can emerge more generally as a linear response coefficient around a non-equilibrium steady-state \cite{banerjee2017odd}.

\subsection{Pseudopolar surface}
For a pseudopolar surface described by the order parameter $\mathbf{p}_{\rm C}$ or by the combination $\epsilon_{\rm PC}$, $\mathbf{p}_{\rm C}$, the constitutive equations read:
\begin{align}
\overline{t}_d^{ij}=&2\eta \tilde{v}^{ij} +\eta_b v_k{}^k g^{ij}+\zeta \Delta \mu g^{ij} +\nu' h_{{\rm C},d}^k p_{{\rm C}, k} g^{ij}\nonumber\\
&+\zeta_{\rm n} \Delta \mu q^{ij}+\frac{\nu}{2}(p_{\rm C}^i h_{{\rm C},d}^j+p_{\rm C}^j h_{{\rm C},d}^i-h_{{\rm C},d}^k p_{{\rm C}, k} g^{ij}) \nonumber\\
&+\epsilon_{\rm PC}\left[\eta_{\rm{PC}} \left(\epsilon^i{}_k v^{kj}+\epsilon^j{}_k v^{ki}\right)+\zeta_{\rm PC{\rm n}}\Delta \mu \epsilon^{ik} q_k{}^j \right.\nonumber\\
&\left.+\frac{\nu_{{\rm PC}}}{2}(\epsilon^{i}{}_{k}h^k_{{\rm C},d} p_{\rm C}^j+\epsilon^{j}{}_{k} h_{{\rm C},d}^k p_{\rm C}^i- \epsilon_{kl}h_{{\rm C},d}^l  p_{\rm C}^k  g^{ij}) \right.\nonumber\\
&\left.+ \nu_{\rm PC}' \epsilon_{kl} p_{\rm C}^k h_{{\rm C},d}^l g^{ij}\right]\\
\overline{m}_d^{ij}=&2 \eta_c D_t \tilde{C}^{ij}+  \eta_{cb} D_t C_{k}{}^{k} g^{ij}\nonumber\\
&+2 \tilde{\zeta_c} \Delta \mu  \tilde{C}^{ij}+\zeta_c'  \Delta\mu C_k{}^k g^{ij}\nonumber\\
&+\tilde{\zeta}_{c{\rm n}}  \Delta \mu  q^{kl} C_{kl} g^{ij}+\tilde{\zeta'_{c{\rm n}}} \Delta \mu q^{ij}C_{k}{}^k   \nonumber\\
&+\epsilon_{\rm PC}\left[\eta_{\rm{cPC}} \left(\epsilon^{ik} D_t C_{k}{}^{j}+\epsilon^{jk} D_t C_{k}{}^{i}\right)\right.\nonumber\\
&\left.+\zeta_{\rm{PC}}\Delta \mu \left(\epsilon^i{}_k C^{kj}+\epsilon^j{}_k C^{ki}\right)\right.\nonumber\\
&\left.+\zeta_{\rm{cPC n}}\Delta\mu(\epsilon^{ik} q_{k}{}^l C_l{}^j+\epsilon^{jk} q_k{}^l C_l{}^i)\right.\nonumber\\
&\left.+\tilde{\zeta}_{\rm{cPC n}}\Delta\mu(\epsilon^{ik} C_{k}{}^l q_l{}^j+\epsilon^{jk} C_k{}^{l} q_l{}^i) \right]\\
m^i_{n,d}&=\kappa \omega^i_n+(\chi_{\rm{C}p} C^{i}{}_{j}  +\chi_{\rm{C}p}' C_k{}^k g^i{}_{j})p_{\rm C}^j  \Delta \mu \nonumber\\
&+\epsilon_{\rm PC}(\kappa_{\rm{PC}} \epsilon^{ij}\omega_{jn} +(\overline{\chi}_{\rm p} \epsilon^{ik} C_{kj} +\overline{\chi}'_{\rm p}  \epsilon_{jk} C^{ki} ) p_{\rm C}^j \Delta \mu) \\
D_t p_{\rm C}^i =&\frac{1}{\gamma} h_{{\rm C},d}^i-\nu \tilde{v}^{ij} p_{{\rm C},j}-\nu' v_k{}^k p_{\rm C}^i+\lambda \Delta\mu p_{\rm C}^i\nonumber\\
&+\epsilon_{\rm PC}\left[\frac{1}{\gamma_{\rm PC}}\epsilon^{i}{}_{j} h_{{\rm C},d}^j+\nu_{\rm{PC}} \epsilon^{ik} \tilde{v}_{kj}p_{\rm C}^j\right.\nonumber\\
&\left.+\nu_{\rm{PC}}'v_k{}^k \epsilon^{ij} p_{{\rm C},j}+\lambda_{\rm{PC}}\Delta\mu \epsilon^{i}{}_{j} p_{\rm C}^j \right]\\
j^{\alpha, i}=&-L^{\alpha\beta} \partial^i \bar{\mu}_d^{\beta}+(\kappa^{\alpha}_{\rm Cp} \epsilon^{ik} C_{kj} +\kappa_{\rm Cp}'^{\alpha}   \epsilon_{jk} C^{ki} ) p_{\rm C}^j \Delta \mu  \nonumber\\
&-\epsilon_{\rm PC} L_{\rm PC}^{\alpha\beta} \epsilon^{ij} \partial_j \bar{\mu}_d^{\beta}\nonumber\\
&+\epsilon_{\rm PC}(\kappa^{\alpha}_{\rm{UD}p} C^{i}{}_{j}  +\kappa_{\rm{UD}p}'^{\alpha} C_k{}^k g^{i}{}_{j})p_{\rm C}^j  \Delta \mu\\
r=&-\zeta v_k{}^k-\zeta_{\rm n} v^{ij} q_{ij}-2\tilde{\zeta}_c \tilde{C}^{ij} D_t \tilde{C}^{ij}- \zeta'_c C_k{}^k D_t C_k{}^k\nonumber\\
&-(\tilde{\zeta}_{c{\rm n}} q^{kl} C_{kl} g^{ij} + \tilde{\zeta}'_{c\rm n} C_k{}^k q^{ij}  ) D_t C_{ij}\nonumber\\
&-(\chi_{\rm Cp}C^{ij}+\chi_{\rm Cp}' C_{k}{}^k g^{i}{}_{j}  )p_{\rm C}^j \omega_{in}+\Lambda\Delta\mu \nonumber\\
&+\lambda p_{{\rm C},i} h_{{\rm C},d}^i
-(\kappa_{\rm Cp}^{\alpha} \epsilon^{ik} C_{kj} 
+\kappa_{\rm Cp}'^{\alpha}\epsilon_{jk} C^{ki}  )p_{\rm C}^j\partial_i \bar{\mu}_d^{\alpha}\nonumber\\
&- \epsilon_{\rm PC}\left[-\lambda_{\rm PC}\epsilon_{ij} p_{\rm C}^j h_{{\rm C},d}^{i}  +2\zeta_{\rm PC} \epsilon_{ik} C^{kj} D_t C^i{}_j\right.\nonumber\\
&\left.+ \zeta_{\rm PC n} \epsilon^{ik} q_k{}^j v_{ij} +(\overline{\chi}_{\rm p} \epsilon^i{}_k C^{kj} +\overline{\chi}'_{\rm p}\epsilon^j{}_k C^{ki} )p_j\omega_{in}\right.\nonumber\\
&\left.+( \kappa^{\alpha}_{\rm UDp}C^{ij}+\kappa_{\rm UDp}'^{\alpha} C_{k}{}^k g^{ij})p_{{\rm C},j}\partial_i \bar{\mu}_d^{\alpha}\right.\nonumber\\
&\left. +(2 \zeta_{c\rm PCn} \epsilon^{ik} q_k{}^l C_l{}^j  +2 \tilde{\zeta}_{c\rm PCn} \epsilon^{ik} C_k{}^l q_l{}^j )D_t C_{ij} \right]
\end{align}

\subsection{Nematic surface}

The contributions to the tension tensor read:
\begin{align}
\overline{t}^{ij}_0=&2\eta \tilde{v}^{ij} +\eta_b v_k{}^k g^{ij}+\zeta \Delta \mu g^{ij} +\nu H_d^{ij}\nonumber\\
& +\nu' Q_{kl} H_d^{kl} g
^{ij}+\zeta_{\rm n} \Delta \mu Q^{ij}\nonumber\\[.2cm]
\overline{t}^{ij}_{\rm{UD}}=&2 \bar{\eta}D_t \tilde{C}^{ij}+\bar{\eta}_b (D_t C_k{}^k) g^{ij}+2 \tilde{\zeta} \Delta \mu\tilde{C}^{ij}\nonumber\\
&+\zeta' \Delta\mu C_k{}^k g^{ij} +\tilde{\zeta_{\rm n}} \Delta \mu Q^{kl} C_{kl} g^{ij}   +\tilde{\zeta'_{\rm n}} \Delta \mu  Q^{ij} C_{k}{}^k 
\nonumber \\[.2cm]
\overline{t}^{ij}_{\rm{C}}=&\eta_{\rm{C}} \left(\epsilon^{ik} D_t C_{k}{}^{j}+\epsilon^{jk} D_t C_{k}{}^{i}\right)\nonumber\\
&+\zeta_{\rm{C}}\Delta\mu  \left(\epsilon^i{}_k C^{kj}+\epsilon^j{}_k C^{ki}\right)\nonumber\\
&+\zeta_{\rm{C n}}\Delta\mu(\epsilon^{ik} Q_{k}{}^l C_l{}^j+\epsilon^{jk} Q_{k}{}^{l} C_l{}^i)\nonumber\\
&+\tilde{\zeta}_{\rm{C n}}\Delta\mu(\epsilon^{ik} C_{k}{}^l Q_l{}^j+\epsilon^{jk} C_k{}^l Q_l{}^i)  \nonumber \\[.2cm]
\overline{t}^{ij}_{\rm{PC}}=&\eta_{\rm{PC}} \left(\epsilon^i{}_k v^{kj}+\epsilon^j{}_k v^{ki}\right)+  \nu_{\rm PC} \epsilon^{i}{}_{k} H_{d}^{kj}\nonumber\\
&+\nu_{\rm PC}' \epsilon^{kl} Q_{km}H_d^{ml} g^{ij}+\zeta_{\rm PC{\rm n}}\Delta \mu  \epsilon^{ik} Q_k{}^j .
	\label{Appendix:ConstitutiveEquationtij_nematic}
\end{align}

The contributions to the moment tensor, in the decomposition indicated in Eq. \ref{GeneralTensorDecompositionSymmetry}, read
\begin{align}
\overline{m}^{ij}_0=&2 \eta_c D_t \tilde{C}^{ij}+  \eta_{cb} D_t C_{k}{}^{k} g^{ij}+2 \tilde{\zeta_c} \Delta \mu  \tilde{C}^{ij}+\zeta_c' \Delta\mu C_k{}^k g^{ij} \nonumber\\
&+\tilde{\zeta}_{c{\rm n}} \Delta \mu Q^{kl} C_{kl} g^{ij}  +\tilde{\zeta'_{c{\rm n}}} \Delta \mu  Q^{ij}C_{k}{}^k  
 \nonumber\\
\overline{m}^{ij}_{\rm{UD}}=&2\bar{\eta} \tilde{v}^{ij}+\bar{\eta}_b v_k{}^k g^{ij}+\beta H_d^{ij}+\beta' Q_{kl} H_d^{kl} g
^{ij}\nonumber\\
&+ \zeta_c\Delta \mu  g^{ij} +\zeta_{c{\rm n}} \Delta \mu Q^{ij}\nonumber\\
\overline{m}^{ij}_{\rm{C}}=&-\eta_{\rm{C}} \left(\epsilon^i{}_k v^{kj}+\epsilon^j{}_k v^{ki}\right)+\beta_{\rm C}  \epsilon^i{}_k H_d^{kj}\nonumber\\
&
+\beta_{\rm C}' \epsilon^{kl} Q_{km}H_d^{ml} g^{ij}+\zeta_{c{\rm C}{\rm n}}\Delta\mu  \epsilon^{ik} Q_k{}^j
\nonumber\\
\overline{m}^{ij}_{\rm{PC}}=&\eta_{\rm{cPC}} \left(\epsilon^{ik} D_t C_{k}{}^{j}+\epsilon^{jk} D_t C_{k}{}^{i}\right)\nonumber\\
&+\zeta_{\rm{PC}}\Delta \mu \left(\epsilon^i{}_k C^{kj}+\epsilon^j{}_k C^{ki}\right)\nonumber\\
&+\zeta_{c\rm{PC n}}\Delta\mu(\epsilon^{ik} Q_{k}{}^l C_l{}^j+\epsilon^{jk} Q_k{}^l C_l{}^i)\nonumber\\
&+\tilde{\zeta}_{c\rm{PC n}}\Delta\mu(\epsilon^{ik} C_{k}{}^l Q_l{}^j+\epsilon^{jk} C_k{}^{l} Q_l{}^i) ~.
	\label{Appendix:ConstitutiveEquationmij_nematic}
\end{align}
Contributions to the tensor $m_n$ read:
\begin{align}
m^i_{n0}&=\kappa \omega_{in} \nonumber\\
m^i_{n\rm{UD}}&=0  \nonumber\\
m^i_{n\rm{C}}&=0 \nonumber \\
m^i_{n\rm{PC}}&=\kappa_{\rm{PC}} \epsilon^{ij}\omega_{jn}
\label{Appendix:ConstitutiveEquationmn}
\end{align}
and the corotational time derivative of the nematic order parameter is:
\begin{align}
	\mathcal{Q}_0^{ij}=&\frac{1}{\gamma}H_d^{ij}-\nu \tilde{v}^{ij}	-\nu' v_k{}^k Q^{ij}+\lambda \Delta\mu \,Q^{ij}\nonumber\\
	\mathcal{Q}_{\rm  UD}^{ij}=&
	-\beta  D_t \tilde{C}^{ij}-\beta'  D_t C_{k}{}^{k} Q^{ij}\nonumber\\&
	+\lambda_{\rm UD}\Delta\mu \,\tilde{C}^{ij}
	+\lambda_{\rm UD}'\Delta\mu C_k{}^k Q^{ij}	
	\nonumber\\[.2cm]
	\mathcal{Q}_{\rm  C}^{ij}=& \beta_{\rm C}\epsilon^i{}_k D_t \tilde{C}^{kj}+\beta_{\rm C}' \epsilon^{i}{}_{k}Q^{kj} D_t C_{l}{}^l\nonumber\\&
	+\lambda_{\rm C}\Delta\mu (\epsilon^i{}_k C^{kj}+\epsilon^j{}_k C^{ki})
	+\lambda_{\rm{C n}}\Delta\mu C_k{}^k \epsilon^{ik} Q_k{}^j\nonumber\\
	\mathcal{Q}_{\rm  PC}^{ij}=&\frac{1}{\gamma_{\rm PC}} \epsilon^{i}{}_{k} H_d^{kj}+\nu_{\rm PC} \epsilon^{ik} \tilde{v}_k{}^j	\nonumber\\
	&+ \nu'_{{\rm PC}}\epsilon^{i}{}_{k}Q^{kj} v_{l}{}^l +\lambda_{\rm PC}\Delta\mu \epsilon^{ik} Q_{k}{}^{j}\label{Appendix:constitutiveequation_dQdt}~.
\end{align}
The flux of species $\alpha=2...N$, relative to the centre of mass has different contributions given by
\begin{align}
j^{\alpha, i}_{0}=&-L^{\alpha\beta} \partial^i \bar{\mu}_d^{\beta} \nonumber\\
j^{\alpha, i}_{\rm{UD}}=&0\nonumber \\
j^{\alpha,  i}_{\rm{C}}=&0 \nonumber\\
j^{\alpha, i}_{\rm PC}=&-L_{\rm PC}^{\alpha\beta} \epsilon^{ij} \partial_j \bar{\mu}_d^{\beta}
~.
\label{Appendix:ConstitutiveEquationfluxj}
\end{align}
Finally, the rate of fuel consumption has the following decomposition:
\begin{align}
r_0=&-\zeta v_{k}{}^k -2\tilde{\zeta}_c \tilde{C}^{ij} D_t \tilde{C}^{ij}- \zeta'_c C_k{}^k D_t C_k{}^k-\zeta_{\rm n} v^{ij} Q_{ij}\nonumber\\
&-(\tilde{\zeta}_{c{\rm n}} Q^{kl} C_{kl} g^{ij} + \tilde{\zeta}'_{\rm cn} C_k{}^k Q^{ij}  )D_t C_{ij}+\lambda Q_{ij} H_d^{ij}+\Lambda\Delta\mu\nonumber\\
r_{\rm UD}=&-\zeta' C_k{}^k v_k{}^k-2\tilde{\zeta}\tilde{C}^{ij}v_{ij}-\zeta_c D_t C_k{}^k\nonumber\\&
-(\tilde{\zeta}_{\rm n} Q^{kl} C_{kl} g^{ij} +\tilde{\zeta}_{\rm n}' Q^{ij} C_k{}^k  )v_{ij} -\zeta_{c{\rm n}} Q^{ij} D_t C_{ij} \nonumber\\
&+(\lambda_{\rm UD} \tilde{C}^{ij}  +\lambda'_{\rm UD} C_k{}^k Q^{ij}   ) {H_d}_{ij}\nonumber\\
r_{\rm C}=&-2\zeta_{\rm C} \epsilon_{ik}C^{kj} v^i{}_j-( 2 \zeta_{\rm Cn} \epsilon^{ik} Q_{kl} C_l{}^j   + 2 \tilde{\zeta}_{\rm Cn} \epsilon^{ik} C_k{}^l Q_l{}^j  )v_{ij} \nonumber\\
&-   \zeta_{\rm cCn}\epsilon^{ik} Q_k{}^j   D_t C_{ij}+2 \lambda_{\rm C} \epsilon_{jk} C^{k}{}_{i}  H_d^{ij}\nonumber\\
&+ \lambda_{\rm{C n}}C_k{}^k \epsilon^{ik} Q_k{}^j{H_d}_{ij} \nonumber\\
r_{\rm PC}=& -2\zeta_{\rm PC} \epsilon_{ik} C^{kj} D_t C^i{}_j- \zeta_{\rm PC n} \epsilon^{ik} Q_k{}^j v_{ij}\nonumber\\
& -(2 \zeta_{c\rm PCn} \epsilon^{ik} Q_k{}^l C_l{}^j  +2 \tilde{\zeta}_{c\rm PCn} \epsilon^{ik} C_k{}^l Q_l{}^j ) D_t C_{ij}\nonumber\\
&+\lambda_{\rm PC} \epsilon_{ik} Q^{k}{}_{j} H_d^{ij}.
\end{align}
As for the polar case, the odd or Hall viscosities $\eta_{\rm PC}$, $\eta_{c{\rm PC}}$, $\kappa_{\rm PC}$, the transport coefficients $L_{\rm PC}^{\alpha\beta}$ and the inverse rotational viscosity $1/\gamma_{\rm PC}$ are odd under time-reversal, and can be non-zero in a system close to equilibrium for instance in the presence of a magnetic field.

\subsection{Pseudonematic surface}

For pseudonematic surfaces whose broken symmetries can be described by the pseudotensor $\mathbf{Q}_{\rm UD}$ and the combination of the pseudoscalar and pseudovector $\{\epsilon_{\rm PC},\mathbf{Q}_{\rm UD} \}$, we find:
\begin{align}
\overline{t}^{ij}_d=&2\eta \tilde{v}^{ij} +\eta_b v_k{}^k g^{ij}+\zeta g^{ij} \Delta \mu\nonumber\\
&+\nu' Q_{{\rm UD},kl} H_{{\rm UD}d}^{kl} g^{ij}+\tilde{\zeta}_{\rm n}  \Delta \mu  Q_{\rm UD}^{kl} C_{kl} g^{ij} \nonumber\\
&+\tilde{\zeta}'_{\rm n} \Delta \mu Q_{\rm UD}^{ij} C_{k}{}^k +\epsilon_{\rm PC}\left[\eta_{\rm{PC}} \left(\epsilon^i{}_k v^{kj}+\epsilon^j{}_k v^{ki}\right)\right.\nonumber\\
&\left.+\nu_{\rm PC}' \epsilon^{kl} Q_{{\rm UD},km}H_{{\rm UD}d}^{ml} g^{ij}\right. \nonumber\\
&\left.+\zeta_{\rm{C n}}\Delta\mu(\epsilon^{i}{}_{k} Q_{\rm UD}^{kl} C_l{}^j+\epsilon^{j}{}_{k} Q_{\rm UD}^{kl} C_l{}^i)\right.\nonumber\\
&\left.+\tilde{\zeta}_{\rm{C n}}\Delta\mu(\epsilon^{ik} C_{kl} Q_{\rm UD}^{lj}+\epsilon^{jk} C_{kl}Q_{\rm UD}^{li}) \right]\\
\overline{m}^{ij}_d=&2 \eta_c D_t \tilde{C}^{ij}+  \eta_{cb} D_t C_{k}{}^{k} g^{ij}+2 \tilde{\zeta_c} \Delta \mu \tilde{C}^{ij}\nonumber\\
&+\zeta_c' \Delta\mu C_k{}^k g^{ij} +\beta H_{{\rm UD}d}^{ij}+\zeta_{c{\rm n}} \Delta \mu Q_{\rm UD}^{ij} \nonumber\\
&+\epsilon_{\rm PC}\left[\eta_{\rm{cPC}} \left(\epsilon^{ik} D_t C_{k}{}^{j}+\epsilon^{jk} D_t C_{k}{}^{i}\right)\right.\nonumber\\
&\left.+\zeta_{\rm{PC}\Delta \mu \left(\epsilon^i{}_k C^{kj}+\epsilon^j{}_k C^{ki}\right)}\right.\nonumber\\
&\left.+\beta_{\rm C}  \epsilon^i{}_k H_{{\rm UD}d}^{kj} +\zeta_{c{\rm C}{\rm n}}\Delta\mu \epsilon^{i}{}_{k} Q_{\rm UD}^{kj}\right]\\
m_{n,d}^i=&\kappa \omega_n^i + \epsilon_{\rm PC} \kappa_{\rm PC} \epsilon^{ij} \omega_{jn}\\
D_t Q^{ij}_{\rm UD}=&\frac{1}{\gamma}H_{{\rm UD}d}^{ij}+\lambda \Delta\mu Q_{\rm UD}^{ij}	+\lambda_{\rm UD}\Delta\mu \,\tilde{C}^{ij}\nonumber\\
&	-\nu' v_k{}^k Q_{\rm UD}^{ij}-\beta  D_t \tilde{C}^{ij}\nonumber\\
	&
	+\epsilon_{\rm PC}\left[ \frac{1}{\gamma_{\rm PC}} \epsilon^{i}{}_{k} H_{{\rm UD}d}^{kj}+\nu'_{{\rm PC}}\epsilon^{i}{}_{k}Q_{\rm UD}^{kj} v_{l}{}^l\right.\nonumber\\
	&\left.+\frac{1}{2} \beta_{\rm C}\,\left(\epsilon^i{}_k D_t C^{kj}+\epsilon^j{}_k D_t C^{ki}\right) \right.\nonumber\\
	&\left.+	\lambda_{\rm C}\Delta\mu (\epsilon^i{}_k C^{kj}+\epsilon^j{}_k C^{ki})+\lambda_{\rm PC}\Delta\mu \epsilon^{i}{}_{k} Q_{\rm UD}^{kj}\right]\\
	j^{\alpha, i}_{0}=&-L^{\alpha\beta} \partial^i \bar{\mu}_d^{\beta} -\epsilon_{\rm PC}L_{\rm PC}^{\alpha\beta} \epsilon^{ij} \partial_j \bar{\mu}_d^{\beta}
~\\
r=&\Lambda\Delta\mu-\zeta v_{k}{}^k -2\tilde{\zeta}_c \tilde{C}^{ij} D_t \tilde{C}^{ij}- \zeta'_c C_k{}^k D_t C_k{}^k\nonumber\\
&+\lambda Q_{\rm UD}^{ij} H_{{\rm UD}d,ij}+\lambda_{\rm UD} \tilde{C}_{ij}      H_{{\rm UD}d}^{ij}-\zeta_{c\rm n} Q_{\rm UD}^{ij} D_t C_{ij}   \nonumber\\
& -(\tilde{\zeta}_{\rm n} Q_{\rm UD}^{kl} C_{kl} g^{ij} +\tilde{\zeta}_{\rm n}' Q_{\rm UD}^{ij} C_k{}^k  )v_{ij}\nonumber\\
&+\epsilon_{\rm PC}\left[-2\zeta_{\rm PC} \epsilon_{ik} C^{kj} D_t C^i{}_j\right.\nonumber\\
&\left.-( 2 \zeta_{\rm Cn} \epsilon^{i}{}_{k} Q_{\rm UD}^{kl} C_{l}{}^{j}   + 2 \tilde{\zeta}_{\rm Cn} \epsilon^{ik} C_{kl} Q_{\rm UD}^{lj}  )v_{ij}\right.\nonumber\\
&\left.-   \zeta_{\rm cCn}\epsilon^{i}{}_{k} Q_{\rm UD}^{kj}   D_t C_{ij} +2 \lambda_{\rm C} \epsilon^{j}{}_{k} C^{ki} H_{{\rm UD}d,ij}\right.\nonumber\\
&\left.+\lambda_{\rm PC} \epsilon_{ik} {Q_{\rm UD}}^{k}{}_{j} H_{{\rm UD}d}^{ij}\right]~.
\end{align}

\section{Glossary of physical realisations of surface symmetries}
\label{appendix:surface_glossary}

In this Appendix we give examples of physical realisations of surfaces with symmetries  that are discussed in the main text (see Tables \ref{isotropic_surfaces_table}, \ref{polar_surfaces_table}, \ref{nematic_surfaces_table} and Fig. \ref{fig:symmetries}). We discuss mostly examples of biological membranes bilayers and epithelia.
\begin{itemize}
\item $D_{\rm \infty h}^{\perp}$, {\bf isotropic}:
Corresponds to the symmetry of pure phospholipid membrane bilayers used for instance in giant unilamellar vesicles \cite{walde2010giant}.  Another example is given by free standing smectic $A$ films \cite{degennesprost}. 

\item $C_{\rm \infty v}^{\perp}$, {\bf isotropic, up-down broken symmetry}:
 Corresponds to the symmetry of a phospholipid bilayer in which the two leaflets have a different composition.  An alternative is provided by a conventional phospholipid bilayer loaded with ion pumps \cite{manneville1999activity}, oriented in a preferred direction.
 
\item $D_{\infty}^{\perp}$, {\bf isotropic, chiral broken symmetry}: 
Corresponds to the symmetry of a phospholipid-cholesterol membrane bilayer \cite{veatch2003separation}, a smectic $A$ free standing film made of chiral molecules, or a free standing cell monolayer film with no apico-basal polarity (since cells are chiral \cite{inaki2016cell}). 

\item $C_{\rm \infty h}^{\perp}$, {\bf isotropic, planar- chiral broken symmetry}: 
Such a symmetry may be obtained with spin polarised molecules or magnetically polarised nanoparticles, with magnetic polarisation along the membrane normal, embedded in a conventional phospholipid membrane.  An alternative way to obtain such a symmetry would be to have an isotropic bilayer, one leaflet with an excess of dextrorotatory chiral molecules, the other with an excess of levorotatory ones, in exactly the opposite concentration. An out of equilibrium version would be obtained by a membrane composed of rotary motors with rotation axis parallel to the membrane normal and with up-down symmetry, embedded in a conventional phospholipid membrane.  Note that motors such as F$_0$F$_1$-ATPase do not have up-down symmetry  \cite{noji1997direct}.  

\item $C_{\infty}^{\perp}$, {\bf isotropic, chiral and up-down broken symmetries}:  
Most biological membranes correspond to this symmetry group. They contain chiral molecules such as cholesterol, oriented ion pumps, different leaflet composition, and are in-plane isotropic.

\item $C_{2 \rm  v}^{||}$, {\bf polar}: 
A simple example is that of a membrane with polar order parallel to the membrane.  We do not know of such a system made of small molecules, neither do we know of more complex realisations. 
		
\item  $C_{\rm s}^{||}$, {\bf polar, up-down broken symmetry}: 
This corresponds to a polar phase with polarity parallel to the membrane with broken up-down symmetry.  The simplest example is that of a tilted phospholipid monolayer.  Such phases are known in Langmuir films \cite{ adamson1967physical, kaganer1999structure}.  An example at a cell level would be a membrane, on one side of which actin polymerization takes place with a tilt with respect to the membrane normal.  The composite membrane-cortex layer has the required symmetry if we neglect actin chirality.  At the tissue scale this phase corresponds to any polar epithelial tissue, either {\it in vivo} or {\it in vitro} on a substrate, the polarity having components both along and perpendicular to the membrane, provided we again neglect the weak chirality of cells. Such tissues have a spontaneous tendency to move homogeneously with respect to their substrate \cite{jain2020role}.
		
\item $C_{2}^{||}$, {\bf polar, chiral broken symmetry}: 
This phase typically corresponds to a free standing chiral smectic $C$ film \cite{degennesprost} or a phospholipid bilayer made of chiral molecules that are tilted with respect to the normal. Such phases have not been described in biological membranes, but could exist.

\item  $C_{\rm s}^{\perp}$, {\bf polar, planar-chiral broken symmetry}: 		
A possible picture is that of a polar membrane, with polarity parallel to the membrane, embedding rotatory motors with up-down symmetry and rotation axis parallel to the membrane normal.

\item  $C_{1}$, {\bf polar, chiral and up-down broken symmetry}: 
Adding chirality to any of the physical realizations of $C_s^{||}$ membranes suppresses mirror planes, and leads to membrane which are not invariant under any point group symmetry operation. This is, for example, the case of tilted chiral phospholipid bilayers with inequivalent leaflets, as well as epithelium with in-plane polar symmetry and apico-basal polarity, since they lack up-down symmetry and cells are chiral. On the cell surface, the actomyosin cortex, with polar and chiral actin filaments polymerizing away from the cell membrane, enters this category \cite{salbreux2012actin}.
		
\item  $C_{2\rm h}^{||}$, {\bf pseudopolar}:  
The simplest physical picture is that of a free standing smectic $C$ film \cite{degennesprost}, or a phospholipid bilayer membrane in which the molecules are tilted with respect to the membrane normal.  An alternative out-of-equilibrium version would correspond to a similar membrane in which rotatory motors with up-down symmetry and rotation axis parallel to the membrane are embedded in the membrane.
 
 \item  $C_i=S_2$, {\bf pseudopolar and planar-chiral broken symmetry}:
This phase corresponds to a $C_{2\rm h}^{||}$, pseudopolar surface, with an additional broken symmetry. We are not aware of an experimental realisation of such a surface.

\item $D_{\rm 2h}^{\perp}$, $D_{2h}^{||}$, {\bf nematic}: 
This phase corresponds to a membrane with in-plane nematic order. To our knowledge there is no simple example with phospholipids or smectic free standing films. Thin spherical nematic shells obtained with a nematic drop containing a smaller aqueous drop would enter this category \cite{fernandez2007novel, lopez2011frustrated}.

\item $C_{\rm 2v}^{\perp}$, {\bf nematic, up-down broken symmetry}: 
A simple representation is of a membrane with nematic order parallel to the membrane and with no up-down symmetry.  Such a membrane type,  not reported to our knowledge in systems made of small molecules, corresponds to many epithelial layers {\it in vivo} or {\it in vitro} when they rest on a substrate and cells spontaneously elongate, whenever cell chirality can be neglected \cite{duclos2017topological, duclos2018spontaneous}.

\item $D_{2}^{\perp}$, $D_{2}^{||}$, {\bf nematic, chiral broken symmetry}:  
  Corresponds to a membrane made of chiral molecules with in-plane nematic order.  We do not know of any example with phospholipids or free standing smectic films.    
  
\item  $C_{\rm 2h}^{\perp}$, {\bf nematic, planar-chiral broken symmetry}:  
A  simple realization would be obtained by dissolving planar pinwheel-shaped molecules in an otherwise nematic membrane with nematic order parallel to the membrane.  We know of no such realization.

 \item $C_{2}^{\perp}$, {\bf nematic, chiral and up-down broken symmetries}:  
This corresponds to an in-plane nematic order, with chiral molecules and broken up-down symmetry.  We do not know of such phases made of phospholipids. 
Chiral microtubule filaments with no coarse-grained polarity but orientational order coated on a lipid vesicle \cite{keber2014topology}, or nematic epithelial layers on substrate or {\it in vivo} provide good examples of such phases \cite{duclos2017topological, duclos2018spontaneous}. 
 
\item $D_{\rm 2d}^{\perp}$, {\bf pseudonematic}:  
Corresponds to a bilayer membrane in which a nematic order parallel to the membrane exists in each monolayer, but with their main axis being orthogonal to each other.  We do not know of any example of such a phase made of small molecules. The supracellular actin fiber system in \emph{Hydra} could provide a beautiful example of such a symmetry, if we ignore chirality and assume that the orthogonal actin network fibers have an equal degree of organisation \cite{Maroudas-Sacks:2021aa}.

\item $S_4^{\perp}$, {\bf pseudonematic and planar-chiral broken symmetry:}
We are not aware of an experimental realisation of such a surface.
\end{itemize}

\section{Quasi one-dimensional confined active film}
\label{appendix:quasi_1D_active_film}

Here we describe calculations for the steady-state solutions of a confined, quasi-dimensional active film. We assume that the system has invariance of translation in the $y$ direction, is confined along the $x$ direction, and is free from external forces and torques. The active film position in space $\mathbf{X}(x,y)$ is denoted using the Monge gauge:
\begin{align}
\mathbf{X}(x,y) = x \tilde{\mathbf{e}}_x + y \tilde{\mathbf{e}}_y + h(x,y) \tilde{\mathbf{e}}_z
\end{align}
with $h$ an height function, and $ \tilde{\mathbf{e}}_x$, $\tilde{\mathbf{e}}_y$, $\tilde{\mathbf{e}}_z$ are unit vectors of a 3D cartesian basis. We assume a weakly deformed surface, $|\partial_x h|\ll 1 $ and $|\partial_y h|\ll1$. To first order in the height function gradients, $g_{ij}\simeq \delta_{ij}$, the Christoffel coefficients vanish and $C_{ij} \simeq -\partial_i \partial_j h$. Because $g_{ij}\simeq \delta_{ij}$, we use indifferently contravariant and covariant tensor components. At low Reynolds number where inertial terms can be neglected, the relevant force balance equations (Eqs. \ref{ForceBalanceTangential}-\ref{TorqueBalanceNormal}) then are, in the weakly deformed limit,
\begin{align}
\label{eq:appendix_UD_tangential_force_balance_1}
\partial_x t_{xx} +C_{xx} \partial_x \bar{m}_{xx}&=0\\
\label{eq:appendix_UD_tangential_force_balance_2}
\partial_x  t_{xy}&=0\\
\label{eq:appendix_UD_normal_force_balance}
\partial_x^2 \bar{m}_{xx}-C_{xx} t_{xx}&=0~.
\end{align}
In both examples discussed below we assume that $t_{xy}(0)=t_{xy}(L)=0$ (no stress tangent to the active film at the boundary), resulting in $t_{xy}=0$.

Using the normal torque balance Eq. \ref{TorqueBalanceNormal}, the total tension tensor can be written:
\begin{align}
\label{eq:appendix_application_tij_sym_antisym}
t^{ij}=& t^{ij}_s - \frac{1}{2} \epsilon^{ij} (\nabla_k m_{n}^k-C_{kl} m^{kl})~.
\end{align}
In this section we consider only equilibrium contributions to $m_n^k$. Using Eq. \ref{eq:application:equilibrium_m}, one finds
\begin{align}
\label{eq:appendix_application_nabla_k_mnk}
\nabla_k m_{n}^k=\nabla_k m_{n,e}^k=\epsilon_{kl}p^k h^l - K \epsilon_{kl} p^k p^j C_{ij} C^{il}~.
\end{align}
Using Eqs. \ref{eq:appendix_application_tij_sym_antisym}-\ref{eq:appendix_application_nabla_k_mnk} and the relation between the symmetric part of the tension tensor $t_s^{ij}$ and $\bar{t}^{ij}$, Eq. \ref{eq:tbardef}; one obtains:
\begin{align}
\label{appendix_application_eq_t_bar_t}
t^{ij} =& \overline{t}^{ij}- \frac{1}{2} \epsilon^{ij}\left[\epsilon_{kl}p^k h^l -K \epsilon^{km} p_k p_l C^{nl} C_{nm}\right] \nonumber\\
&- \frac{1}{2}\left(\overline{m}^{ik}C_k{}^j+\overline{m}^{jk}C_k{}^i-C_{kl} \bar{m}^{km}\epsilon_m{}^l  \epsilon^{ij}\right)~.
\end{align}
To find the total tension tensor (Eqs. \ref{eq:tension_tensor_application_flat_PC} and \ref{eq:application_UD_tij}), the expression of the modified equilibrium tension tensor, which follows from Eqs. \ref{eq:tbardef}, \ref{eq:application:equilibrium_t} and \ref{eq:application:equilibrium_m}, is useful:
\begin{align}
\label{eq:appendix_application_bartije}
\bar{t}^{ij}_e =&\left[\gamma+\frac{\kappa}{2} (C_k{}^k)^2 + \frac{K}{2} (\partial_i \mathbf{p})^2 \right] g^{ij} - K( \nabla^i p^k)( \nabla^j p_k)~.
\end{align}

Using Eq. \ref{eq:application_total_molecular_field}, the tangential components of the total molecular field have explicit expression:
\begin{align}
\label{eq:appendix_application_hi}
h^i = -\lambda_p p^i +K\left(\nabla_j \nabla^j p^i -C^{ij} C_{jk} p^k\right)~,
\end{align}
which gives with the simplifications considered here:
\begin{align}
h_x = -\lambda_p p_x + K \partial_x^2 p_x\label{eq:appendix_application_hx}~,\\
h_y = -\lambda_p p_y + K \partial_x^2 p_y\label{eq:appendix_application_hy}~.
\end{align}

\subsection{Flat planar chiral active film}
\label{appendix:application:PC}
We first discuss a flat planar film confined between two walls, such that $v_x(x=0)=v_x(x=L)=0$. Since we consider an incompressible fluid, one has $\partial_x v_x+\partial_y v_y=0$. Translational invariance in the $y$ direction then implies $\partial_x v_x=0$ which then results in $v_x=0$. 
The normal component of the vorticity $\boldsymbol{\omega}=\frac{1}{2}\boldsymbol{\nabla}\times\mathbf{v}$ (defined in Eq. \ref{eq:expression_vorticity}) is $\omega_n=\partial_x v_y/2$. Using the constitutive equation  tension tensor, Eq. \ref{eq:tension_tensor_application_flat_PC},
one obtains:
\begin{align}
\label{eq:application_PC:txy}
t_{xy} =&2 \eta \omega_n + \frac{\nu}{2}(p_x h_y +p_y h_x) + \frac{\nu_{\rm PC}}{2}(h_y p_y - h_x p_x) \nonumber\\
&+ \zeta_{\rm n} \Delta\mu p_x p_y +\frac{1}{2}\zeta_{{\rm PC n} } \Delta\mu(p_y^2-p_x^2)-\frac{1}{2}\left(p_x h_y- p_y h_x\right)~.
\end{align}

Writing $p_x=\cos\theta$, $p_y=\sin\theta$, the polarity equation \ref{eq:polarity_dynamics_application_flat_PC} results in an equation for the time evolution of the angle $\theta$:
\begin{align}
\label{eq:appendix_application_PC_dttheta}
\partial_t \theta =&\frac{K}{\gamma}\partial_x^2 \theta + \omega_n (1-\nu \cos2\theta -\nu_{\rm PC}\sin2\theta)-\lambda_{\rm{PC}}\Delta\mu~,
\end{align}
and in an expression for the Lagrange multiplier $\lambda$ obtained by imposing $\mathbf{p}\cdot \partial_t \mathbf{p}=0$:
\begin{align}
\label{eq:appendix_application_lambda}
\lambda_p= -K \left(\partial_x\theta\right)^2 -\gamma \omega_n (\nu  \sin2\theta-\nu_{\rm PC}\cos2\theta)~.
\end{align}
Using Eq. \ref{eq:appendix_application_lambda} allows to evaluate the molecular field $\mathbf{h}$, Eqs. \ref{eq:appendix_application_hx}-\ref{eq:appendix_application_hy}, and the tension component $t_{xy}$, Eq. \ref{eq:application_PC:txy}.
Imposing $t_{xy}=0$ one then obtains for $\omega_n$:
\begin{align}
\omega_n=&\frac{\partial_x v_y}{2}=\frac{1}{(4 \eta + \gamma (\nu\sin 2\theta-\nu_{\rm {PC}} \cos2\theta)^2)}\times\nonumber\\
&\bigg(K  \partial_x^2  \theta (1-\nu  \cos2\theta-\nu_{\rm PC}\sin2\theta)\nonumber\\
&- \zeta_{\rm n} \Delta\mu \sin2\theta+\zeta_{{\rm PC n} } \Delta\mu\cos 2\theta\bigg)~,
\end{align}
which can be plugged in Eq. \ref{eq:appendix_application_PC_dttheta} to obtain a dynamic equation for $\theta$, Eq. \ref{eq:application:dttheta:PC}.

In the following we consider situations for which $\nu$ and $\nu_{\rm PC}$ can be neglected, $\nu=\nu_{\rm PC}=0$. In that case the homogeneous steady-state solutions for the angle $\theta=\theta_s$ satisfy:
\begin{align}
\sin(2\theta-\theta_p)=-\frac{4\eta\lambda_{\rm PC} \cos\theta_p}{\zeta_{\rm n}}~,
\end{align}
where $\theta_p$ is defined in Eq. \ref{eq:def_theta_p}. The corresponding steady-state solutions are given in Eq. \ref{eq:PC_solution_theta_s_1}.

To study the stability of homogeneous steady-states, we now linearise Eq. \ref{eq:application:dttheta:PC} around the steady-state solutions, $\theta=\theta_s+\delta\theta$, and obtain:
\begin{align}
\label{eq:application_PC:lineardtheta}
\partial_t \delta\theta = K\left(\frac{1}{\gamma}+\frac{1}{4 \eta }\right)\partial_x^2 \delta \theta - \frac{\zeta_{\rm n} \Delta\mu}{2 \eta \cos\theta_p} \cos(2\theta_s-\theta_p) \delta \theta~.
\end{align}
We note that the steady-state solutions $\theta_s^1$ and $\theta_s^2$ satisfy
\begin{align}
\cos(2\theta_s^1-\theta_p) =-\sqrt{1-\frac{(4\eta\lambda_{\rm PC})^2}{ \zeta_{{\rm PC n}}^2+\zeta_{\rm n}^2}}\\
\cos(2\theta_s^2-\theta_p) =\sqrt{1-\frac{(4\eta\lambda_{\rm PC})^2}{\zeta_{{\rm PC n}}^2+\zeta_{\rm n}^2}}.
\end{align}
such that when these solutions exist, $\cos(2\theta_s^1-\theta_p)\leq 0$ and $\cos(2\theta_s^2-\theta_p)  \geq 0$. Considering Eq. \ref{eq:application_PC:lineardtheta}, this implies that the stability of solutions for $L\rightarrow \infty$ depends on $\zeta_{\rm n}$: $\theta_s^1$ is stable for $\zeta_{\rm n}<0$, $\theta_s^2$ is stable for $\zeta_{\rm n}>0$.

In the regime where $\theta$ is spatially homogeneous and rotates ($L\rightarrow \infty$ and $|\frac{\zeta_{\rm n}}{4 \eta \lambda_{\rm{PC}}\cos\theta_p }|<1$, corresponding to the opposite regime to Eq. \ref{eq:application_PC_threshold_rotation}), integration of Eq. \ref{eq:application:dttheta:PC} shows that the period of rotation is given by
\begin{align}
T=&\left | \int_0^{2\pi} \frac{d\theta}{\lambda_{\rm PC}+\frac{\zeta_{\rm n} \sin(2\theta-\theta_p)}{4\eta\cos\theta_p}} \right|\nonumber\\
=&\frac{2\pi }{|\lambda_{\rm PC}|\sqrt{1-(\frac{\zeta_{\rm n} }{4\eta \lambda_{\rm PC}\cos\theta_p})^2}}~.
\end{align}

In the regime where the polarity is fixed with planar anchoring at the confinement boundaries ($\theta(0)=\theta(L)=\frac{\pi}{2}$), and in the limit $\zeta_{\rm n}=\zeta_{\rm PC n}=\nu=\nu_{\rm PC}=0$, the steady-state profile of $\theta$ is:
\begin{align}
\theta(x)=\frac{\pi}{2}+\frac{2\eta\gamma \lambda_{\rm PC}\Delta\mu}{K(\gamma+4\eta)} x(x-L)~.
\end{align}
The number of full spatial rotations of the polarity can be estimated from $n\simeq 2|\theta(x=L/2)-\theta(x=0)|/(2\pi)$, giving here $n\simeq \frac{\eta\gamma |\lambda_{\rm PC}|\Delta\mu L^2}{2\pi K(\gamma+4\eta)} $.

\subsection{Weakly deformed active film with broken up-down symmetry}
\label{Appendix:weakly_deformed_active_film}

Here we consider a weakly deformed active film whose boundaries are enforced to be straight ($\partial_y v_x(x=0)=\partial_y v_x(x=L)=0$ and $\partial_y v_n(x=0)=\partial_y v_n(x=L)=0$) but whose boundaries position can still freely adjust, without external force $t_{xx}(x=0)=t_{xx}(x=L)=t_{xy}(x=0)=t_{xy}(x=L)=0$ or bending moment ($\bar{m}_{xx}(x=0)=\bar{m}_{xx}(x=L)=0$). The tangential and normal force balance equations \ref{eq:appendix_UD_tangential_force_balance_1} - \ref{eq:appendix_UD_normal_force_balance} then imply $t_{xx}=t_{xy}=\bar{m}_{xx}=0$. The incompressibility condition $v_k{}^k=0$ becomes here $\partial_x v_x + v_n C_{xx}=0$. Since calculations are done to first order in the velocity field $\mathbf{v}$ and in the surface height gradient $|\partial_x h|$, this condition becomes $\partial_x v_x\simeq0$ and, without loss of generality since this fixes a condition on rigid translation of the system, $v_x=0$. This implies that at this order, the change in system size $dL/dt=v_x(x=L)-v_x(x=0)$ can be neglected.

Using Eqs. \ref{eq:application_UD_tij}-\ref{eq:application_UD_mij} and the molecular field expression Eqs \ref{eq:appendix_application_hx}-\ref{eq:appendix_application_hy}, we obtain to first order in the curvature tensor and in the flow field $v_y$ the expression for the components $t_{xy}$ and $\bar{m}_{xx}$:
\begin{align}
t_{xy}
 =& 2\eta\omega_n \nonumber\\
 &+ \frac{\nu-\beta C_{xx}}{2}(-(\lambda_p+K (\partial_x\theta)^2) \sin2\theta +K  \cos2\theta \partial_x^2 \theta ) \nonumber\\
 &+ \frac{(\zeta_n+(\tilde{\zeta}'_n - \zeta_{c{\rm n}}) C_{xx}) \Delta\mu}{2} \sin2\theta-\frac{1}{2}K \partial_x^2 \theta ~,\label{eq:application_appendix_txy}\\
 \bar{m}_{xx} = & (\eta_c +\eta_{cb})\partial_t C_{xx} +(\kappa + K \cos^2\theta) C_{xx} \nonumber\\
 &+\zeta_c\Delta\mu+\frac{\zeta_{cn}}{2} \Delta\mu \cos 2\theta\nonumber\\
&+\beta \left(-(\lambda_p+K(\partial_x \theta)^2)\frac{\cos 2\theta}{2} - \frac{K}{2} \sin 2 \theta \partial_x^2 \theta \right)\label{eq:application_appendix_mxx}~.
\end{align}
where we have used the notation $p_x=\cos\theta$, $p_y=\sin\theta$. We have also used that to first order in the velocity field and in the curvature tensor, $D_t C_{xx}\simeq \partial_t C_{xx}$.
 Using the polarity dynamic equation, Eq. \ref{eq:application_UD_dpdt}, we obtain:
\begin{align}
\partial_t \theta =&\frac{K}{\gamma}\partial_x^2 \theta  + \omega_n (1-\nu \cos2\theta )  -\frac{1}{2}\lambda_{\rm UD} \Delta\mu \sin 2\theta C_{xx}\nonumber\\
&+\beta \frac{\sin2\theta}{4}  \partial_t C_{xx} \label{eq:application_appendix_dttheta}~,
\end{align}
and
\begin{align}
\lambda_p=&-K\left(\partial_x \theta\right)^2-\gamma\omega_n \nu \sin2\theta- \frac{1}{2}\gamma  \beta\cos^2\theta  \partial_t C_{xx}\nonumber\\
&+\gamma\lambda_{\rm UD} \Delta \mu C_{xx} \cos^2\theta   ~.
\end{align}
Setting $t_{xy}=0$ in Eq. \ref{eq:application_appendix_txy} gives an expression for the gradient of flow $\omega_n$, while setting $\bar{m}_{xx}=0$ in Eq. \ref{eq:application_appendix_mxx} gives a dynamic equation for the curvature time derivative, $\partial_t C_{xx}$. Combined with Eq. \ref{eq:application_appendix_dttheta}, one obtains two coupled dynamic equations for $\theta$ and $C_{xx}$. These equations are lengthy but take a simpler form in the limit $\nu=\beta=0$ and $\kappa \gg K$(Eqs. \ref{eq:weakly_deformed_dtCxx}-\ref{eq:weakly_deformed_dttheta}).

We now discuss the stability of uniform, steady-state solutions of Eqs. \ref{eq:weakly_deformed_dtCxx}-\ref{eq:weakly_deformed_dttheta}. This analysis is carried out more easily in the limit of fast shape relaxation, $\eta_c\rightarrow 0$, $\eta_{cb}\rightarrow 0$. In that case $C_{xx}$ relaxes quasi-statically to the steady-state solution of Eq. \ref{eq:weakly_deformed_dtCxx} and the polarity equation becomes:
\begin{align}
\partial_t \theta 
=&\frac{K(4\eta+\gamma)}{4\eta\gamma}\partial_x^2\theta+\sin2\theta \bar{\lambda}_{\rm UD} \frac{2 \bar{\zeta}_c+\zeta_{cn} \cos 2\theta }{4\kappa}\Delta\mu^2 ~,
\end{align}
where for convenience we have introduced the reduced parameters $\bar{\lambda}_{\rm UD}=\lambda_{\rm UD}+(\tilde{\zeta}_{\rm n}'-\zeta_{c{\rm n}})/(2\eta)$ and $\bar{\zeta}_c=\zeta_c - \zeta_{\rm n}\kappa/(2\eta \bar{\lambda}_{\rm UD} \Delta\mu)$.

We now linearize around the steady-state solutions, $\theta=\theta_s+\delta \theta $, and consider only uniform perturbations. For $\theta_s=\pm\frac{\pi}{2}$ one obtains:
\begin{align}
\partial_t \delta \theta=- \frac{\bar{\lambda}_{\rm UD}  (2 \bar{\zeta}_c-\zeta_{c\rm n})(\Delta\mu)^2 }{2 \kappa } \delta \theta~,
\end{align}
so that the steady-state solutions $\theta_s=\pm\frac{\pi}{2}$ are stable for $\bar{\lambda}_{\rm UD} (2 \bar{\zeta}_c -\zeta_{c\rm n})>0$. For $\bar{\lambda}_{\rm UD} >0$, this corresponds to  $\zeta_{c\rm n} <2 \bar{\zeta_c} $.  For $\bar{\lambda}_{\rm UD} <0$, this corresponds to  $\zeta_{c\rm n} >2\bar{ \zeta}_c $. 

Around $ \theta_s=0$ or $\theta_s=\pi$ one obtains:
\begin{align}
\partial_t \delta \theta= \frac{ \bar{\lambda}_{\rm UD} (2 \bar{\zeta}_c+\zeta_{c\rm n})(\Delta\mu)^2 }{2 \kappa} \delta \theta~,
\end{align}
which is stable for $\bar{\lambda}_{\rm UD} (2\bar{ \zeta}_c+\zeta_{c\rm n})<0$. For $\bar{\lambda}_{\rm UD} >0$, this corresponds to  $\zeta_{c{\rm n}} <-2\bar{\zeta}_c$.  For $\bar{\lambda}_{\rm UD} <0$, this corresponds to  $\zeta_{c\rm n}  >-2\bar{\zeta}_c$. 

Around $\theta_s=\theta_0$ with $\theta_0$ defined by $\cos 2\theta_0 =- 2 \bar{\zeta}_c/\zeta_{c\rm n}$ (see Eq. \ref{eq:application_UD_first_steady_state_solution}; a solution exists only if $|\zeta_{c\rm n}/\bar{\zeta}_c|>2$), one obtains:
\begin{align}
\partial_t \delta \theta = &- \frac{\bar{\lambda}_{\rm UD} \zeta_{c\rm n} \left(1-\frac{4\bar{\zeta}_c^2}{\zeta_{c\rm n}^2}\right) (\Delta\mu)^2 }{2 \kappa }\delta \theta
\end{align}
Here the corresponding steady-state solutions are stable for $\bar{\lambda}_{\rm UD}\zeta_{ c\rm n}  \left(\left(\frac{\zeta_{ c\rm n}}{\bar{\zeta}_c}\right)^2 - 4\right)>0$.

For an inhomogeneous $\theta$ perturbation around $\theta_s=\frac{\pi}{2}$, one has:
\begin{align}
\partial_t \delta \theta=\frac{K(4\eta+\gamma)}{4\eta\gamma}\partial_x^2\delta \theta- \frac{\bar{\lambda}_{\rm UD}  (2 \bar{\zeta}_c-\zeta_{c\rm n})(\Delta\mu)^2 }{2 \kappa } \delta \theta~.
\end{align}
Considering a confined system with planar anchoring boundary conditions $\theta(0) = \theta(L) = \frac{\pi}{2}$, one has $\delta \theta(0) = \delta\theta(L) =0$. The most unstable mode $\delta\theta = A\sin ( \pi \frac{x}{L})$ then has an amplitude $A$ with the dynamics:
\begin{align}
\partial_t A = -\left[\frac{K (4\eta+\gamma)\pi^2}{4 \gamma \eta L^2}+\frac{\bar{\lambda}_{\rm UD}  (2 \bar{\zeta}_c-\zeta_{c\rm n})(\Delta\mu)^2 }{2 \kappa } \right]A~.
\end{align}
If $\bar{\lambda}_{\rm UD}  (2 \bar{\zeta}_c-\zeta_{c\rm n})>0$, this mode is always stable. if $\bar{\lambda}_{\rm UD}  (2 \bar{\zeta}_c-\zeta_{c\rm n})<0$, the system becomes unstable above a critical length $L>L^*$, with $L^*$ given by
\begin{align}
L^* = \sqrt{\frac{ \kappa K \pi^2 (4\eta+\gamma)}{2 \eta\gamma\bar{\lambda}_{\rm UD}  (2 \bar{\zeta}_c-\zeta_{c\rm n})(\Delta\mu)^2}}~.
\end{align}
\bibliographystyle{unsrt}
\bibliography{ActiveSurfaces}
\end{document}